\documentclass[fleqn,usenatbib]{mnras}

\usepackage{newtxtext,newtxmath}
\usepackage{float}
\usepackage{pdflscape}

\usepackage[T1]{fontenc}

\usepackage{graphicx}	
\usepackage{amsmath}	
\usepackage{multirow}
\usepackage{soul}
\usepackage{flexisym}
\usepackage{breqn}
\usepackage{pifont}%
\newcommand{\xmark}{\ding{55}}

\DeclareRobustCommand{\VAN}[3]{#2}
\let\VANthebibliography\thebibliography
\def\thebibliography{\DeclareRobustCommand{\VAN}[3]{##3}\VANthebibliography}

\title{Searching for Magnetar Binaries Disrupted by Core-Collapse Supernovae}

\author[M. B. Sherman et al.]{
Myles B. Sherman,$^{1}$\thanks{E-mail: msherman@caltech.edu}
Vikram Ravi,$^{1}$
Kareem El-Badry,$^{1}$
Kritti Sharma,$^{1}$
Stella Koch Ocker,$^{1,2}$
\newauthor{Nikita Kosogorov,$^{1}$
Liam Connor,$^{1}$
Jakob T. Faber$^{1}$}
\\
$^{1}$Cahill Center for Astronomy and Astrophysics, MC 249-17 California Institute of Technology, Pasadena CA 91125, USA.\\
$^{2}$The Observatories of the Carnegie Institution of Washington, Pasadena, CA 91101, USA.\\
}

\date{Accepted XXX. Received YYY; in original form ZZZ}

\pubyear{2024}

\begin{document}
\label{firstpage}
\pagerange{\pageref{firstpage}--\pageref{lastpage}}
\maketitle




\begin{abstract}
Core-collapse Supernovae (CCSNe) are considered the primary magnetar formation channel, with 15 magnetars associated with supernova remnants (SNRs). A large fraction of these should occur in massive stellar binaries that are disrupted by the explosion, meaning that $\sim45\%$ of magnetars should be nearby high-velocity stars. Here we conduct a multi-wavelength search for unbound stars, magnetar binaries, and SNR shells using public optical ($uvgrizy-$bands), infrared ($J-$, $H-$, $K-$, and $K_s-$bands), and radio ($888$\,MHz, $1.4$\,GHz, and $3$\,GHz) catalogs. We use Monte Carlo analyses of candidates to estimate the probability of association with a given magnetar based on their proximity, distance, proper motion, and magnitude. In addition to recovering a proposed magnetar binary, a proposed unbound binary, and 13 of 15 magnetar SNRs, we identify two new candidate unbound systems: an OB star from the \textit{Gaia} catalog we associate with SGR\,J1822.3-1606, and an X-ray pulsar we associate with 3XMM\,J185246.6+003317. Using a Markov-Chain Monte Carlo simulation that assumes all magnetars descend from CCSNe, we constrain the fraction of magnetars with unbound companions to $5\lesssim f_u \lesssim 24\%$, which disagrees with neutron star population synthesis results. Alternate formation channels are unlikely to wholly account for the lack of unbound binaries as this would require $31\lesssim f_{nc} \lesssim 66\%$ of magnetars to descend from such channels. Our results support a high fraction ($48\lesssim f_m \lesssim 86\%$) of pre-CCSN mergers, which can amplify fossil magnetic fields to preferentially form magnetars.
\end{abstract}



\begingroup
\let\clearpage\relax
\endgroup
\newpage


\section{Introduction}\label{sec:intro}

Magnetars are highly magnetized ($B \gtrsim 10^{14}$\,G) neutron stars which emit X-ray, gamma ray, or radio emission driven by their magnetic fields \citep{kaspi2017magnetars}. At the time of submission, there are 24 confirmed magnetars and 7 magnetar candidates identified based on their high-energy X-ray and gamma ray bursts \citep{richardson2023high,2000ApJ...531..407C,lamb2003sgr,mereghetti2012magnetar,torii1998discovery,sakamoto2011probing,leahy2007distance}\footnote{While a number of Gamma Ray Bursts (GRBs) in nearby galaxies have been proposed as magnetar Giant Flares \citep[e.g.][]{trigg2023grb,burns2021identification,svinkin2021bright,mazets2008giant,frederiks2007possibility,minaev2024grb,mereghetti2024magnetar}, we focus this work on magnetars and candidates within the Galaxy and Magellanic Clouds.}. With burst events initially categorized as either Anomalous X-ray Pulsars (AXPs) or Soft Gamma Repeaters (SGRs), magnetars were later identified as the common source of their emission \citep{gavriil2002magnetar}. 

It is unclear whether magnetars constitute a unique class of neutron star or instead that neutron stars exhibit a continuum of magnetar-like features depending on their magnetic fields and environments. The latter case would be supported by the magnetar-like activity (glitches, X-ray outbursts, and high magnetic field strengths) observed in some pulsars \citep[e.g.][]{2008Sci...319.1802G,rea2016magnetar,
archibald2016magnetar,beniamini2023evidence}. In this work, we will consider magnetars to be a class of highly-magnetized neutron star distinguished by their magnetic field decay-powered emission.

Magnetars, like pulsars, are believed to be formed when massive stars undergo Core-collapse Supernovae (CCSNe, formed from Type-II, Type-Ib, and Type-Ic supernovae), with 15 magnetars or candidates having confirmed or potential associations with Supernova Remnants (SNRs) \citep[e.g.,][]{baade1934super,wheeler1966superdense,
olausen2014mcgill,nomoto1995evolution,thielemann1996core,barker2023inferring,truelove1999evolution}. However, what factors determine whether a CCSN leaves behind a magnetar or a normal neutron star remain unclear \citep[e.g.,][]{giacomazzo2013formation, giacomazzo2015producing, fuller2022spins, revnivtsev2016magnetic, popov2020high}. Furthermore, the CCSN model may be inconsistent with trends in the growing neutron star population. \citet{keane2008birthrates} discuss the birth rates of various species of galactic neutron star, including radio and X-ray pulsars, Rotating Radio Transients (RRATs), X-ray Dim Isolated
Neutron Stars (XDINs), and magnetars. They find that while the total neutron star birth rate is between $5-10$\,century$^{-1}$, the CCSN rate is only $1.9 \pm 1.1$\,century$^{-1}$, suggesting either our understanding of evolutionary links among neutron star species, our CCSN model, or our birth rate estimate is incomplete \citep[e.g.][]{popov2010isolated,keane2008birthrates,
kaspi2010grand}. CCSNe are also unlikely to account for the population of young pulsars in Galactic globular clusters \citep[e.g.][]{2023MNRAS.525L..22K}.


Magnetar birth rates $\beta \approx N_{\rm mag}\mathcal{T}^{-1}\sim0.3$\,century$^{-1}$, are difficult to constrain due to less confident age $\mathcal{T}$ and population size $N_{\rm mag}$ estimates. Most notably, the characteristic spindown age $\mathcal{T}_{\rm rot} \propto P(2\dot{P})^{-1}\sim10$\,kyr$-10$\,Myr used for radio pulsars (where $P$ is the pulse period and $\dot{P}$ is the spindown rate) does not incorporate the magnetic field decay on $\sim10$\,kyr timescales thought to drive magnetar emission\footnote{In this work, we adopt the notation $\mathcal{T}_{\rm rot}$ for the characteristic age as opposed to the more widely used $\tau_c$ to distinguish it from estimates of the angular travel time used in Section~\ref{sec:gaia} and Appendix~\ref{app:pvals_GAIA}.}. Therefore, $\mathcal{T}_{\rm rot}$ can significantly overestimate the true age \citep[e.g.][]{nakano2015suzaku,ferrario2008origin,pons2007magnetic}. While this is the major effect, `glitches', or abrupt decreases in rotation period $P$ \citep[e.g.][]{tong2020magnetar,younes2023magnetar,livingstone2011spin}, can also complicate magnetar age estimation when data is limited \citep[e.g. for transient magnetars as discussed in ][]{rea2016magnetar,ibrahim2024xray,lower20232022,scholz2017spin}, as can `anti-glitches' and `glitch recovery' for which $P$ increases \citep[e.g.][]{scholz2014long,archibald2013anti,olausen2014mcgill}. For these reasons, magnetic field decay ages $\mathcal{T}_{\rm decay}\sim1-100$\,kyr derived from X-ray luminosity and magnetic field strength may be preferred, although these are model-dependent, and may not account for rapid initial spindown models \citep[see Appendix~\ref{app:decay} in this work, and e.g.][for examples of decay models]{ferrario2008origin,mondal2021life,arras2004magnetars}. The ages of associated SNRs $\mathcal{T}_{\rm SNR}\sim1$\,kyr$-100$\,kyr are more accurate, but can cover a wide range and are only known for a small fraction of magnetars \citep[][]{suzuki2021quantitative,borkowski2017expansion,lyman2022fast,gaensler1999new}. 

While \citet{muno2008search} estimate that $N_{\rm mag}\lesssim540$ magnetars reside in the Galaxy, a number of factors make this uncertain. First, it is unclear how magnetar beaming affects the observed population, particularly due to the small sample size \citep[e.g.][]{keane2008birthrates}. Furthermore, 
the detection of a globular cluster (M81) Fast Radio Burst (FRB), thought to originate in a magnetar's magnetosphere, offers a counterexample to our assumption that $N_{\rm mag}$ scales with star formation, and supports the notion of multiple magnetar formation channels \citep{kirsten2022repeating,lu2022implications,petroff2022fast}. The evolutionary connections among magnetars, pulsars, XDINs, RRATs, and isolated neutron stars are also unclear, leading to potential double-counting errors in the \citet{keane2008birthrates} birthrates. Thus, the continued search for and recovery of CCSNe and SNRs will help inform our models and improve age estimates.




SNR radio emission surrounding a magnetar is the most direct indicator of a CCSN progenitor \citep[e.g.,][]{1996ASPC..105..375K}. Supernova shells are detectable in radio and X-rays expanding from a young ($1-10$\,kyr) magnetar's birth site at its center; most of the known magnetar-SNR associations were identified by their remnant shell radio emission \citep[e.g.][]{2007ApJ...667.1111G,2012ApJ...751...53A}. The significant increase in data from radio all-sky surveys will boost the sensitivity considerably. The Rapid ASKAP Continuum Survey (RACS), for example, uses the 36-antenna Australian Square Kilometre Array Pathfinder (ASKAP) telescope to survey the southern sky (declination range $-80^\circ < \delta < +30^\circ$) from 700–1800 MHz \citep{mcconnell2020rapid}. The Karl G. Jansky Very Large Array (VLA) Sky Survey (VLASS) and National Radio Astronomy Observatory (NRAO) VLA Sky Survey (NVSS) similarly cover the northern sky above $\delta > -40^\circ$ at 1.4 and 3\,GHz \citep{2020RNAAS...4..175G,2021ApJS..255...30G,2023ApJS..267...37G,condon1998nrao}, while the Low Frequency Array (LOFAR) covers lower frequencies near 135\,MHz \citep{shimwell2022lofar,vo:LoTSS_PDR_images,condon1998nrao,2020RNAAS...4..175G,2021ApJS..255...30G,2023ApJS..267...37G}. Identifying supernova shells through radio emission relies on the magnetar being young enough that the shell has not yet dissipated ; for expansion rates up to $\sim1000$\,km\,s$^{-1}$, SNRs with ages $\lesssim100\,$kyr may be detected \citep[e.g.,][]{2012ApJ...751...53A,truelove1999evolution,sedov2018similarity}.

In recent years, the presence (or absence) of binary magnetar systems has become a powerful probe of CCSN formation channels for the broader neutron star population. \citet{chrimes2022magnetar}, \citet{renzo2019massive}, \citet{kochanek2019stellar}, and \citet{moe2017mind} conducted population synthesis simulations with massive stellar binaries to estimate what fraction of primary stars would remain bound to their secondaries as neutron stars following their CCSN. \citet{moe2017mind} estimate that $F_0\sim84\%$ of massive stars should initially be in binaries, while \citet{kochanek2019stellar} predicts that $f_m\sim48\%$ of these merge prior to core-collapse\footnote{Note this fraction includes only mergers where both stars are massive OB stars. For a discussion of CCSNe from mergers of intermediate mass stars, see \citet{zapartas2017delay}.}; the other $1-f_m\sim52\%$ undergo CCSNe as non-merged binaries. \citet{renzo2019massive} suggest a lower merger rate $f_m\sim22^{+26}_{-9}\%$. They also find that $\sim14^{+22}_{-10}\%$ of the binaries that do not merge remain bound following the CCSN. Thus only $f_b\sim 6-9\%$ of all neutron stars should have massive binary companions. \citet{chrimes2022magnetar} affirm these results, obtaining a bound neutron star fraction $f_b\sim 5\%$. Notably, the binary statistics of magnetars may deviate from those above if they form from distinct progenitor populations, such as those with enhanced rotation or magnetic fields, though this has not been explored extensively through population synthesis \citep[e.g.][]{popov2006progenitors}\footnote{We caution that the results of \citet{popov2006progenitors} may be biased by their definition of `magnetars', which they define as evolutionary paths that include ``Coalescence prior to NS formation" or ``Roche lobe overflow", rather than using an independent condition on their magnetic field strength. Therefore, the merger fractions and binary fractions from Tables~1 and 2 of \citet{popov2006progenitors} should not be directly compared to those from \citet{chrimes2022magnetar}, \citet{renzo2019massive}, \citet{kochanek2019stellar}, \citet{moe2017mind}, or this work.}.



Taking a census of young, non-degenerate neutron star-massive star binaries, 7 pulsars with bound OB star companions have been reported in the Australia Telescope National Facility\footnote{\url{https://www.atnf.csiro.au/research/pulsar/psrcat/}} (ATNF) catalog, while SGR-0755-2933 is a tentatively confirmed magnetar binary \citep{chrimes2022magnetar,richardson2023high,doroshenko2021sgr}. A candidate stellar companion was also proposed for magnetar CXOU\,J171405.7-381031 using Infrared (IR) Hubble Space Telescope (HST) images, though the IR counterpart's stellar nature is unconfirmed \citep{chrimes2022magnetar}. The two proposed binaries out of 31 magnetars are marginally consistent with the $f_b\sim5-9\%$ expected to be in bound systems, while the 9 total neutron stars with massive companions out of the total catalog of $\sim2950$ neutron stars marginally underestimate the binary fraction due to selection effects \citep{lorimer2008binary}\footnote{Our young neutron star census includes 2903 slow (period $P>30$\,ms) pulsars from the ATNF catalog, $\sim10$ thermally discovered isolated neutron stars including the \textit{Magnificent Seven} ROSAT sample \citep{pires2009isolated,schwope1998isolated,haberl2007magnificent,turolla2009isolated,tetzlaff2011origin,avakyan2023xrbcats}, and 31 magnetars and magnetar candidates from the McGill magnetar catalog. We acknowledge this may be incomplete; for example, not all pulsars have sufficient timing solutions to rule out binary companions, and some may be old enough that their companions have undergone supernovae, as well. However, this provides a reasonable estimate of the number of known young neutron stars to date. We exclude older populations such as High- and Low-Mass X-ray Binaries (HMXBs, LMXBs) and millisecond pulsars since they are only observable in binary systems, and are therefore subject to selection effects \citep[e.g.][]{lorimer2008binary}. }.


However, one expects a much larger fraction ($f_u\sim 38-56\%$) of `unbound' neutron star binaries disrupted by the CCSN, motivating a search for `walkaway' or `runaway' stellar companions \citep{renzo2019massive}. Considering a physical scenario, a massive companion ejected at $\sim 30$\,km\,s$^{-1}$ \citep{renzo2019massive} could travel $\sim 0.03$\,pc over $\sim1$\,kyr. At a distance of $5-10$\,kpc, this should appear only $\sim1.2-2.5\arcsec$ away from its previously bound magnetar. \citet{kochanek2019stellar} and \citet{kochanek2018cas, kochanek2021supernovae,kochanek2023search} have used the \textit{Gaia} optical catalog to search for both runaway and bound stars nearby 43 SNRs with compact objects, including central pulsars, HMXBs, LMXBs, radio sources, and gamma ray sources. Only the pulsar PSR\,J0538+2817 was confidently associated with a runaway companion star \citep{dinccel2015discovery,kochanek2021supernovae}. \citet{clark2014vlt} note that the Wolf-Rayet star WR\,77F (Wd1-5) may be a runaway unbound companion of magnetar CXOU\,164710.2-455216, but the absence of a magnetar proper motion measurement make this difficult to confirm. More general searches for runaway stars with high velocities relative to field stars have also been carried out, but haven't identified magnetar or pulsar companions \citep[e.g.][]{carretero2023galactic}. With $\sim2950$ neutron stars and 31 magnetars, one expects $\sim1100-1700$ and $\sim11-18$ unbound companions, respectively\footnote{See Table~\ref{table:magstat} for definitions of $f_b$, $f_m$, $f_u$, and other relevant fractions. These will be discussed in detail in Section~\ref{sec:stats}.}. 

A comprehensive search of archived data can leverage catalogs like \textit{Gaia}, the Two Micron All Sky Survey (2MASS), and Panoramic Survey Telescope and Rapid Response System (Pan-STARRS or PS1), as well as surveys such as the SkyMapper Southern Survey (SMSS), UKIRT Infrared Deep Sky Survey (UKIDSS) and the Visible and Infrared Survey Telescope for Astronomy (VISTA) Variables in the V{\'i}a L{\'a}ctea (VVV) to find stellar companions \citep{Flewelling_2020, 2006AJ....131.1163S, chambers2019panstarrs1, brown2021gaia, lawrence2007ukirt, kaiser2002pan, nikzat2022vvv,wolf2018skymapper,onken2024skymapper}. For example, \citet{chrimes2023searching} predict the ability of \textit{Gaia} Data Release 3 (DR3), as well as future releases from HST, the James Webb Space Telescope (JWST), \textit{Euclid}, and the Nancy Grace Roman Space Telescope (NGRST) to detect unbound stellar objects. They find that while \textit{Gaia} is limited by extinction ($A_V \lesssim 10$), its absolute reference capability and proper motion measurements could increase the number of unbound companion discoveries by $\sim8\times$ when combined with next generation IR telescope data. \textit{Gaia} data have also been used effectively to search for unbound companions of White Dwarfs and neutron stars \citep{kochanek2019stellar,el2023fastest,2018ApJ...865...15S}.

In this paper, we conduct a search for unbound binary companions of magnetars using the \textit{Gaia} catalog, supplemented by a search for bound companions and CCSN remnants. In Section~\ref{sec:sample} we describe the magnetar sample used for the search. Section~\ref{sec:gaia} describes the search and trajectory analysis of sources in the \textit{Gaia} archive for unbound magnetar companions and results. In Section~\ref{sec:bound} we describe the methods and results of the optical and IR search for bound magnetar companions. In Section~\ref{sec:RACS} we summarize the radio image search using the RACS, VLASS, and NVSS catalogs. Section~\ref{sec:discussion} discusses the sensitivity, completeness, and implications of our search. Section~\ref{sec:conclusion} concludes with a summary and outlook on next steps for analyzing magnetar formation. Section~\ref{sec:data} provides information regarding data availability. Details on the statistical analysis are included in the Appendices.

\section{Magnetar Sample}\label{sec:sample}

In Table~\ref{table:magnetars}, we present the sample of 24 magnetars listed in the McGill magnetar catalog\footnote{\url{http://www.physics.mcgill.ca/~pulsar/magnetar/main.html}}, as well as 7 unconfirmed magnetar candidates \citep{olausen2014mcgill}. For each we list the Right Ascension $(\alpha)$, declination $(\delta)$, distance ($d$), and proper motion ($\mu$) from the available literature. We also provide the $V$-band dust extinction estimates along each magnetar's line-of-sight, which we derive using the \textit{Bayestar19}\footnote{\url{https://dustmaps.readthedocs.io/en/latest/modules.html}} dust map for magnetars above declination $\delta \gtrsim -30^{\circ}$ \citep{green2015three,schlafly2015three,green2018galactic,2019ApJ...887...93G}. For magnetars outside this range we use the \citet{predehl1995x} relation between the X-ray column density, $N_H$, and $A_V$ if an $N_H$ measurement is available. If not, the average photometric extinction from \textit{Gaia} Data Release 3 (DR3) sources within $1^\circ$ is given \citep{2016A&A...595A...1G, gaiacollaboration2022gaia}. This will likely underestimate the true extinction, since most \textit{Gaia} sources are in the foreground. 12 magnetars and 3 magnetar candidates have secure or proposed SNR associations.


\begin{table*}
 \begin{center}
 \scriptsize
 \caption{Magnetars and Magnetar Candidates from the McGill Catalog (revised position errors are boldfaced for clarity)}
 \label{table:magnetars} 
 \begin{tabular}{c|c|c|c|c|c|c|c|c|c}
  \hline
  \textbf{Magnetar} & $\alpha$ & $\sigma_\alpha (\arcsec)$ & $\delta$ & $\sigma_\delta (\arcsec)$ & \begin{tabular}{@{}c@{}} \textbf{Distance} \\ \textbf{(kpc)} \\ \end{tabular} & \begin{tabular}{@{}c@{}} $\mu_{\alpha}{\rm cos}(\delta)$ \\ \textbf{(mas\,yr$^{-1}$)} \\ \end{tabular} & \begin{tabular}{@{}c@{}} $\mu_{\delta}$ \\ \textbf{(mas\,yr$^{-1}$)} \\ \end{tabular} & $A_V$ & \textbf{Refs.}\\
  \hline
    CXOU\,J010(043.1-721134) & $\rm01^{\rm h}00^{\rm m}43\fs14$ & $0.60$ & $-72^\circ11\arcmin33\farcs80$ & $0.60$ & $62.4 \pm 1.6$ & -- & -- & ${0.35^{+0.11}_{-0.09}}^\ddagger$ & 1,2\\
    4U\,0142(+61) & $\rm01^{\rm h}46^{\rm m}22\fs407$ & $0.50$ & +61$^\circ45\arcmin03\farcs19$  & $0.50$ & $3.6\pm 0.4$ & $-4.1\pm1.0$& $1.9\pm1$ & $3.81^{+0.51}_{-0.16}$ & 3,4,5\\
    SGR\,0418(+5729) & $\rm04^{\rm h}18^{\rm m}33\fs867$ & $0.35$ &  +57$^\circ32\arcmin22\farcs91$ & $0.35$ & $\sim 2$ & -- & -- & $1.60$ & 6\\
    \textbf{SGR\,0501(+4516)}$^{*}$ & $\rm05^{\rm h}01^{\rm m}08\fs00$ & $0.11$ & +45$^\circ16\arcmin31\farcs00$ & $0.11$ & $2.0\pm0.3$ & -- & -- & $2.82^{+0.10}_{-0.42}$ & 7,8,9,73\\
    \textbf{SGR\,0526(-66)}$^{*}$ & $\rm05^{\rm h}26^{\rm m}00\fs89$ & $0.60$ & -66$^\circ04\arcmin36\farcs30$ & $0.60$ & $53.6\pm1.2$ & -- & -- & ${3.37^{+0.32}_{-0.33}}^\ddagger$ & 1,10,11\\ 
    1E\,1048.1(-5937) & $\rm10^{\rm h}50^{\rm m}07\fs14$ & $0.60$ & -59$^\circ53\arcmin21\farcs40$ & $0.60$ & $9.0 \pm 1.7$ & -- & -- & $5.42\pm0.06^\ddagger$ & 3,12,13,76,77,78,79\\
    \textbf{1E\,1547.0(-5408)} & $\rm15^{\rm h}50^{\rm m}54\fs12$ & $0.0056$ & -54$^\circ18\arcmin24\farcs11$ & $0.0020$ & $3.91\pm0.07$ & $4.8\pm0.5$ & $-7.9\pm0.3$ & $17.88\pm1.12^\ddagger$ & 14,15,16 \\
    \textbf{PSR\,J1622(-4950)} & $\rm16^{\rm h}22^{\rm m}44\fs89$ & $0.80$ & -49$^\circ50\arcmin52\farcs70$& $0.80$ & $9.00\pm1.35$ & -- & -- & ${30.17^{+8.94}_{-7.82}}^\ddagger$ & 17,18\\
    \textbf{SGR\,1627(-41)} & $\rm16^{\rm h}35^{\rm m}51\fs84$ & $0.60$ & -47$^\circ35\arcmin23\farcs31$ & $0.60$ & $11.0\pm0.3$ & -- & -- & $55.87\pm11.17^\ddagger$ & 19,20\\
    CXOU\,J1647(10.2-455216 )& $\rm16^{\rm h}47^{\rm m}10\fs20$ & $0.30$ & -45$^\circ52\arcmin16\farcs90$ & $0.30$ & $3.9 \pm 0.7 $ & -- & -- & $13.35\pm0.28^\ddagger$ & 21,22\\
    1RXS\,J1708(49.0-400910) & $\rm17^{\rm h}08^{\rm m}46\fs87$ & $0.80$ & -40$^\circ08\arcmin52\farcs44$ & $0.80$ & $3.8 \pm 0.5$ & -- & -- & $7.60\pm0.22^\ddagger$& 3,23 \\
    \textbf{CXOU\,J1714(05.7-381031)} & ${\rm17^{\rm h}14^{\rm m}05\fs74}$ & $0.60$ & -38$^\circ10\arcmin30\farcs90$& $0.60$ & $9.8\pm1.5$ & -- & -- & ${22.07^{+0.84}_{-0.78}}^\ddagger$ & 24,25,26,69\\
    \textbf{SGR\,J1745(-2900)}$^{*}$ & $\rm17^{\rm h}45^{\rm m}40\fs16$ & $0.02$ & -29$^\circ00\arcmin29\farcs82$ & $0.09$ & $8.3\pm0.3$ & $2.44\pm0.33$ & $5.89\pm0.11$ & $3.01$ & 27,28,29,30,74,75\\
    SGR\,1806(-20)& $\rm18^{\rm h}08^{\rm m}39\fs34$ & $0.04$ & -20$^\circ24\arcmin39\farcs85$ & $0.04$ & $8.7_{-1.5}^{+1.8}$ & $-4.5\pm1.4$ & $-6.9\pm2.0$ & $6.66\pm0.06$ & 31,32,33\\
    XTE\,J1810(-197)& $18^{\rm h}09^{\rm m}51\fs08$ & $0.0003$ & $-19^\circ43\arcmin52\farcs14$ & $0.0005$ & $2.5_{-0.3}^{+0.4}$ & $-3.79\pm0.04$ & $-16.2\pm0.1$ & $3.71^{+0.03}_{-0.10}$ & 34,35,70,87\\
    Swift\,1818(.0-1607) & $\rm18^{\rm h}18^{\rm m}03\fs70$ & $2.70$ & -16$^\circ07\arcmin31\farcs80$ & $2.70$ & $\sim 4.8$ & $-3.54\pm0.05$ & $-7.65\pm0.09$ & $\sim 3.78$ & 36,37,38\\
    Swift\,J1822(.3-1606) & $\rm18^{\rm h}22^{\rm m}18\fs00$ & $1.80$ & -16$^\circ04\arcmin26\farcs80$ & $1.80$ &  $1.6\pm 0.3$ & -- & -- &$2.02^{+1.47}_{-1.09}$ & 39,40\\
    SGR\,1833(-0832) & $\rm18^{\rm h}33^{\rm m}44\fs37$ & $0.30$ & -08$^\circ31\arcmin07\farcs50$ & $0.30$ & -- & -- & -- & $\sim 3.16^\dagger$ & 41 \\
    \textbf{Swift\,J1834(.9-0846)}& $\rm18^{\rm h}34^{\rm m}52\fs12$ & $0.60$ &-08$^\circ45\arcmin56\farcs02$ & $0.60$ & $4.2\pm0.3$ & -- & -- & $4.03\pm0.06$ & 42,43\\
    \textbf{1E\,1841(-045)} & $\rm18^{\rm h}41^{\rm m}19\fs34$ & $0.30$ & -04$^\circ56\arcmin00\farcs69$ & $0.30$ & $8.5\pm1.3$ & -- & -- & $3.76\pm0.06$& 44,45,46,80,81 \\ 
    \textbf{3XMM\,J1852(46.6+003317)} & $\rm18^{\rm h}52^{\rm m}46\fs67$ & $2.4$ & +00$^\circ33\arcmin17\farcs80$ & $2.4$ & $\sim 7.1$ & -- & -- & $\sim 6.30$ & 47 \\
    SGR\,1900(+14) & $\rm19^{\rm h}07^{\rm m}14\fs33$ & $0.15$ & +09$^\circ19\arcmin20\farcs1$ & $0.15$ & $12.5\pm1.7$ & $-2.1\pm0.4$ & $-0.6\pm0.5$ & $10.94^{+0.06}_{-0.03}$ & 48,49,33 \\
    \textbf{SGR\,1935(+2154)} & $\rm19^{\rm h}34^{\rm m}55\fs60$ & $0.70$ & +21$^\circ53\arcmin47\farcs79$ & $0.70$ & $9.0\pm2.5$ & -- & -- & $8.58^{+0.16}_{-1.98}$ & 50,51,52,82,83\\
    \textbf{1E\,2259(+586)} & $\rm23^{\rm h}01^{\rm m}08\fs30$ & $0.70$ & +58$^\circ52\arcmin44\farcs45$ & $0.70$ & $3.2\pm2.0$ & $-6.4\pm0.6$ & $-2.3\pm0.6$ & $2.59^{+0.38}_{-1.79}$ & 53,54,55,5,84\\
    \hline
    \textit{SGR\,0755(-2933)} & $\rm07^{\rm h}55^{\rm m}42\fs48$ & $180$ & -29$^\circ33\arcmin53\farcs46$ & $180$ & $3.5\pm2.0$ & -- & -- & $1.82^{+0.32}_{-1.18}$ & 56,57,58,72\\ 
    \textbf{\textit{SGR 1801(-23)}}$^{*}$ & $\rm18^{\rm h}00^{\rm m}59s$ & $13680$ & -22$^\circ56\arcmin48\farcs00$ & $13680$ & -- & -- & -- & $\sim 1.95^\dagger$ & 59\\
    \textit{SGR\,1808(-20)} & $\rm18^{\rm h}08^{\rm m}11\fs20$ & $414$ & -20$^\circ38\arcmin49\farcs00$& $414$ & -- & -- & -- & $\sim 2.82^\dagger$ & 60\\
    \textit{AX\,J1818(.8-1559)} & $\rm18^{\rm h}18^{\rm m}51\fs38$ & $0.60$ & -15$^\circ59\arcmin22\farcs62$ & $0.60$ & -- & -- & -- & $20.11\pm2.79^\ddagger$ & 61\\
    \textbf{\textit{AX\,J1845(-0258)}} & $\rm18^{\rm h}44^{\rm m}54\fs70$ & $0.60$ & -02$^\circ56\arcmin53\farcs10$ & $0.60$ & $8.5$ & -- & -- &$\sim 5.15$ & 62,63,64\\
    \textit{SGR\,2013(+34)} & $\rm20^{\rm h}13^{\rm m}56\fs90$ & $2.7$ & +34$^\circ19\arcmin48\arcsec$ & $2.7$ & $8.5\pm1.32$ & -- & -- & $\sim 4.38$ & 65,71 \\
    \textbf{\textit{PSR\,J1846(-0258)}} & $\rm18^{\rm h}46^{\rm m}24\fs94$ & $0.15$ & -02$^\circ58\arcmin30\farcs10$ & $0.20$ & $6.0^{+1.5}_{-0.9}$ & -- & -- &  $5.00^{+0.13}_{-0.06}$ &66,67,68,85,86\\
  \hline
 \end{tabular}
\small
\flushleft
Abbreviated magnetar names are given with full names in parentheses, which are used throughout this work. RA ($\alpha$), declination ($\delta$), and distance are taken from the McGill Magnetar Catalog \citep{olausen2014mcgill}, while proper motions ($\mu_\alpha {\rm cos}(\delta)$ and $\mu_{\delta}$) were found in the available literature. Italicized entries are unconfirmed magnetar candidates and boldfaced entries have confirmed or proposed supernova remnant associations. Proposed associations are indicated by a $^*$. $A_V$ extinction estimates are from the \textit{Bayestar19} map for magnetars with $\delta \gtrsim -30^\circ$. Magnetars outside this range have $A_V$ from the X-ray column density $N_H$ (indicated by a $^\ddagger$) using the \citet{predehl1995x} relation: $A_V\approx(5.59\times10^{-22}\,{\rm cm}^{2})N_H$. If no $N_H$ measurement is available, we list the median $A_V$ from \textit{Gaia} GSP photometry for massive sources within $1^\circ$ (indicated by a $^\dagger$; see Section~\ref{sec:gaia} for details).

Relevant citations referenced in the rightmost column are below: (1) \citet{2012AJ....144..107H} (2) \citet{2002ApJ...574L..29L} (3) \citet{2006ApJ...650.1070D} (4) \citet{hulleman2004anomalous} (5)\citet{tendulkar2013proper} (6) \citet{2010ApJ...711L...1V} (7) \citet{2011ApJ...739...87L} (8) \citet{2010ApJ...722..899G} (9) \citet{gaensler2008sgr} (10) \citet{2003ApJ...585..948K} (11) \citet{2004ApJ...609L..13K} (12) \citet{2002ApJ...579L..33W} (13) \citet{2005ApJ...620L..95G} (14) \citet{2010ApJ...710..227T} (15) \citet{2012ApJ...748L...1D} (16) \citet{2007ApJ...667.1111G} (17) \citet{2010ApJ...721L..33L} (18) \citet{2012ApJ...751...53A} (19) \citet{1999ApJ...526L..29C} (20) \citet{2004ApJ...615..887W} (21) \citet{kothes2008distance} (22) \citet{2006ApJ...636L..41M} (23) \citet{2003ApJ...589L..93I} (24) \citet{2012MNRAS.421.2593T} (25) \citet{2010ApJ...710..941H} (26) \citet{2010ApJ...725.1384H} (27) \citet{2014ApJ...780L...2B} (28) \citet{2013ApJ...770L..23M} (29) \citet{2013MNRAS.435L..29S} (30) \citet{bower2015proper} (31) \citet{2008MNRAS.386L..23B} (32) \citet{israel2005discovery} (33) \citet{tendulkar2012proper} (34) \citet{2008ApJ...676.1189M} (35) \citet{2007ApJ...662.1198H} (36) \citet{karuppusamy2020detection} (37) \citet{stamatikos2020sgr} (38) \citet{ding2020probing} (39) \citet{2012ApJ...761...66S} (40) \citet{pagani2011swift} (41) \citet{2010ApJ...718..331G} (42) \citet{2008AJ....135..167L} (43) \citet{2012ApJ...748...26K} (44) \citet{2008ApJ...677..292T} (45) \citet{2004ApJ...615..887W} (46) \citet{1997ApJ...486L.129V} (47) \citet{2014ApJ...781L..16Z} (48) \citet{2009ApJ...707..844D} (49) \citet{1999Natur.398..127F} (50) \citet{gaensler2014grb} (51) \citet{2016MNRAS.457.3448I} (52) \citet{zhong2020distance} (53) \citet{2012ApJ...746L...4K} (54) \citet{1981Natur.293..202F} (55) \citet{2001ApJ...563L..49H} (56) \citet{2016ATel.8831....1B} (57) \citet{doroshenko2021sgr} (58) \citet{richardson2023high} (59) \citet{2000ApJ...531..407C} (60) \citet{lamb2003sgr} (61) \citet{mereghetti2012magnetar} (62) \citet{1998ApJ...503..843T} (63) \citet{1999ApJ...526L..37G} (64) \citet{2006ApJ...652..548T} (65) \citet{2011AdSpR..47.1346S} (66) \citet{leahy2007distance} (67) \citet{2000ApJ...542L..37G} (68) \citet{2003ApJ...582..783H} (69) \citet{blumer2019x} (70) \citet{2012ApJ...756...27L} (71) \citet{roming2008first} (72) \citet{2016ATel.8831....1B} (73) \citet{2023MNRAS.523.4949J} (74) \citet{ponti2015xmm} (75) \citet{yalinewich2017evolution} (76) \citet{1982ApJ...255L..45C} (77) \citet{park2012x} (78) \citet{park2020nustar} (79) \citet{sano2023alma} (80) \citet{borkowski2017expansion} (81) \citet{zhou2019spatially} (82) \citet{kothes2018radio} (83) \citet{zhou2020revisiting} (84) \citet{nakano2017study} (85) \citet{leahy2008distance} (86) \citet{2023ApJ...942..103S} (87) \citet{ding2020magnetar}
\end{center}
\end{table*}

\section{\textit{Gaia} Catalog Search For Unbound Companions}\label{sec:gaia}

\subsection{Gaia Data}

The \textit{Gaia} DR3 catalog was queried to search for unbound magnetar companions ejected by the CCSN \citep{2016A&A...595A...1G, gaiacollaboration2022gaia}. As is highlighted in \citet{chrimes2023searching}, the \textit{Gaia} catalog is a unique tool given its absolute reference frame which allows for comparison of proper motion data. We build upon previous magnitude-based searches \citep[e.g.][]{chrimes2022magnetar} by analyzing the trajectories for each GDR3 source in reference to the magnetars'. \citet{chrimes2023searching} note the \textit{Gaia} survey depth $m_{G,{\rm min}} = 20.7$ and proper motion uncertainty ranging from $9.6\le \mu_{\rm min} \le 801.6$\,$\mu$as\,yr$^{-1}$ for $13 \le m_G \le 20.7$ limit the survey's detection of walkaway or runaway stars to those with low extinction \citep[$A_V\lesssim10$;][]{lindegren2021gaia}\footnote{\url{https://www.cosmos.esa.int/web/gaia/science-performance} Section 1}. Despite this, our combination of magnitude analysis with proper motion data will still make a marked improvement over previous searches, seen for example in \citet{el2023fastest} and inferred through simulations by \citet[][Figure\,8 row 1]{chrimes2023searching}. 


The \textit{Gaia} query is limited to potential O and B type stars, as we expect many of the companions of magnetars to be massive stars, and removing dim stars significantly reduces contamination by false positives. To do this we impose an upper limit on the absolute $V$-band magnitude $M_V<-2.5{\rm log}_{10}(L_{OB}/L_\odot) + 4.68 = 1.185$, where we use $L_{OB} \approx 25L_\odot$, and color $G_{BP}-G_{RP}<2$. Appendix~\ref{app:ADQL} contains the Astronomical Data Query Language (ADQL) query used to compile the initial \textit{Gaia} sample.

Following the initial query, extinction corrections and cutoffs were re-applied using the \textit{Bayestar19}\footnote{\url{https://dustmaps.readthedocs.io/en/latest/modules.html}} dust map \citep{green2018galactic,2019ApJ...887...93G} for any \textit{Gaia} sources above declination $\delta \gtrsim -30^{\circ}$. From \textit{Bayestar19} we obtain the $B-V$ color excess\footnote{Note that \textit{Bayestar19} reports reddening in arbitrary units and is converted to $E_{B-V}$ by multiplying by $0.995$ as described at \url{http://argonaut.skymaps.info/usage} \citep{schlafly2011measuring}. We proceed assuming this factor is approximately 1, which is accurate to the first decimal place.}, $E_{B-V}$, which is converted to $V-$band extinction with $A_V\approx3.2E_{B-V}$ and the $G-V$ excess using the Johnson-Cousins Relations\footnote{See the \textit{Gaia} DR3 Documentation at \url{https://gea.esac.esa.int/archive/documentation/GDR3/} Section 5.5.1 Table 5.9 for details.}. These are used along with the parallax distance $d$ from the \citet{bailer2023estimating} catalog \citep[see also][]{2021AJ....161..147B} to obtain the absolute $G-$ and $V-$band magnitudes, $M_G$ and $M_V$. Details on conversions between photometric systems are in Appendix~\ref{app:ADQL}.

For sources with $\delta \lesssim -30^\circ$, $M_V$ is computed with Equations~\ref{eq:deriv5}-\ref{eq:deriv8} using $A_G$ and $E_{G_{BP}-G_{RP}}$ reported in \textit{Gaia}, which are derived from $G_{BP}-G_{RP}$ spectra in the GSP-Aeneas library \citep{vallenari2022gaia}. However, we proceed cautiously with the GSP extinction estimates, which are known to be unreliable within the Galactic Plane where the effective temperature is overestimated \citep[see][Section 3.4, and the \textit{Gaia} DR3 Documentation Section 11.3.3 for details]{andrae2022gaia, fouesneau2022gaia}. 

\begin{figure*}
 \includegraphics[width=\linewidth]{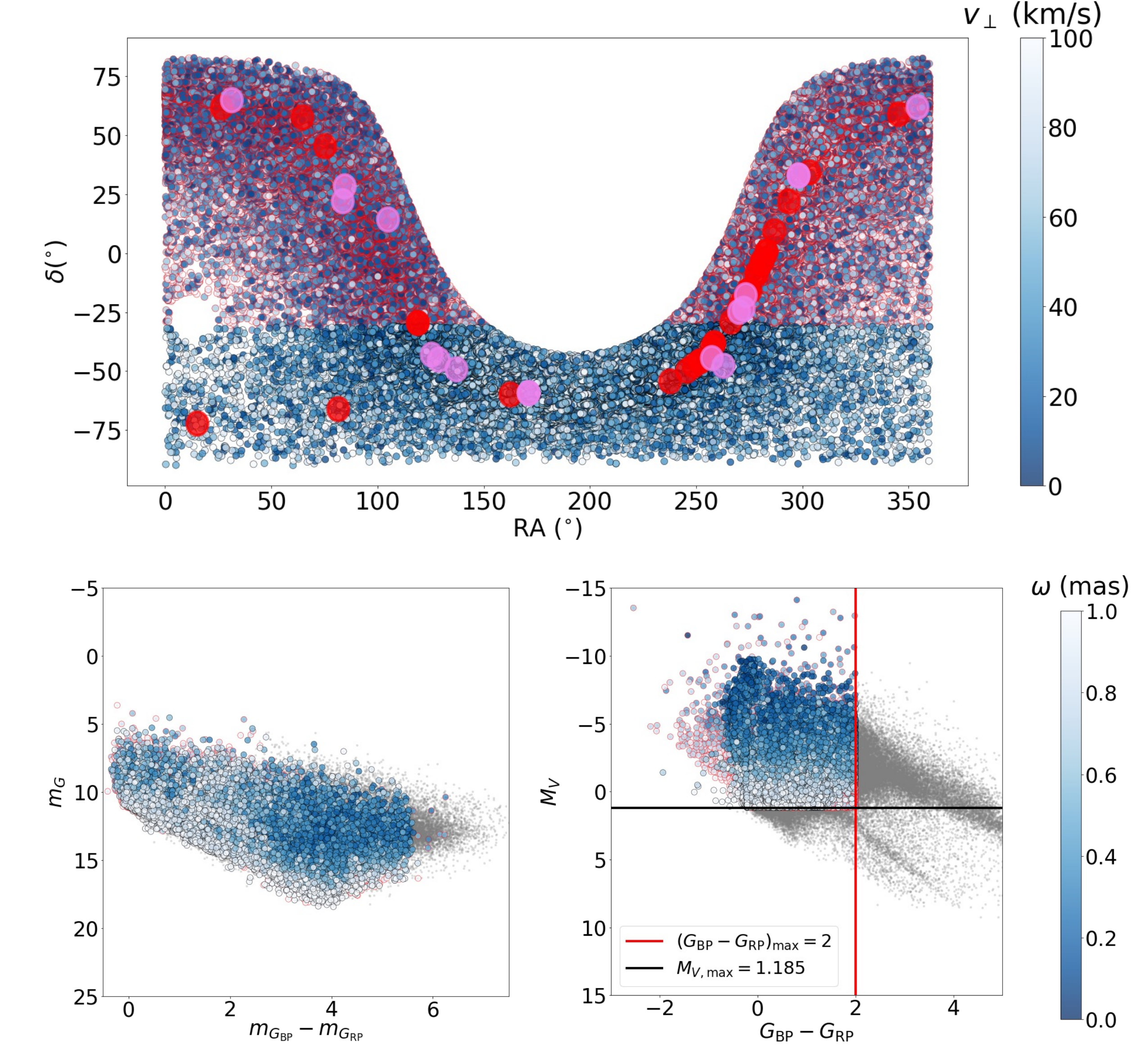}
 \caption{ (\textit{Top}) Locations of \textit{Gaia} search sample in Celestial coordinates. The tangential velocity magnitude in km\,s$^{-1}$ is indicated by the colorbar. The \textit{Bayestar19} Dust Map is used for sources at $\delta > -30^{\circ}$ (red outline), and the \textit{Gaia} photometric extinction is used for $\delta < -30^{\circ}$ \citep[black outline;][]{vallenari2022gaia,green2019revised}. The $5^\circ$ radius cones queried around each magnetar are shown in red, and those around each pulsar used for the test sample (see Section~\ref{sec:gaiatest}) are shown in violet. Lower transverse velocities are found in the Galactic Plane (GP), meaning that the unbound companions of GP magnetars can be found at smaller angular separations. (\textit{Bottom}) Color-Magnitude Diagrams for the \textit{Gaia} search sample. The \textit{left} plot shows the apparent $G-$band magnitude $m_G$ plotted against the apparent $m_{G_{BP}}-m_{G_{RP}}$ color. The \textit{right} plot shows absolute $V-$band magnitude $M_V$ for each source against the absolute color $G_{BP}-G_{RP}$. $M_V$ is derived using the Johnson-Cousins relationships in Equations~\ref{eq:deriv1}-\ref{eq:deriv8} (see Appendix~\ref{app:ADQL}). Sources with $\delta>-30^\circ$ are again outlined in black while those with $\delta\ge-30^\circ$ are outlined in red. The parallax of each source is indicated by the colorbar in milliarcseconds (mas). Sources from the initial query (see Appendix~\ref{app:ADQL}) which were cut by the magnitude (black) and color (red) limits are shown in grey. GSP-Aeneas extinction estimates may be unreliable in the Galactic plane according to the \textit{Gaia} DR3 Documentation (Section 11.3.3), resulting in a rough color cutoff near $G_{\rm BP}-G_{\rm RP}\approx-0.5$ \citep{vallenari2022gaia,andrae2022gaia,fouesneau2022gaia}.}
 \label{fig:sampleHR}
\end{figure*}

Sources were constrained to have parallax $0.008 < \omega \pm \sigma_\omega < 1.25\,$mas based on the distances of the farthest (CXOU\,J010043.1-721134, $62.4\pm1.6$\,kpc) and nearest (Swift\,J1822.3-16.06, $1.6\pm3$\,kpc) magnetars in the McGill catalog, placing an upper limit $\sim125\,$kpc on the sample. The query was further constrained to Galactic longitudes $-80^\circ \le b\le 20^\circ$ based on magnetar locations. Figure~\ref{fig:sampleHR} shows the resulting \textit{Gaia} sample of 156572 potential OB star sources. 

We use a Monte Carlo simulation to estimate p-values for the \textit{Gaia} search. The criteria for association for a given sampled trajectory are below:

\begin{enumerate}
    \item When the magnetar and source are extrapolated back in time over a sampled travel time (between $1-10$\,kyr), their initial angular separation $\Theta \approx 0^\circ$.
    \item The absolute $V-$band magnitude of the \textit{Gaia} source is consistent with that of an OB (possibly A) star ($-10\lesssim M_V \lesssim 3$)\footnote{While we expect most OB stars to have $M_V \lesssim0$, we use a wider range that also includes A stars to avoid missing dim companions.}.
    \item The difference in 2D travel time for the magnetar and source to travel from their past intersection point to their present positions is $\Delta \tau_{\rm 2D}\approx0$.
    \item The magnetar's physical travel time $\mathcal{T}_{\rm 3D, mag}$ is consistent with its age $\mathcal{T}$.
    \item The source's parallax distance is consistent with magnetar's distance, such that $\Delta d \approx 0$.
\end{enumerate}

\noindent which are used to define p-values $p_1, p_2$, $p_3$, and $p_4$ respectively (the first two criteria are combined to define $p_1$). Note that the \textit{Gaia} radial velocity measurements are not used in our search since they have not been confirmed through spectroscopic follow-up and would over-constrain the sampled velocity vectors. Appendix~\ref{app:pvals_GAIA} describes the definition of $p_1, p_2$, $p_3$, and $p_4$ in more detail, which are combined as shown below:

\begin{equation}\label{eq:pvaltot}
    p = 1 - (1-p_1)(1-p_2)(1-p_3)(1-p_4)
\end{equation}

\noindent We choose this definition (as opposed to e.g. the Fisher statistic, Lancaster statistic, or a direct average) for its intuitive application to this search, as all four criteria must be met with $p_i<<5\%$ to obtain $p<5\%$. This also makes it straightforward to interpret p-values by tracing back each $p_i$ to its criterion.

Each magnetar's age is estimated as the age of its associated SNR $\mathcal{T}_{\rm SNR}$ \citep{suzuki2021quantitative}. If no SNR is associated, its X-ray luminosity decay age $\mathcal{T}_{\rm decay}$ is used, which adopts the \citet{ferrario2008origin} field model and is derived in Appendix~\ref{app:decay}. For magnetars with no persistent X-ray emission or magnetic field estimates, the characteristic spin-down age $\mathcal{T}_{\rm rot}$ is used \citep{olausen2014mcgill}. In Table~\ref{table:magtimeparams} we provide the derived or reported timescales for each magnetar.

\begin{table}
 \begin{center}
 \scriptsize
 \caption{Magnetar Age Upper Limits Used for the \textit{Gaia} Search}
 \label{table:magtimeparams} 
 \begin{tabular}{c|c|c|c}
  \hline
  \textbf{Magnetar} & \textbf{Age Method} & $\mathbf{\mathcal{T}_{\rm decay/rot/SNR}}$ & \textbf{Reference}\\
  \hline
    CXOU\,J010 & Decay & $7.7$\,kyr & This work\\
    4U\,0142 & Decay & $3.8$\,kyr &  This work\\
    SGR\,0418& Decay & $8.3$\,Myr & This work\\
    SGR\,0501 &Decay & $59.4$\,kyr & This work\\
    SGR\,0526 &Decay &  $2.9$\,kyr  & This work\\ 
    1E\,1048.1 &Decay &  $8.9$\,kyr & This work\\
    1E\,1547.0 &Decay &  $30.4$\,kyr & This work \\
    PSR\,J1622 &Decay &  $44.2$\,kyr & This work\\
    SGR\,1627 &Decay &  $37.0$\,kyr  & This work\\
    CXOU\,J1647&Decay &  $2.1$\,Myr  & This work\\
    1RXS\,J1708 &Decay &  $8.1$\,kyr & This work \\
    CXOU\,J1714 & SNR  & $6.2$\,kyr & \citet{blumer2019x} \\
    SGR\,J1745 &Decay &  $64.9$\,kyr & This work\\
    SGR\,1806 &Decay &  $1.3$\,kyr  & This work\\
    XTE\,J1810 &Decay &  $81.1$\,kyr & This work\\
    Swift\,1818 & Spindown & $0.86$\,kyr & \citet{rajwade2022long}\\
    Swift\,J1822 &Decay &  $0.2$\,Myr  & This work\\
    SGR\,1833 & Spindown & $34$\,kyr & \citet{2011MNRAS.416..205E} \\
    Swift\,J1834 &Decay &  $0.2$\,Myr &This work\\
    1E\,1841 & SNR & $1.8$\,kyr & \citet{borkowski2017expansion} \\ 
    3XMM\,J1852 &Decay &  $0.7$\,Myr & This work \\
    SGR\,1900 &Decay &  $4.1$\,kyr & This work\\
    SGR\,1935 & SNR & $16.0$\,kyr & \citet{lyman2022fast}\\
    1E\,2259 &Decay &  $62.0$\,kyr & This work\\
    \hline
    \textit{SGR\,0755} & -- &  -- & --\\
    \textit{SGR 1801} & -- & -- & --\\
    \textit{SGR\,1808} &-- &  -- &  --\\
    \textit{AX\,J1818} & -- & -- & --\\
    \textit{AX\,J1845} & SNR & $1.4$\,kyr & \citet{gaensler1999new}\\
    \textit{SGR\,2013} & -- & -- &  -- \\
    \textit{PSR\,J1846} & Spindown &  $0.73$\,kyr  & \citet{2011ApJ...730...66L}\\
  \hline
 \end{tabular}
\small
\flushleft
$\mathcal{T}_{\rm decay}$ is estimated in this work \citep[see Appendix~\ref{app:decay} in this work and][]{ferrario2008origin} using the magnetic field $B_d$ and X-ray luminosity $L_X$ reported in the McGill Magnetar Catalog \citep[][and references therin]{olausen2014mcgill}. Spindown ages are those from the McGill catalog, and are approximated as the characteristic age $\mathcal{T}_{\rm rot}= P(2\dot{P})^{-1}$. SNR ages $\mathcal{T}_{\rm SNR}$ are used only for magnetars with confident SNR associations. References for the spindown and SNR ages are in the rightmost column. 
\end{center}
\end{table}

\subsection{\textit{Gaia} Search Test Using High Velocity Pulsars with Supernova Remnants}\label{sec:gaiatest}

In order to test the search pipeline, we applied the Monte Carlo simulation to a sample of 15 pulsars with known SNR associations and proper motion measurements compiled from the ATNF catalog \citep{vanderwateren2023psr,andersen2023chime,lyne2015binary,stairs2001psr,dinccel2015discovery,kochanek2021supernovae}. Within this sample, only PSR\,J0538+2817 is known to have a runaway companion B0.5V-type star, as discovered by \citet{dinccel2015discovery} and confirmed by \citet{kochanek2021supernovae}. The upper limit on the travel time $\tau$ is set to 100\,kyr, the maximum age of SNR shells before they dissipate \citep[e.g.][]{frail1994radio}, given that pulsars can have ages $>100$\,kyr \citep[e.g.][]{lorimer2008binary,kaspi2017magnetars}. The initial Monte Carlo simulation (criteria 1 and 2) identifies between 2 and 800 potential candidates with $p_1<5\%$ for each of the 15 pulsars. Following the trajectory analysis, only 1 pulsar, PSR\,B1951+32 remains with $p<5\%$ and a sufficient fraction of valid trajectories\footnote{Three other pulsars had candidates with $p<5\%$, but used only one valid trajectory, for which the source and pulsar paths intersect, out of 100 trials, and they are considered false positives.}. The companion of PSR\,J0538+2817 is not initially recovered. In this section we discuss the likelihood of PSR\,B1951+32's candidate companion, and the reasons for PSR\,J0538+2817's omission. Figure~\ref{fig:magnetarcontoursnew0} shows the trajectory plots for both pulsars.

\subsubsection{PSR J0538+2817}

For PSR\,J0538+2817, we initially do not recover the runaway companion star, HD\,37424, which was identified by \citet{dinccel2015discovery} and \citet{kochanek2021supernovae}. The corresponding \textit{Gaia} source, \textit{Gaia}\,DR3\,3441732292729818752, was assigned a p-value $p>99\%$\footnote{The minimum and maximum p-values are set by the number of trials, which varies for each magnetar depending on what constraints on its position, distance, and proper motion are available. 1000 and 100 trials were used for PSR\,B1951+32 and PSR\,J0538+2817, respectively.} and therefore was not identified as a candidate. Upon further inspection, we find the high p-value is the result of a high false positive rate; among the simulated sources used to estimate the \textit{H0} distribution, $\sim95\%$ are found with all Monte Carlo samples within the $\{\Theta,M_V\}$ association region ($\hat{f}=0$; see Appendix~\ref{app:GAIAFPs}). We attribute this to the large association region in $\{\Theta,M_V\}$ space which extends to $\gtrsim1\arcmin$ (see Figure~\ref{fig:magnetarcontoursnew0}). While pulsars have larger confidence regions in $\{\Theta,M_V\}$ space than magnetars due to the higher age limit and lower distances, the high proper motions of PSR\,J0538+2817, PSR\,J1809-2332, and PSR\,B2334+61 further extend the region, resulting in unusually high false positive rates $\gtrsim99\%$, $\gtrsim91\%$, and $\gtrsim76\%$, respectively. 

We therefore increase the p-value thresholds for these three sources to match their minimum false positive rates. We use a threshold $p_1<100\%$ for PSR\,J0538+2817 since the false positive rate is $\gtrsim99\%$. For PSR\,B2334+61 and PSR\,J1809-2332, no additional sources are identified with the new threshold, whereas HD\,37424 is identified as a candidate for PSR\,J0538+2817 after increasing the threshold. This scenario also justifies our use of the search statistic $\hat{f}<\hat{f}_c$ to identify candidates initially, as it implicitly accounts for this false-positive rate bias. Our conditional recovery of PSR\,J0538+2817's runaway star demonstrates the accuracy and sensitivity of our search method, but indicates we must be wary of false positives and sources along unique sightlines. We address completeness of the magnetar search in detail in Section~\ref{sec:discussion}.

\subsubsection{PSR B1951+32 }

One \textit{Gaia} companion candidate is identified for PSR\,B1951+32 with $p<0.1\%$, Gaia\,DR3\,2035109887486067456 \citep[catalogued as TYC\,2673-414-1;][]{2000A&A...355L..27H}. The parallax distance $1.13\pm0.02$\,kpc is only marginally consistent with the pulsar's $\sim3$\,kpc distance, which has an uncertainty that is not well defined \citep{kulkarni1988fast}. The pulsar's dispersion measure (DM) and more recent estimates of the associated SNR\,CTB\,80 suggest a closer $\sim2$\,kpc distance, which would be more consistent with TYC\,2673-414-1 \citep{kramer2000pulsar,migliazzo2002proper,yao2017new}. From the 2MASS counterpart, 2MASS\,J19544429+3327472, we estimate absolute $H-$band magnitude $M_H\approx-4.5$ and $J-H$ color $J-H\approx-0.1$ when placed at the pulsar distance $3$\,kpc. While these are reasonable, they do fall outside of the expected ranges of IR magnitude and color from \citet{chrimes2022magnetar}'s simulations. From the valid trajectory samples, TYC\,2673-414-1 is required to have velocity $|\overrightarrow{v}|\approx75^{+14}_{-6}$\,km\,s$^{-1}$, which would be consistent with a more rare runaway star if ejected from the pulsar's CCSN \citep{renzo2019massive}. While pulsar companions are not the focus of this search, we encourage follow-up of TYC\,2673-414-1 as a possible unbound companion of PSR\,B1951+32.

\subsection{\textit{Gaia} Search Results}\label{sec:MCresults}

\begin{figure*}
 \includegraphics[width=\linewidth]{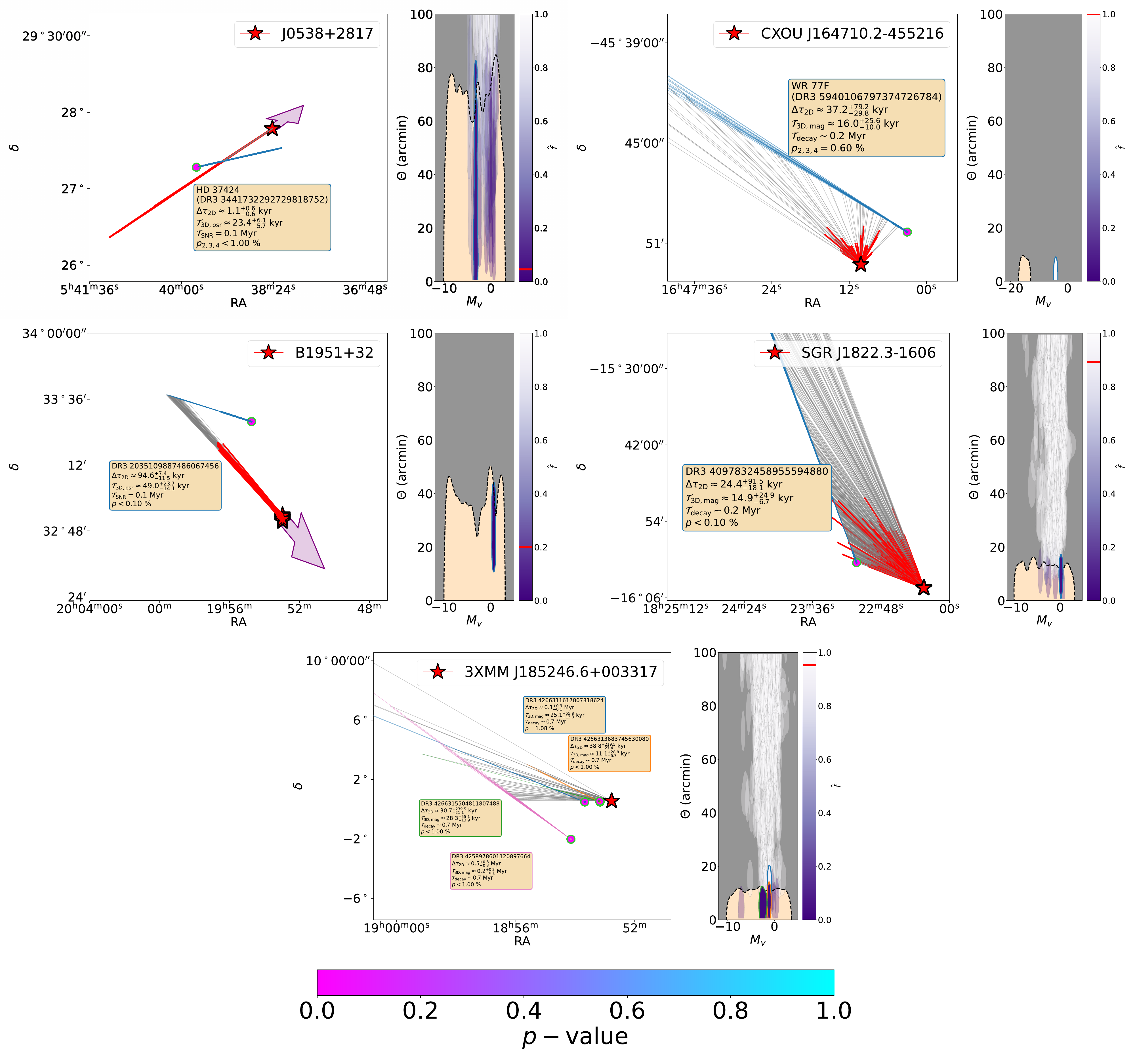}
 \caption{(\textit{Left}) Monte Carlo sampled trajectories for candidate \textit{Gaia} DR3 sources with p-values $p<5\%$ possibly associated with 3 magnetars (\textit{right}, \textit{bottom}), PSR\,J0538+2817 (\textit{top left}), and PSR\,B1951+32 (\textit{middle left}). Each figure includes only \textit{Gaia} OB star candidates with association p-value $p<5\%$, including the unbound companion of CXOU\,J164710.2-455216 proposed by \citet{clark2014vlt}, WR\,77F (\textit{top right}). Present-day magnetar/pulsar and GDR3 source locations are shown as red stars and circular markers, respectively; the latter has colorbar corresponding to the total p-value $p$, except for PSR\,0538+2817 and CXOU\,J164710.2-455216, for which we exclude $p_1$ (see Section~\ref{sec:gaia}). Red lines show the initial magnetar/pulsar trajectory samples while thin lines extrapolate the magnetar/pulsar and source trajectories to their past intersection. Note for these initial trajectories extend farther for pulsars (\textit{top left}, \textit{middle left}) than for magnetars (\textit{right}, \textit{bottom}) since travel time samples were drawn from a uniform distribution up to $1$\,Myr for pulsars in contrast to $10$\,kyr for magnetars. If proper motion measurements are available, the proper motion direction is indicated by a purple arrow. Each GDR3 source is shown with its source ID, 2D travel time difference $\Delta \tau_{\rm 2D}$, the magnetar's/pulsar's 3D travel time $\mathcal{T}_{{\rm 3D,mag/psr}}$, an upper limit age estimate $\mathcal{T}_{\rm SNR/decay/rot}$, and the relevant p-value. (\textit{Right}) Contour plot of Monte Carlo samples in $\{\Theta-M_V\}$ space. The $95\%$ contour region is in beige, and candidates with search statistic $\hat{f}$ less than the critical threshold, $ \hat{f}_c$, have colorbar corresponding to $\hat{f}$ (see Appendix~\ref{app:pvals_GAIA}). $\hat{f}_c$ is indicated by a red horizontal line on each colorbar. We recover the proposed companions of CXOU\,J164710.2-455216 (WR\,77F; \textit{upper right}) and PSR\,J0538+2817 (HD\,37424; \textit{top left}) with $p<5\%$. We also identify potential runaway companions of SGR\,1822.3-1606 and PSR\,B1951+32 (\textit{middle row}) with $p<5\%$. The companions shown for 3XMM\,J185246.6+003317 (\textit{bottom}) are most likely false positives as discussed in Appendix~\ref{app:GAIAFPs}.} 
 \label{fig:magnetarcontoursnew0}
\end{figure*}

Through the \textit{Gaia} Monte Carlo search, we identify potential candidates with $p<5\%$ for 2 magnetars. In Figure~\ref{fig:magnetarcontoursnew0}, we show the trajectory plots and contour plots of the Monte Carlo samples for these, along with CXOU\,J164710.2-455216, whose proposed companion WR\,77F we recover with $p=0.6\%$ \citep{clark2014vlt}. We rule out one of the two candidates, 3XMM\,J185246.6+003317, as a false positive based on detailed analysis of its candidates' velocities and distances, which we discuss in Appendix~\ref{app:GAIAFPs}. Having not recovered the proposed binaries SGR\,0755-2933 and CXOU\,J171405.7-38131, we find this search method is relatively insensitive to bound stellar companions due to their position within magnetars' error ellipses \citep{chrimes2022magnetar,richardson2023high,doroshenko2021sgr}. However, it is a useful tool to identify which candidates are likely to be bound companions (see Section~\ref{sec:boundresults}). In the following sections, we focus on the unbound companion candidates for SGR\,J1822.3-1606 and CXOU\,J164710.2-455216.


\subsubsection{SGR J1822.3-1606}


The primary candidate for an unbound system is SGR\,J1822.3-1606 and \textit{Gaia}\,DR3\,4097832458955594880, which is assigned a p-value $p<0.1\%$. Its distance, $d = 1.00^{+0.03}_{-0.02}$\,kpc, is consistent with that of the magnetar, $1.6\pm0.3$\,kpc, at the $2\sigma$ level, and from the IR counterpart, 2MASS\,18230515-1600296, we estimate absolute magnitudes $M_J =-3.9\pm0.5$ and $M_H = -4.6^{+0.5}_{-0.4}$, and color $J-H = 0.7\pm0.1$\footnote{Uncertainties are estimated from the $16^{\rm th}$ and $84^{\rm th}$ percentiles of the magnitude and color samples, which take into account distance and apparent magnitude errors.}, which are both reasonable for a low-mass B-type star. SGR\,J1822.3-1606 does not have a measured proper motion, and our simulation suggests $\mu_{\alpha}{\rm cos}(\delta) = -29^{+15}_{-24}$\,mas\,yr$^{-1}$, $\mu_{\delta} = -25^{+13}_{-19}$\,mas\,yr$^{-1}$ 
are required for the magnetar to have a valid trajectory\footnote{Estimates of the proper motion and velocity are derived from Monte Carlo samples found to produce `valid' trajectories which are found to intersect through our iterative analysis. Uncertainties are from the 16$^{\rm th}$ and 84$^{\rm th}$ percentiles of these `valid' samples.}. A wide range of magnetar velocities are permitted for this association, with $|\overrightarrow{v}_{\rm mag}| = 413^{+184}_{-175}$\,km\,s$^{-1}$. 
Notably, the permitted stellar velocity $|\overrightarrow{v}| = 60^{+7}_{-5}$\,km\,s$^{-1}$ would classify this as a rare `runaway' companion star \citep{renzo2019massive}. This is roughly consistent with the source's high transverse velocity, $\sim55$\,km\,s$^{-1}$, making the association achievable without a significant radial component. While it is unlikely that such velocities can be reached from CCSN disruption alone, other mechanisms may contribute to the source's and magnetar's space motion, such as dynamical ejection from an OB association \citep[e.g.][]{1967BOTT....4...86P,tetzlaff2011origin}. Therefore, we identify this as a promising candidate, and further characterization from follow-up observations would be of great interest.


\subsubsection{CXOU J164710.2-455216}
Next, we consider the Wolf-Rayet star WR\,77F (or Wd1-5), which \citet{clark2014vlt} suggest as an unbound companion of CXOU\,164710.2-455216 based on the position, radial velocity and chemical content of the star. WR\,77F is associated with \textit{Gaia} DR3 5940106797374726784 which was not included in the initial \textit{Gaia} query, possibly due to an underestimated \textit{Gaia} extinction from GSP photometry. We add this candidate to the sample and run the simulation with an increased 1000 trials after finding a low fraction of valid trajectories with 100 samples. Given that the candidate is an evolved Wolf-Rayet star rather than a main-sequence OB star, and that magnetar CXOU\,J164710.2-455216 lies along a high extinction sightline with $A_V=13.35\pm0.28$, we omit $p_1$ from this analysis (which initially was $p_1>99.9\%$). With this adjustment, we recover a p-value $p_{2,3,4}=0.6\%$. 
The magnetar's derived physical travel time $\mathcal{T}_{\rm 3D, mag} \approx16.0^{+25.6}_{-10.0}$\,kyr, is consistent with its decay age $\mathcal{T}_{\rm decay}\approx0.2$\,Myr within $\sim1\sigma$. Finally, the valid trajectories require velocities $|\overrightarrow{v}_{\rm mag}|\approx 390\pm170$\,km\,s$^{-1}$ and $|\overrightarrow{v}|\approx80^{+20}_{-13}$\,km\,s$^{-1}$ for CXOU\,164710.2-455216 and WR\,77F, respectively. The former is consistent with \citet{2005MNRAS.360..974H} while the latter suggests that WR\,77F is a runaway star \citep[$|\overrightarrow{v}|\gtrsim30$\,km\,s$^{-1}$][]{renzo2019massive}. However, the localization of CXOU\,J164710.2-455216 and WR\,77F to the Westerlund-1 massive star cluster implies that dynamical ejection may have contributed to the large observed velocities, as well. We conclude it is likely that CXOU\,J164710.2-455216 are WR\,77F are an unbound magnetar binary disrupted by the former's CCSN, potentially supplemented by dynamical ejection, supporting the results of \citet{clark2014vlt}.

\section{Optical and IR Search for Bound Companions}\label{sec:bound}

In this section, we search for bound OB star-magnetar binaries using optical and near-IR images. As discussed in Section~\ref{sec:intro}, bound companions should have negligible ($\lesssim1\arcmin$) separation from the magnetar in high resolution ($\sim0.25\arcsec-0.4\arcsec$) images. \citet{chrimes2022new} conducted a similar search using HST Wide Field Camera 3\footnote{See the WFC3 Instrument Handbook for details: \url{https://hst-docs.stsci.edu/wfc3ihb/chapter-7-ir-imaging-with-wfc3/7-8-ir-sensitivity}} images with $\sim27$ magnitude depth and $0.1265\arcsec$ resolution. They place tight constraints with milliarcsecond precision on $\sim15$ magnetar NIR counterpart candidates using $H-$band HST data, laying the foundation for the \citet{chrimes2022magnetar} search.

While \citet{chrimes2022magnetar} discuss that searching in NIR is more effective than optical due to the $\sim7\times$ lower extinction, searching optical data is still beneficial. The massive, un-evolved OB star companions we expect are intrinsically fainter in the NIR than optical wavelengths. The magnitude depth of the PS1 ($\sim24$) and SkyMapper ($\sim18$) catalogs therefore makes them powerful resources for this search \citep{maund2016possible,2004SPIE.5489...11K,wolf2018skymapper,onken2024skymapper}. We also note that since the resolution ($0.4\arcsec$, $0.34\arcsec$, and $3\arcsec$, respectively) and depth ($\sim18$, $\sim15$, and $\sim19.3$, respectively) of the UKIDSS, VVV, and 2MASS images are far exceeded by HST, any additional data to supplement the breadth of the search is valuable.

\subsection{Pan-STARRS and SkyMapper Data and Optical Search Method}\label{sec:PS1}



The Pan-STARRS Data Release 2 (PS1 DR2) catalog contains optical images of the northern sky (above $\delta > -30^\circ$) in 5 passbands ($g-$,$r-$,$i-$,$z-$,and $y-$bands, or `$grizy$') with $0.25\arcsec$ pixels \citep{Flewelling_2020, chambers2019panstarrs1}. $1\arcmin\times1\arcmin$ image cutouts were obtained from the PS1 Image Server\footnote{\url{https://ps1images.stsci.edu/ps1image.html}} at the position of each magnetar and candidate. 17 magnetars and all 7 magnetar candidates fall within the PS1 survey range. For the remaining 7 with $\delta \le -30^{\circ}$, we use images from the SkyMapper Southern Survey Data Release 4 (SkyMapper DR4\footnote{\url{https://skymapper.anu.edu.au/}, \url{doi.org/10.25914/5M47-S621}}), which records in  $u-$,$v-$,$g-$,$r-$,$i-$, and $z-$bands (`$uvgriz$') with $0.5\arcsec$ pixels \citep{wolf2018skymapper,onken2024skymapper}. Note the point source detection limit is $\sim18.0$ in comparison to $\sim 21.5$ for PS1. 



In addition to image cutouts, the PS1 and SkyMapper point source catalogs were queried to obtain known sources within a $30\arcsec$ search cone of each magnetar. We also include any bright sources identified `by-eye' within the image that do not appear to coincide with catalogued point sources (see Appendix~\ref{app:RACSbyeye}). We then use the catalogued magnitudes and position of each source to conduct a Monte Carlo simulation. Similarly to the \textit{Gaia} search we define the association p-value using three criteria:

\begin{enumerate}
    \item The PS1 source and magnetar are located at the same RA and declination.
    \item If we assume the source absolute magnitude is that of an OB star ($-10 \lesssim M_V \lesssim 3$), the distance is equal to the magnetar's distance.
    \item The difference of the Kron and point source (PSF) magnitude ($M_{x,{\rm Kron}} - M_{x,{\rm PSF}}$) exceeds the \citet{farrow2014pan} point source cutoff for x = $`uvgrizy'$ bands as available \citep{kron1980photometry}:
    
    \begin{dmath}
        \Delta M_{\rm min}(M_{x,{\rm Kron}}) = -0.319 + 0.129(M_{x,{\rm Kron}} - 21) + 0.007(M_{x,{\rm Kron}} - 21)^2
    \end{dmath} 
\end{enumerate}

\noindent for which we define p-values $p_1, p_2,$ and $p_3$, respectively. The three p-values are combined to get the total association p-value:

\begin{equation}
    p = 1 - (1-p_1)(1-p_2)(1-p_3)
\end{equation}

\noindent An explicit derivation of $p_1, p_2,$ and $p_3$ is provided in Appendix~\ref{app:pvals_PS1}. 

The method above was applied only for magnetars with errors defined for the RA, declination and distance, and only for point sources with at least $g-$ and $r-$ band PSF and Kron magnitudes\footnote{This unfortunately means that p-values are not estimated for `by-eye' sources in the optical and IR searches. However, most lie at a large angular separation relative to the magnetars' $\sim1\arcsec$ position errors making them unlikely candidates.}, and errors defined for RA and declination. Note that our evaluation of $p_2$ contrasts with the \citet{chrimes2022magnetar} search, which uses the magnetar distance to estimate extinction values for nearby candidate stars. However, magnetar distances are not well-constrained, with errors of order $\sim$\,kpc. In this work, we invert this process to obtain a distance estimate whose uncertainty is driven by the expected absolute magnitude range. Through this, we avoid making assumptions about the source distance beyond the requirement that it be an OB star.


\subsection{UKIDSS, VVV, and 2MASS Data and IR Search Method}\label{sec:UKIDSS}

Infrared images were obtained from the UKIDSS survey catalog, which is a superset of the Large Area Survey (LAS), Galactic Plane Survey (GPS), Galactic Clusters Survey (GCS), Deep Extragalactic Survey (DES), and Ultra Deep Survey (UDS) catalogs \citep{2006MNRAS.372.1227D}. Collectively, this covers the declination range from $-60^\circ <\delta<+40^\circ$, including sightlines for 12 magnetars and 6 magnetar candidates from this sample. We obtained $1\arcmin\times1\arcmin$ image cutouts with $0.2\arcsec$ pixels in the $J-$, $H-$, and $K-$bands ($390-1220$\,nm) for each magnetar to search for binary companions. For 8 magnetars in the southern portion of the Galactic disk, the VISTA Variables in the Via Lactea (VVV) survey was queried for $J-$, $H-$, and $K_s-$band images with $0.2\arcsec$ pixels \citep{nikzat2022vvv}. Images for the five remaining magnetars were obtained from the Two-Micron All-Sky Survey (2MASS)\footnote{\url{https://irsa.ipac.caltech.edu/Missions/2mass.html}}, which surveys the full sky in the $J-$, $H-$, and $K_s-$bands at much lower $\sim 3\arcsec$ resolution \citep[$\sim2\arcsec$ pixels;][]{2006AJ....131.1163S}. The UKIDSS Galactic Plane Survey (UGPS)\footnote{\url{https://vizier.cfa.harvard.edu/viz-bin/VizieR?-source=II/316}} catalog, 2MASS point source catalog, and VVV Infrared Astrometric Catalogue (VIRAC)\footnote{\url{https://vizier.cfa.harvard.edu/viz-bin/VizieR?-source=II/364}} were queried in a cone search for discrete objects within $0.5\arcmin$ of each magnetar \citep{2006AJ....131.1163S,nikzat2022vvv,lucas2008ukidss,2012yCat.2314....0L,2013yCat.2319....0L,smith2018virac}. Additional sources were identified `by-eye' in each image to include in the search.

The criteria used for the IR search are below:

\begin{enumerate}
    \item The source and magnetar are located at the same RA and declination.
    \item If we assume the source's absolute $H-$band magnitude and $J-H$ color are stellar in nature and resemble that of a bound OB star ($-2.76\lesssim M_H \lesssim 3.34$, $-0.14\lesssim J-H \lesssim 0.31$ are $95\%$ confidence ranges from \citet{chrimes2022magnetar} population synthesis), the distance is equal to the magnetar's distance.
\end{enumerate}

\noindent for which we define p-values $p_1$ and $p_2$, respectively. Kron magnitudes are not available from the included IR catalogs. The p-values are combined using:

\begin{equation}
    p = 1 - (1-p_1)(1-p_2)
\end{equation}

\noindent An explicit derivation of these p-values is provided in Appendix~\ref{app:pvals_2MASS}. Note the p-value is only evaluated for magnetars and sources with position errors, and only for 2MASS and VVV sources with at least $H-$ and $J-$ band magnitudes. \citet{chrimes2022magnetar} revealed that magnetar surface IR emission can be easily mistaken for a companion star, with IR magnitudes $M_H\gtrsim 2.5$ \citep[see][Figure\,5]{chrimes2022magnetar}. Although interacting binaries such as SGR\,0755-2933 can be far brighter (its companion CPD-29\,2076 has $M_H\sim-4$), passive binaries like PSR\,J0210+5845 are much fainter in the absence of mass transfer \citep[its companion B0V star has $M_H\sim0.5$;][]{2020MNRAS.492.5878T}. Both magnitude and color must be considered to distinguish massive stellar companions from surface emission, which we account for in $p_2$.


\subsection{Optical and IR Search Results}\label{sec:boundresults}

\begin{figure*}
 \includegraphics[width=0.93\linewidth]{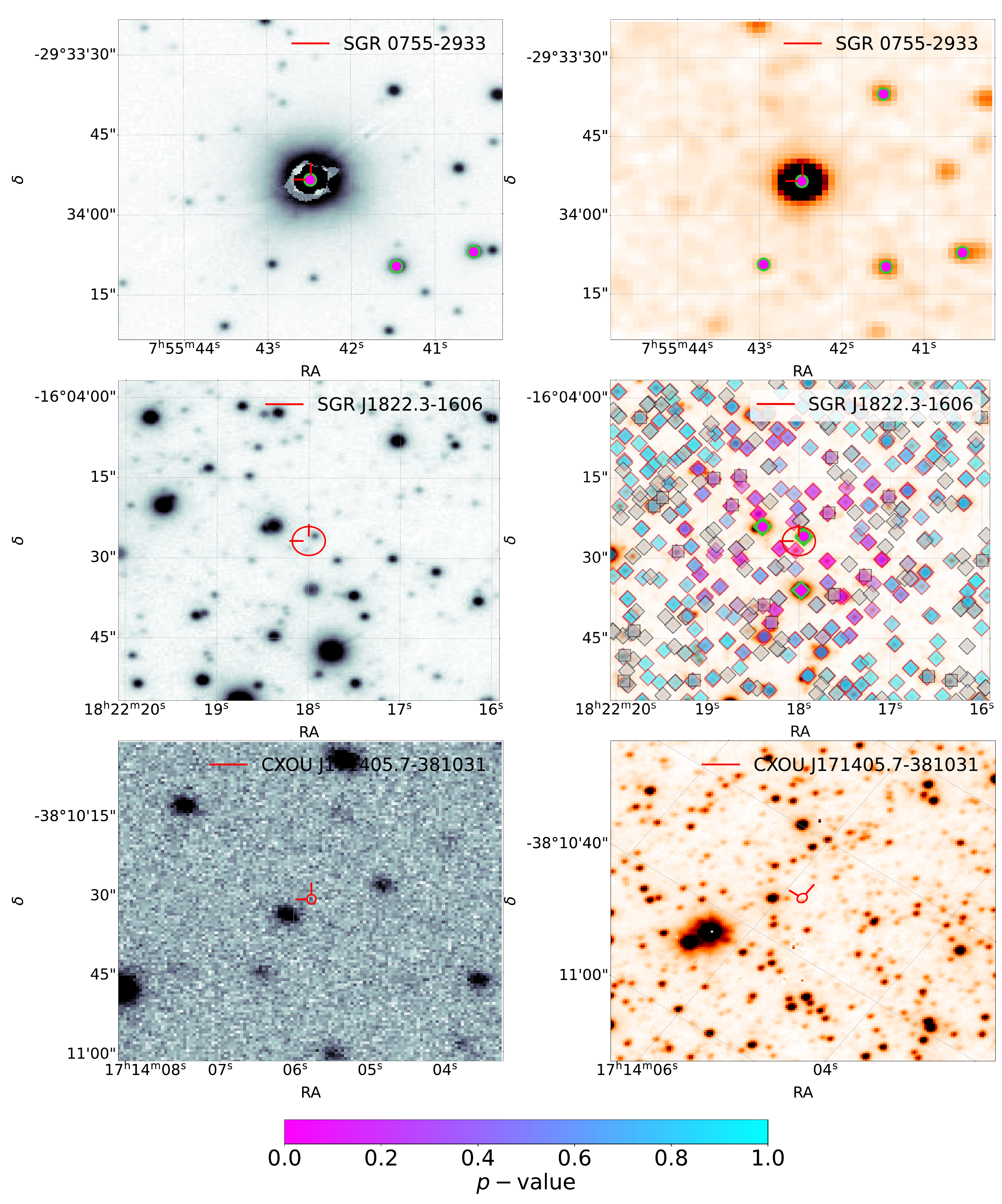}
 \caption{Optical (\textit{Left}) and Infrared (\textit{Right}) $1\arcmin\times1\arcmin$ image cutouts for 2 magnetars and 1 magnetar candidate. Each image is in grayscale where RGB channels are combined using $0.2989R + 0.5870G + 0.1140B$. Optical images use the $g-$, $i-$, and $y-$band filters for PS1 images (CXOU\,J1714) and the $g-$, $r-$, and $i-$band filters for SkyMapper images (SGR\,0755 and SGR\,J1822). Infrared images use the $J-$, $H-$, and $K-$band filters for UKIDSS images (SGR\,1822) and the $J-$, $H-$, and $K_s-$band filters for VVV (CXOU\,J1714) and 2MASS (SGR\,0755) images. The magnetar location is indicated by red cursors and the $3\sigma$ error ellipse is shown in red. Note the error ellipse is much larger than the image for SGR\,0755-2933. Each marker indicates a point source from the optical or infrared catalogs with colorbar corresponding to the association p-value, $p$. Sources with insufficient data to determine a p-value are grey. Sources with $p<5\%$ are outlined in green, while those with $p>5\%$ are outlined in red. (\textit{Left}) PS1 sources are shown as circles, SkyMapper sources are shown as triangles, and sources identified `by-eye' are shown as squares. (\textit{Right}) 2MASS sources are shown as circles, VVV sources are shown as triangles, UKIDSS sources are shown as diamonds, and sources idenified `by-eye' are shown as squares. Sources with no corresponding marker were not identified in the catalogs considered, and could not be well-fit with an ellipse. The Be-star companion of SGR\,0755 is recovered in the optical and infrared searches, and candidates are identified for SGR\,1822. The proposed companion of CXOU\,J1714 is not recovered, but cannot be ruled out as a bound companion due to uncertainty in the extinction estimate. Note that the axes for CXOU\,J1714 differ between optical and infrared due to the use of different World Coordinate System (WCS) projections (ZPN) for VVV data \citep[e.g.][]{greisen2002representations}.}
 \label{fig:bound}
\end{figure*}

Using a 95\% confidence level, we recover the proposed Be star bound companion of magnetar candidate SGR\,0755-2933, CPD-29 2176, through the optical (PS1) and IR (UKIDSS) searches with $p<0.1\%$\footnote{For the optical and IR searches we use 1000 trials, setting lower an upper limits of $0.1 < p < 99.9\%$ on the p-value.} \citep{chrimes2022magnetar,richardson2023high}. In the following sections, we discuss the association of a PS1 source with the companion of SGR\,0755-2933's error circle and the non-recovery of the proposed counterpart of CXOU\,J171405.7-38131. Figure~\ref{fig:bound} shows optical and IR images for these three magnetars. Candidate companions were identified with $p<5\%$ for PSR\,J1622-4950 and SGR\,J1822.3-1606, but are most likely false positives as discussed in Appendix~\ref{app:PS1FPs}.


\subsubsection{PS1 Counterpart of SGR 0755-2933's Be Star Companion}

Five IR candidates and three optical candidates were recovered with $p<5\%$ as potential companions of SGR\,0755-2933. The abundance of sources is due to its large $3\arcmin$ position error \citep{2016ATel.8831....1B}. This sample includes the known 2MASS counterpart to the proposed companion star CPD-29\,2176, 2MASS\,J07554248-2933535. We also recover a PS1 source within the 2MASS source's $60$\,mas error circle, PSO\,J118.9270-29.5649, and therefore conclude this is the optical counterpart of CPD-29\,2176. 

However, we note that the association of SGR\,0755-2933 with CPD-29\,2176 is tenuous due to the former's lack of activity; only one burst localized to a $3\arcmin$ region by the \textit{Swift} telescope has been detected \citep{2016ATel.8831....1B}. While a revised position uncertainty $0.4\arcsec$ was derived by \citet{richardson2023high} by assuming its association with CPD-29\,2176, \citet{doroshenko2021sgr} identify 8 persistent X-ray sources within the $3\arcmin$ error circle. This prevents a confident association of the burst with any persistent source, including CPD-29\,2176. Although \citet{richardson2023high}'s spectral analysis confirms the presence of an HMXB containing CPD-29\,2176 (formally 2SXPS\,J075542.5-293353), a second burst localized to the binary is required to confirm that it contains SGR\,0755-2933.

\subsubsection{CXOU J171405.7-38131}

We do not recover the infrared source, VIRAC\,384090103, proposed by \citet{chrimes2022magnetar} as a potential stellar companion to CXOU\,J171405.7-38131 \citep{smith2018virac,nikzat2022vvv}. If placed at the distance of the magnetar, $3.8\pm0.5$\,kpc, the source has an $H-$band magnitude $M_H\approx-0.9\pm0.1$ consistent with an OB star and its color $J-H\approx-0.5\pm0.3$ differs only marginally. We note that the magnetar lies along a highly extincted sightline outside of the \textit{Bayestar19} range, making its $A_V\approx22.1\pm0.8$ from the $N_H-A_V$ relation an upper limit \citep{predehl1995x}. This biases the Monte Carlo p-value test since the extinction cannot be estimated as a function of distance. Therefore, the location of the magnetar and source make this analysis inconclusive, and we cannot rule out VIRACS\,384090103 as a potential companion of CXOU\,J171405.7-38131.

\section{Radio Search for SNR Shells}\label{sec:RACS}

\subsection{RACS, VLASS, and NVSS Data and Search Method}

While numerous continuum radio searches for SN shells have been conducted \citep[see e.g.][for comprehensive reviews]{green2019revised,lorimer1998search, dubner2015radio} as have searches for compact objects near SNRs \citep[e.g.][]{kochanek2019stellar,kochanek2018cas,kochanek2021supernovae,kochanek2023search}, few targeted searches for SNRs near compact objects have been carried out beyond individual sources. Leveraging the young ages \citep[$\lesssim1$\,Myr, e.g.][]{suzuki2021quantitative} we search for evidence of SNRs among the magnetar sample. 


The RACS-low Data Release 1 (DR1) contains images with $15\arcsec$ resolution at 888 MHz, with a limiting sensitivity of $0.25$\,mJy\,beam$^{-1}$ \citep{mcconnell2020rapid, hale2021rapid}. Public images are smoothed to a common pixel scale $25\arcsec$. 20 magnetars and 6 magnetar candidates without confirmed SNRs fall within the survey region $-80^\circ < \delta < +30^\circ$. We obtained $1^{\circ}\times1^{\circ}$ radio image cutouts via the RACS Data Access Portal\footnote{\url{https://data.csiro.au/}}. For two magnetars (SGR\,0501+4516 and 1E\,2259+586) with $\delta > + 30^\circ$, images from the Karl G. Jansky Very Large Array Sky Survey (VLASS), which covers $\delta > -40^\circ$ with $0.12$\,mJy sensitivity, were obtained using the Canadian Initiative for Radio Astronomy Data Analysis (CIRADA\footnote{\url{https://cirada.ca/vlasscatalogueql0}}) cutout service \citep{2020RNAAS...4..175G,2021ApJS..255...30G,2023ApJS..267...37G}. VLASS Quick-Look images\footnote{Epochs 3.1 and 2.2 are used for magnetars SGR\,0501+4516 and 1E\,2259+586, respectively.} are taken at 3\,GHz with $1\arcsec$ pixel scale, which we bin to $25\arcsec$ to search for extended emission. For three magnetars covered by neither VLASS nor RACS (4U\,0142+61, SGR\,0418+5729, and SGR\,2013+34), we obtain cutouts from the 1.4\,GHz National Radio Astronomy and VLA Sky Survey (NVSS)\footnote{\url{https://www.cv.nrao.edu/nvss/postage.shtml}}, a precursor to VLASS \citep{condon1998nrao}. This catalog contains images with $15\arcsec$ pixels and RMS noise $0.5$\,mJy\,beam$^{-1}$, or $\sim0.8$\,mJy\,beam$^{-1}$ for $25\arcsec$ pixels\footnote{VLASS data is taken in the B configuration which is sensitive to angular scales between $2.1-58\arcsec$. NVSS data is taken in the D and DnC configurations which are sensitive to angular scales between $14-970\arcsec$ and $46-970\arcsec$, respectively. See \url{https://science.nrao.edu/facilities/vla/docs/manuals/oss/performance/resolution} for details.}. One magnetar, SGR\,1627-41, is not covered by any of the radio surveys discussed \footnote{While this magnetar does fall in the survey range of RACS-DR1, no continuum images are available that contain SGR\,1627-41 in its field-of-view.}.

The Set of Identifications, Measurements and Bibliography for Astronomical Data (SIMBAD\footnote{\url{https://simbad.u-strasbg.fr/simbad/sim-fbasic}}) database was queried for known SNRs, SNR candidates, and unidentified radio sources with `ISM' object-types\footnote{See \url{https://simbad.cds.unistra.fr/guide/otypes.htx} for details on SIMBAD object-types.} within the same $1^{\circ}\times1^{\circ}$ region for each magnetar \citep{egret1991simbad}. If a previously proposed SNR candidate was not found in the initial SIMBAD query, it was added manually to the sample for the association test (for example, SNR\,HBH\,9 is associated with SGR\,0501+4516, but falls more than $1.3^\circ$ away and was missed in the query). Additional sources which may be SNRs are identified `by eye' to include in the search (see Appendix~\ref{app:RACSbyeye}). Sources identified as known radio galaxies or AGN were removed from the sample. While a radio image is not available for SGR\,1627-41, we proceed with the association test using SIMBAD radio sources within $1^\circ$. Through a Monte Carlo search, p-values are estimated with the criteria below:


\begin{enumerate}
    \item The SNR and magnetar are located at the same RA and declination.
    \item The magnetar's proper motion direction is consistent with the direction required to intersect the SNR.
    \item The magnetar's proper motion and age are consistent with the observed angular separation of the magnetar and SNR.
    \item The SNR's dynamical age is consistent with an SNR shell rather than a point source.
\end{enumerate}

\noindent for which we define p-values $p_1, p_2$, $p_3$, and $p_4$, respectively. Appendix~\ref{app:pvals_RACS} contains a derivation of these p-values explicitly. Unlike in the binary companion searches, criteria (i)-(iii) relate to two independent cases: first that the magnetar is located within the SNR shell (i), and then that the magnetar has travelled some distance away from the SNR during its lifetime (ii and iii). Therefore, unless $p_1 < 0.1\%$ or $1 - (1-p_2)(1-p_3) < 0.1\%$\footnote{The minimum p-value is set by the 1000 trials used for the SNR search.}, we combine $p_1, p_2$, and $p_3$ using the Fisher statistic\footnote{\url{https://en.wikipedia.org/wiki/Fisher\%27s\_method}}:

\begin{equation}\label{eq:pvaltotradio1}
    \hat{\chi}^{2_4} = -2({\rm ln}(p_1) + {\rm ln}(1 - (1-p_2)(1-p_3)))
\end{equation}

\noindent This follows a $\chi^2$ distribution with 4 degrees of freedom, from which we compute the p-value $p_{\chi}$ using the inverse survival function. We then combine this with $p_4$ using:

\begin{equation}\label{eq:pvaltotradio2}
    p = 1 - (1-p_{\chi})(1-p_4)
\end{equation}

\noindent Note that $p_1$ requires definition of the RA and declination errors for the magnetar and either the RA and declination errors or angular size and position angle for the SNR. $p_2$ and $p_3$ additionally require definition of the magnetar proper motion errors and the X-ray decay or characteristic age. $p_4$ requires the angular size or position error to be defined.

\subsection{Radio SNR Search Test with Known Magnetar-SNR Associations}

8 magnetars and 2 magnetar candidates with confirmed SNR associations are used to test the radio image search method. With the analysis above, we accurately recover the associated SNRs with $p<5\%$ for 7 magnetars and both magnetar candidates. SGR\,1627-41 is the only confirmed SNR association that we fail to recover in the radio search \citep[SNR\,G337.0-00.1; ][]{1999ApJ...526L..29C}, which is likely due to the large angular offset $\sim15.5\arcmin$. No proper motion is known for SGR\,1627-41, but a velocity $\sim1275$\,km\,s$^{-1}$ would be required given its $\mathcal{T}_{\rm decay}\approx37.0$\,kyr, which is in marginal agreement with the \citet{2005MNRAS.360..974H} distribution. Given that 9 of the 10 trials were successful in recovering the associated SNRs, we proceed with confidence that the image-plane and Monte Carlo search methods are effective. Radio images for five magnetars with proposed SNR associations are shown in Figure~\ref{fig:TMPREF}.

\subsection{Radio SNR Search Results}\label{sec:racsresults}

Among the 13 magnetars and magnetar candidates with no previous SNR associations, we identify two candidates which have $p<5\%$. One of these, associated with SGR\,0755-2933 is most likely a false positive which we discuss in Appendix~\ref{app:RACSFPs}. For SGR\,1808-20, we find two viable SNR association candidates which require proper motion and distance measurements to properly assess. Of the eight magnetars and magnetar candidates with proposed but unconfirmed or rejected SNRs we recover five (and marginally recover a sixth with $p=5.6\%$), one of which is nearby the Galactic Center magnetar SGR\,J1745-2900 and another of which would suggest 3XMM\,J185246.6+003317 has an ejected pulsar companion.  In this section, we discuss the SNR candidates for SGR\,1808-20 and the recovered unconfirmed SNR candidates in detail.


\subsubsection{SGR 1808-20}

For SGR\,1808-20, we find multiple candidate radio sources (29) with $p<0.1\%$, 10 of which were selected for analysis `by-eye' rather than through the SIMBAD query. The large number of sources comes from the loose position ( $\sim7\arcmin$ error) and distance (no measurement) constraints, as only one burst has been detected from the magnetar since its discovery \citep{lamb2003sgr}. $2\arcmin\times2\arcmin$ VLASS cutouts at $1\arcsec$ scales were obtained for each candidate, revealing that 6 of the 29 sources were resolved and unlikely to be SNRs. One source is a known Young Stellar Object (YSO) while another is an unidentified sub-millimetric source, both of which are unlikely to be SNRs \citep{2017MNRAS.469.2163E}. 14 are labelled as `dense cores', or InfraRed Dark Clouds (IRDCs) which are pre-star forming regions that may be the birthplaces of massive stars \citep{jiao2023fragmentation}. IRDCs are distinguished from SNRs based on their high masses ($\sim100M_\odot$) and hydrogen column densities ($N_H\sim10^{23}$\,cm$^{-2}$), essentially ruling out these candidates \citep[e.g.][]{guver2009relation}. 

Six of the seven remaining sources are `by-eye' candidates; we query the SIMBAD database within $1\arcmin$ of each source to identify potentially missed sources of radio emission. One is nearby PSR\,J1808-2057 which has a radio flux $2.3$\,mJy and is likely responsible for the observed emission \citep{hobbs2004parkes}. Another is coincident with a $\sim532M_\odot$ molecular cloud discovered through sub-millimeter ($\sim0.1-1$\,THz) emission, making it a plausible radio source \citep{2021MNRAS.500.3027D}. One is coincident with an infrared-bright `bubble'; these are star-forming sites in HII regions and are therefore also plausible radio sources \citep{2012MNRAS.424.2442S}. Two others are coincident with stars, one an AGB star and the other an eclipsing binary system, both of which are potentially radio bright, though no radio flux has yet been catalogued \citep{2018AJ....156..241H,spangler1977radio,2003tmc..book.....C}. While these five sources cannot be confidently ruled out, we will assume they are not associated given the possible contamination by nearby objects. This leaves two sources as candidates for association: SNR\,G009.7-00.0 and an unidentified radio source, which we formally label S24-RACS\,J180842-202849, with no SIMBAD objects within $1\arcmin$. For the remaining discussion, we will refer to the latter source as `Source\,1'.


Source\,1 was identified by eye at $\rm18^{\rm h}08^{\rm m}42^{\rm s}-20^\circ28\arcmin45\arcsec$ and fit with an ellipse of approximate major and minor axes $a\sim b \sim 18\arcsec$. This lies very near the position of another magnetar, SGR\,1806-20, and the W31 HII region \citep[e.g.][]{frail1989kinematic,corbel2004connection,moises2011spectrophotometric}. The distance to W31 has been disputed in the literature; if we adopt the distance of SGR\,1806-20, $8.7^{+1.8}_{-1.5}$\,kpc, Source\,1 would have a radius $\sim0.8$\,pc. At this size the SNR would be in the free-expansion phase, and assuming $n_0\sim1$\,cm$^{-3}$, we estimate the age of the remnant to be between $0.3-1.2$\,kyr \citep{sedov2018similarity}. This is roughly consistent with $\mathcal{T_{\rm rot}}$ and $\mathcal{T}_{\rm decay}$ for magnetars, but the lack of data for SGR\,1808-20 prevents a direct comparison. A detailed composition analysis is required to confirm Source\,1 as a supernova remnant, though its morphological similarity to other false positives implies it is unlikely to be an SNR.

SNR\,G009.7-00.0, on the contrary, was detected by \citet{frail1994radio} and confirmed through its OH maser by \citet{hewitt2009discovery}, which constrained its distance to $\sim4.7$\,kpc. \citet{2016ApJ...827...41Y} find its distance is inconsistent with W31 and SGR\,1806-20. At $\sim4.7$\,kpc, the $\sim11-15\arcmin$ angular size corresponds to a radius $\sim8-10$\,pc, suggesting a free-expansion age of $\sim10$\,kyr. Both Source\,1 and SNR\,G009.7-00.0 appear to be viable candidates for association with SGR\,1808-20, but a magnetar distance measure and more tightly constrained position are necessary to investigate further. The radio image is shown in Figure~\ref{fig:TMPREF}.

\subsubsection{3XMM J185246.6+003317 and CXOU\,J185238.6+004020 as an Unbound Double Neutron Star Binary}

For 3XMM\,J185246.6+003317 we recover the Kes\,79 SNR (SNR\,G033.6+00.1) with $p<0.1\%$, which is associated with a central X-ray pulsar CXOU\,J185238.6+004020 \citep[PSR\,J1852+0040; e.g.][]{2005ApJ...627..390G,2005ATel..501....1M,2014ApJ...790...94B}. \citet{2014ApJ...790...94B} derive a Sedov-Taylor expansion age of $\sim5$\,kyr for the SNR which contrasts with both the decay age $\mathcal{T}_{\rm decay}\approx0.7$\,Myr\footnote{More recent constraints on the magnetic field from ray-tracing simulations by \citet{delima2024evidence} suggest $\mathcal{T}_{\rm decay}\gtrsim1.2$\,Myr, though they caution not to over-interpret their initial estimates.} and spindown age $\mathcal{T}_{\rm rot}\gtrsim1.3$\,Myr of 3XMM\,J185246.6+003317.\footnote{While the spindown age of CXOU\,J185238.6+004020 is $\sim192$\,Myr, `anti-magnetars' or `compact central objects' (CCO) like this are known to have overestimated spindown ages due to the complex relation of their low magnetic fields to a dynamo process \citep{2008AIPC..983..320G,2007ApJ...665.1304H,2013ApJ...765...58G}.} Magnetars are typically expected to be unobservable beyond $\sim10^{4}$\,years; however,the spindown age $\mathcal{T}_{\rm rot}$ is a well-constrained lower limit derived by \citet{2014ApJ...781L..17R} from an X-ray timing solution. They find a pulse period $P=11.55871346(6)$\,s with no discernible spindown, leading to upper limits of $\dot P < 1.4\times10^{-13}$\,s\,s$^{-1}$ and $B_{\rm dip}\lesssim10^{13}$\,G. on the spindown rate and magnetic field. While these paramters are reminiscent of typical radio pulsars, \citet{2014ApJ...781L..16Z} noted in their initial discovery that the X-ray luminosity ($\sim10^{33-34}$\,erg\,s$^{-1}$) far exceeds its spindown luminosity ($\sim10^{30}$\,erg\,s$^{-1}$) making its magnetar nature robust. Therefore we take both the spindown age and the decay age to be reasonable age estimates. This suggests that Kes\,79 is associated with CXOU\,J185238.6+004020, but allows a scenario wherein the pulsar and magnetar were gravitationally bound in the past \citep{2014ApJ...781L..16Z}.

\begin{figure*}
 \includegraphics[width=\linewidth]{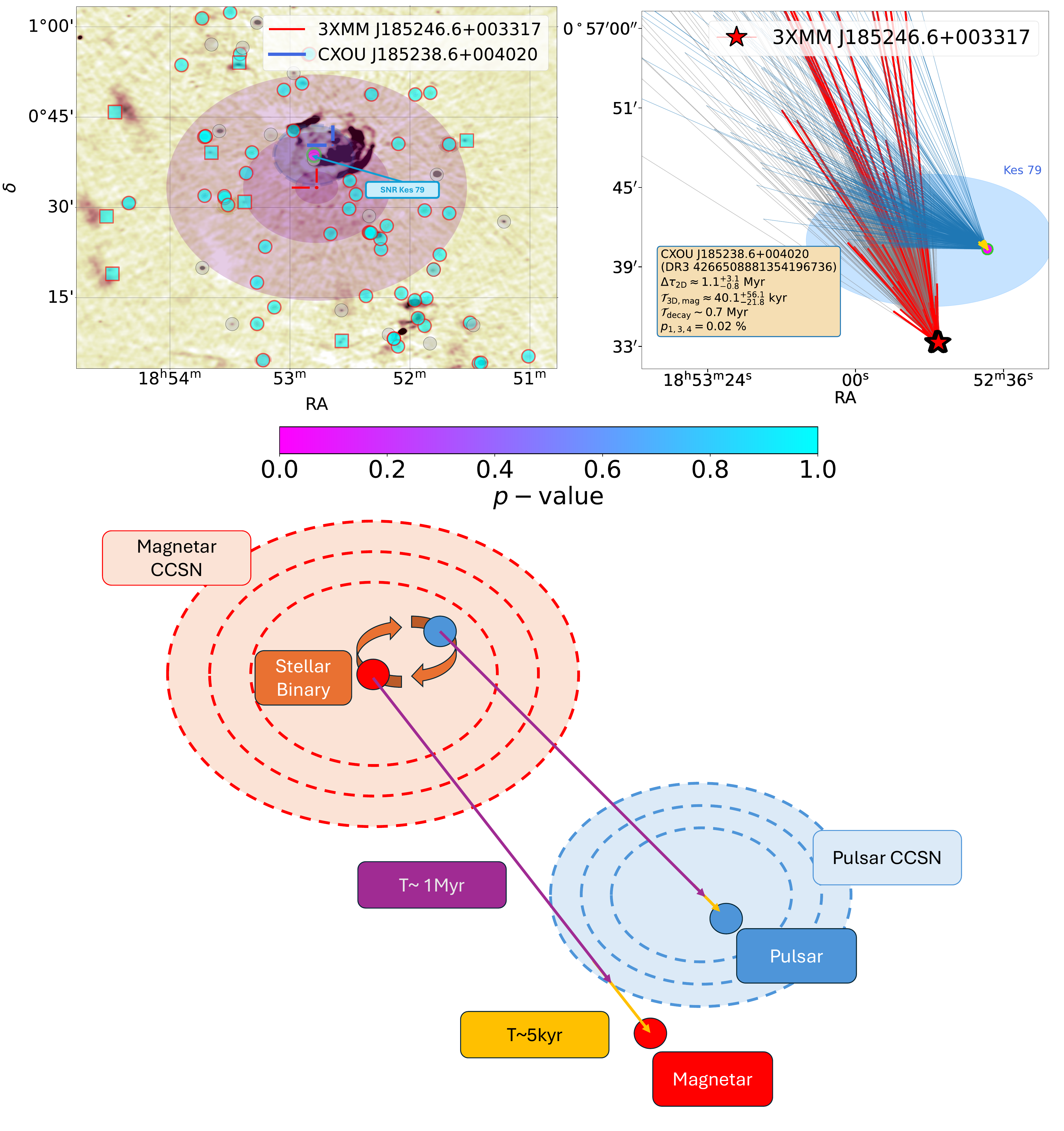}
 \caption{Radio Image and Trajectory Analysis of 3XMM\,J185246.6+00317. (\textit{Top left}) A RACS DR1 888\,MHz continuum $1^\circ\times1^\circ$ image cutout is shown for 3XMM\,J185246.6+00317 with $25\arcsec\times25\arcsec$ pixels. The magnetar and pulsar locations are indicated by red and blue pointers, while circular and square markers show known and `by-eye' identified radio sources, respectively. Note that no markers are shown for catalogued Active Galactic Nuclei or radio galaxies, as they were not included in this search. Other extended sources without markers were not identified in SIMBAD and could not be well-fit with an ellipse. The colorbar corresponds to the association p-value. The purple shaded region indicates the $1$, $2$, and $3\sigma$ past travel distances based on the \citet{2005MNRAS.360..974H} velocity distribution. The angular extent of Kes\,79 (labelled) is shown as a blue circle. (\textit{Bottom}) The cartoon demonstrates how a bound massive stellar binary (orange) may evolve into an unbound magnetar (red) and pulsar (blue) following two CCSNe. (\textit{Top right}) The trajectory plot for 3XMM\,J185246.6+00317 and the X-ray pulsar CXOU\,J185238.6+004020 is shown with 468 valid trajectories (out of 10000 trials). Blue traces show the pulsar's trajectory prior to the second CCSN, and yellow traces show its trajectory after the CCSN. The blue circle indicates the location and extent of Kes\,79.}
 \label{fig:3XMM}
\end{figure*}

Considering this case, let the initially bound massive star progenitors of 3XMM J185246.6+003317 and CXOU\,J185238.6+004020 be called stars 1 and 2, respectively. When the CCSN of star 1 disrupts the binary\footnote{The alternate case in which star 2 undergoes the first CCSN is equally likely, but we we focus the discussion on this scenario which is more consistent with the age estimate for 3XMM\,J185246.6+003317.}, the remnant magnetar (3XMM J185246.6+003317) and star 2 are ejected at $\sim300$\,km\,s$^{-1}$ and $\sim6$\,km\,s$^{-1}$, respectively \citep{2005MNRAS.360..974H,renzo2019massive}. After travelling for $\sim0.7$\,Myr based on the magnetar's age, the CCSN of star 2 would then kick the remnant pulsar (CXOU\,J185238.6+004020) at $\sim300$\,km\,s$^{-1}$, leaving behind the remnant shell Kes\,79. This sequence of events is shown in Figure~\ref{fig:3XMM}.
We investigate this scenario by applying a modified \textit{Gaia} unbound companion search pipeline to the magnetar and CXOU\,J185238.6+004020's \textit{Gaia} counterpart, \textit{Gaia}\,DR3\,4266508881354196736. We split the companion's trajectory into two parts with different velocities to account for the two CCSN events, assume the absolute magnitude of the companion is $M_V > 0$, and neglect $p_2$ since the magnetar's and pulsar's travel times no longer coincide directly. 10000 trials (rather than 100) are used since the magnetar has no proper motion measurment.

The resulting trajectory plot is shown in Figure~\ref{fig:3XMM}; we find a p-value $p_{1,3}=0.02\%$, suggesting that this scenario is likely. From the 497 valid samples out of 10000, the scenario requires the magnetar to have velocity $|\overrightarrow{v}_{\rm mag}|\approx398^{+208}_{-165}$\,km\,s$^{-1}$ (proper motion $\mu_{\alpha}{\rm cos}(\delta)\approx-2.1^{+1.2}_{-5.8}$\,mas\,yr$^{-1}$, $\mu_{\delta}\approx-9.8^{+4.9}_{-20.8}$\,mas\,yr$^{-1}$) and a travel time $\mathcal{T}_{\rm 3D,mag}\approx40.1^{+56.1}_{-21.8}$\,kyr, which is consistent with expected magnetar ages, but disagrees with \citet{2014ApJ...781L..17R}, \citet{2014ApJ...781L..16Z}, and $\mathcal{T}_{\rm decay}$. Despite the robustness of the X-ray timing solution, magnetars are known to glitch, anti-glitch, and slowly recover from glitches during outbursts such as the one used for timing analysis \citep[e.g.][]{tong2020magnetar,younes2023magnetar,scholz2014long,archibald2013anti}. Coupled with \citet{2014ApJ...781L..17R} observing the outburst at a late time, this implies that 3XMM\,J185245.6+003317 may have experienced a glitch event leading to its undetectable spindown. If we impose $\mathcal{T}_{\rm 3D,mag}$ as an age constraint, then using the period $P=11.55871346(6)$\,s we derive a revised spindown rate of $\dot{P}\approx4.6_{-2.7}^{+5.4}\times10^{-12}$\,s\,s$^{-1}$, and dipolar magnetic field $B_{\rm dip}\approx2.3^{+1.1}_{-0.2}\times10^{14}$\,G, both of which are more consistent with the magnetar sample. 

The pulsar requires a velocity $|\overrightarrow{v}_{\rm psr,1}|\approx8^{+11}_{-4}$\,km\,s$^{-1}$ following the first CCSN, and $|\overrightarrow{v}_{\rm psr,2}|\approx422^{+192}_{-172}$\,km\,s$^{-1}$ following the second. The most supportive evidence for this scenario is the position of the pulsar's CCSN, estimated from valid samples as $\rm18^{\rm h}52^{\rm m}38\fs60+00^{\circ}40\arcmin20\farcs09$ with $1\sigma$ errors of $0.2\arcmin$ and $0.4\arcmin$ in RA and declination respectively. This is roughly $8\sigma$ away from the centroid of Kes\,79, $\rm18^{\rm h}52^{\rm m}48^{\rm s}+00^{\circ}41\arcmin00\arcsec$, but is well within its $\sim5\arcmin$ radius. We obtain an estimate of $\rm18^{\rm h}52^{\rm m}56\fs5+00^{\circ}43\arcmin57\farcs5$ with errors $\sigma_{\alpha}\sim3\arcmin$ and $\sigma_{\delta}\sim6\arcmin$ for the position of the first CCSN. This also lies within the extent of Kes\,79, and suggests that its unique multi-shell structure may result from interactions between the magnetar and pulsar SNR shock fronts. Other single-SNR scenarios have been presented to explain this structure as, for example, the interaction of the SNR with a molecular cloud or the progenitor's stellar wind \citep{1991AJ....102..676V,2009A&A...507..841G,2004ApJ...605..742S,2016PASJ...68S...8S,2016ApJ...831..192Z}.
We conclude that it is likely that the progenitors of 3XMM\,J185246.6+003317 and CXOU\,J185238.6+004020 began in a bound system prior to the former's CCSN, and a proper motion measurement for 3XMM\,J185246.6+003317 would further constrain their formation history.

\subsubsection{SGR J1745-2900, the Galactic Center Magnetar}

SGR\,J1745-2900 is a unique magnetar at the Galactic Center (GC) of the Milky Way \citep{2013ATel.5020....1M}. The radio emission from the supermassive black hole Sag\,A$^*$ and high density of radio sources in this region make it difficult to identify any associations with confidence. The black hole would also distort an SNR shell from SGR\,J1745-2900 \citep[e.g.][]{yalinewich2017evolution}. However, \citet{ponti2015xmm} used high resolution X-ray observations to catalog extended regions near the GC, identifying the GC X-ray/radio lobes as a warped and mirrored SNR. SGR\,J1745-2900 lies near the intersection of the two lobes, and \citet{yalinewich2017evolution} use numerical simulations to demonstrate that an `engulfing' SNR explosion at this location could produce the observed SNR structure. 

\begin{figure*}
 \includegraphics[width=0.92\linewidth]{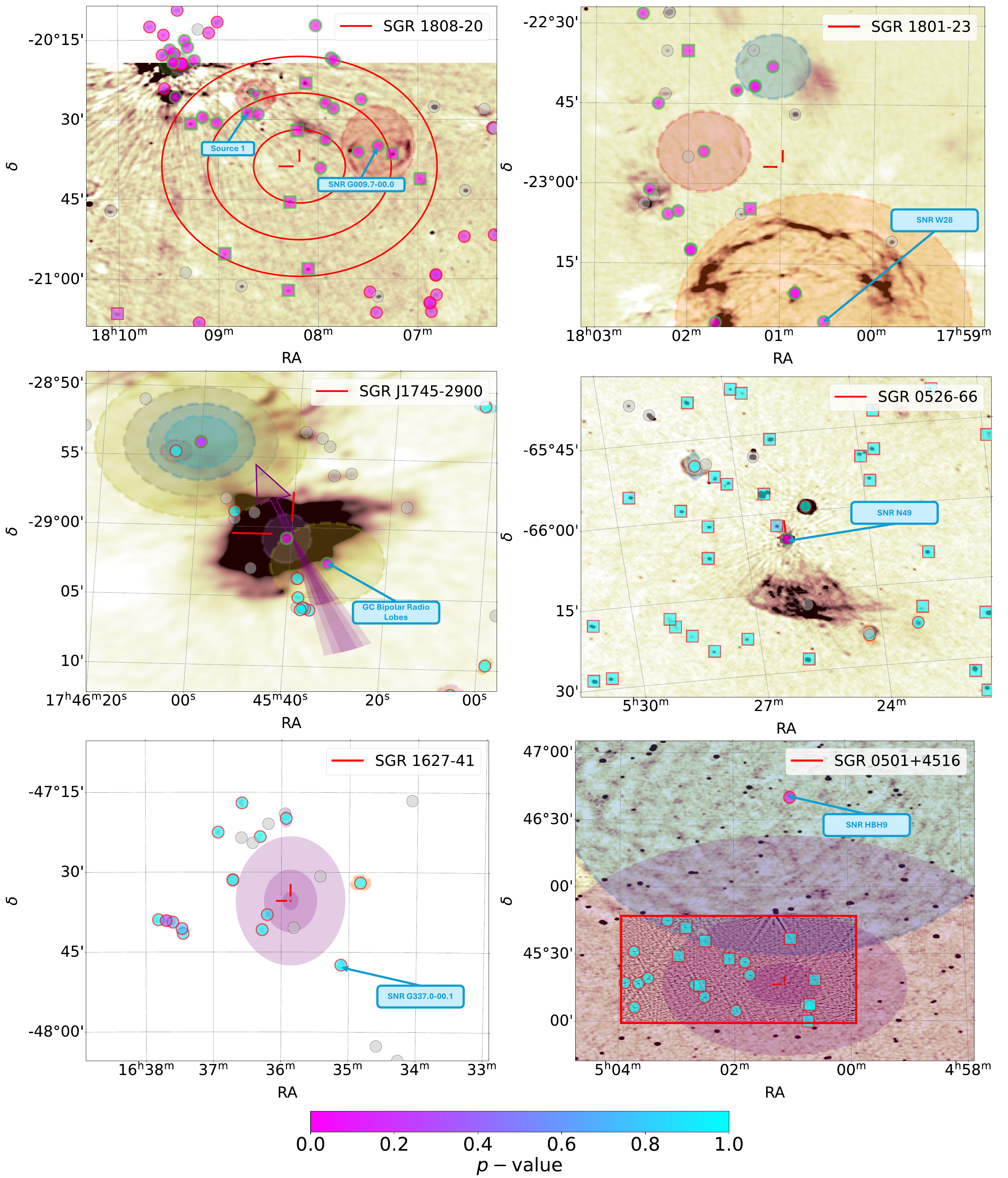}
 \caption{Radio continuum image cutouts for four magnetars (\textit{middle, bottom}), and two magnetar candidates (\textit{top}) with possible SNR associations. Red pointers indicate each magnetar's location and the $1\sigma$, $2\sigma$, and $3\sigma$ magnetar error ellipses are shown in red (although they are too small to be visible except for SGR\,1808-20 and SGR\,1801-23. For the latter, the $3.8^\circ$ error ellipse is too large to be effectively shown.). Note the difference in axis extents for each image. RACS 888\,MHz continuum images with $25\arcsec\times25\arcsec$ pixels are shown for SGR\,1808-20, SGR\,1801-23, SGR\,J1745-2900, and SGR\,0526-66 (\textit{top, middle}). For SGR\,0501+4516, a 3\,GHz $1\arcmin\times1\arcmin$ VLASS image with $1\arcsec\times1\arcsec$ pixels (smoothed to $25\arcsec\times25\arcsec$ with a boxcar kernel) is shown enclosed in a red box indicating the queried region, while the remainder of the image is from the 1.4\,GHz NVSS survey to show SNR\,HBH\,9 \citep[e.g.][]{2010ApJ...722..899G}. However, NVSS is only sensitive to angular sizes $<970\arcsec$ ($\lesssim16.7\arcmin$), making SNR\,HBH9 ($120-140\arcmin$) unobservable in the image shown. No radio images from RACS, VLASS, or NVSS were available at SGR\,1627-41's position (\textit{bottom left}). Circular markers indicate the locations of known radio sources while square markers indicate unknown sources identified by-eye. Note that no markers are shown for catalogued Active Galactic Nuclei or radio galaxies, as they were not included in this search. Other extended sources without markers were not identified in SIMBAD and could not be well-fit with an ellipse. The colorbar corresponds to the association p-value; sources outlined in red (green) have $p>5\%$ ($p\leq5\%$). The most likely or previously proposed SNRs (see Section~\ref{sec:racsresults}) are labelled. `Source\,1' (\textit{top left}) is S24-RACS\,J180842-202849. The angular extents of SNRs, if available, are shown as ellipses. For SGR\,1745-2900 (\textit{middle left}), the proper motion direction is marked by a purple arrow, and the purple shaded region indicates the $1\sigma$, $2\sigma$, and $3\sigma$ past travel direction and extent; for SGR\,1627-41, SGR\,0501+4516, and SGR\,0526-66 (\textit{bottom, middle right}), the \citet{2005MNRAS.360..974H} velocity distribution is used to show possible past travel direction and extent.}
 \label{fig:TMPREF}
\end{figure*}

Given this analysis, the radio lobes were added to the simulation using the centroid position ($\rm17^{\rm h}45^{\rm m}32\fs3-29^\circ2\arcmin32\farcs5$) and angular diameter $\sim5.8\arcmin$ from \citet{ponti2015xmm}. Through the Monte Carlo simulation, we recover the radio lobes, as well as a gamma-ray source 0FGL\,J1746.0-2900 found not to be an SNR \citep{2016ApJS..224....8A} and an HII region \citep{2015ApJ...812....7N}, with $p<0.1\%$. In our simplified model of the lobes, we do not account for the irregular structure of the SNR, but instead assume the SNR is centered at the lobes' centroid  with angular size equal to a single lobe. This simplification should not greatly impact the results since only a lower limit is imposed on the SNR's dynamical age based on the angular size. The position of the radio lobes' centroid is marginally consistent with the trajectory of SGR\,J1745+1900 (see Figure~\ref{fig:TMPREF}). With a proper motion magnitude $\sim6.4$\,mas\,yr$^{-1}$, the $\sim2.7\arcmin$ separation suggests an SNR age $\sim25$\,kyr, which is slightly lower than the decay age estimate $\mathcal{T}_{\rm decay}\approx64.9$\,kyr. Further interpretation requires a more detailed analysis of the GC lobe geometry, but our results support a scenario where the bipolar radio lobes are the SNR of SGR\,J1745+1900. 



\subsubsection{Other Magnetars with Unconfirmed SNRs}

Five other magnetars in the sample were identified in previous searches as potential associations of nearby supernovae, but have not yet been confirmed: SGR\,1900+14, SGR\,1806-20, SGR 0526-66, SGR\,0501+4516, and SGR\,1801-23. SGR\,1900+14's association with SNR\,G042.8+00.6 (for which we find $p=5.6\%$) was rejected by \citet{2000ApJ...533L..17V} given the former's more likely association with a star cluster. We identify one additional by-eye source, S24-RACS\,J190707+091737, as an SNR candidate with $p<0.1\%$, but rule it a false positive as it disagrees with the proper motion direction of SGR\,1900+14. SGR\,1806-20's association with AX J1808.6-2024 (SNR\,G10.0-0.3) was also rejected by \citet{1999ApJ...523L..37H}, suggesting it is not an SNR but a radio nebula powered by its central LBV star. While there is insufficient position data to estimate a p-value for the nebula, we do recover it when using the angular size of its potentially associated gamma ray emission, Fermi\,J1808.2-2029 \citep{2018A&A...612A..11H}. Finally, we do not recover XTE\,J1810-197's association with SNR\,G11.0-0.0, supporting the age considerations which ruled out the candidate in \citet{ding2020magnetar}.



SGR\,0526-66 is located within the Large Magellanic Cloud (LMC) near SNR\,N49 \citep[SNR\,J052559-660453;][]{2004ApJ...609L..13K}. However, its association with SNR\,N49 could not be confirmed as the magnetar's proper motion is unknown, which is necessary to localize its birth place to SNR\,N49's host star cluster, SL\,463. With our pipeline, we obtain a p-value $p<0.1\%$ for this association. Notably, five other `by-eye' radio sources were recovered with $p<5\%$; however, each of these fall within $1\arcmin$ of possible known radio sources that are not within the LMC. While SGR\,0526-66 does not have a proper motion measurement, its position error of $0.6\arcsec$ compared to the $1.4\arcmin$ angular size result in a highly confident association. The large distance $53.6\pm1.2$\,kpc suggests N49 has a tranverse size of roughly $22$\,pc, while the $0.6\arcsec$ position error for the magnetar corresponds to $\sim0.1\,$pc. Therefore it is likely that N49 contains SGR\,0526-66, supporting \citet{2004ApJ...609L..13K}'s claim of association. 

For SGR\,0501+4516, \citet{gaensler2008sgr} localized the magnetar to the outer rim of SNR\,HBH\,9 (HB9 or SNR\,G160.9+2.6) at a position $\sim15\arcmin$ from the next nearest SNR. The large angular separation $\sim1.3^\circ$, however, initially excluded SNR\,HBH\,9 from the test sample. After adding this as a bye-eye radio source, we obtain a p-value $p=7.0\%$, providing insufficient evidence to confirm the association. This is most likely due to the large angular separation $\sim1.3^\circ$; at the magnetar's distance $2\pm0.3$\,kpc, this corresponds to a transverse separation of $\sim45$\,pc. For the magnetar's decay age $\mathcal{T}_{\rm decay}\approx59.4$\,kyr, this would require a velocity $\sim730$\,km\,s$^{-1}$ for the magnetar, which agrees well with the \citet{2005MNRAS.360..974H} distribution expected for neutron stars. Therefore, given the marginal p-value recovered and the achievable kick velocity, we find this association remains possible, and a proper motion measurement is required to confirm or reject it.

\section{Discussion}\label{sec:discussion}


Through this multiwavelength search of 31 magnetars and magnetar candidates, we identify one magnetar with a potential unbound OB star companion, SGR\,J1822.3-1606, but require proper motion data to explore further. We note that magnetar 3XMM\,J185246.6+003317 may be an unbound system whose progenitor was bound to that of an X-ray pulsar, CXOU\,J185238.6+004020. While we recover the Be star companion of SGR\,0755-2933 proposed by \citet{richardson2023high} and the unbound Wolf-Rayet star companion of CXOU\,J164710.2-455216 proposed by \citet{clark2014vlt},  we are unable to confidently associate the IR counterpart of CXOU J171405.7–381031 proposed as a bound companion by \citet{chrimes2022magnetar}. Through the radio search, we recover nine of ten confirmed SNR associations, propose scenarios supporting four of five unconfirmed SNR associations, and identify a new candidate SNR associated with SGR\,1808-20 which requires a proper motion and distance measurement for confirmation. 

Table~\ref{table:results} summarizes our results, which prompt a discussion of the CCSN model for magnetar formation. The recovery of only two unbound binary candidates is inconsistent with  \citet{chrimes2022magnetar} and \citet{renzo2019massive} simulations predicting that $f_u\sim 38-56\%$ of magnetars should be nearby unbound companion stars. We begin this section by analyzing our search's sensitivity before exploring the implications of the search on magnetars' massive star progenitors. We then consider potential limitations in our work and investigate the possibility of more exotic magnetar formation channels.


\subsection{Completeness and Sensitivity Considerations}\label{sec:complete}

For each search, we estimate the expected ranges of intrinsic parameters and derive the ranges of their corresponding observables in order to compare to our sensitivity thresholds. For the bound companion search, we search for OB stars with intrinsic magnitudes in the following ranges:

\begin{enumerate}
    \item $-2.76 \lesssim M_H \lesssim 4.34$ \citep{chrimes2022magnetar}
    \item $-2.61 \lesssim M_i \lesssim 5.55$ \citep[derived by converting the ][ constraints on $M_H$ and $J-H$ using the 2MASS relationships\footnote{See Section 5.5 of \textit{Gaia} DR3 Documentation (\url{https://gea.esac.esa.int/archive/documentation/GDR3/}).}]{chrimes2022magnetar}
\end{enumerate}

\noindent Similarly in the \textit{Gaia} search for unbound companions we expect:

\begin{enumerate}
    \item $-10 \lesssim M_V \lesssim 1.185$
    \item $v_{\rm mag}$ following the \citet{2005MNRAS.360..974H} distribution
    \item $v_{\rm src}$ following the \citet{renzo2019massive} distribution
\end{enumerate}

\noindent and for the SNR search:
\begin{enumerate}
    \item Radio luminosity $L\approx10^{42}$\,erg\,s\,$^{-1}$ in the Rayleigh-Jeans limit \citep{barker2022connecting,barker2023inferring,hamuy2003observed}
    \item Angular size $25\arcsec \lesssim \Delta \Theta \lesssim 1^\circ$ (from observed SNR sizes in the \citet{green1987catalogue} catalog and considerations of free- and Sedov-Taylor-expansion in \citet{truelove1999evolution})
\end{enumerate}

\begin{figure}
 \includegraphics[width=0.88\linewidth]{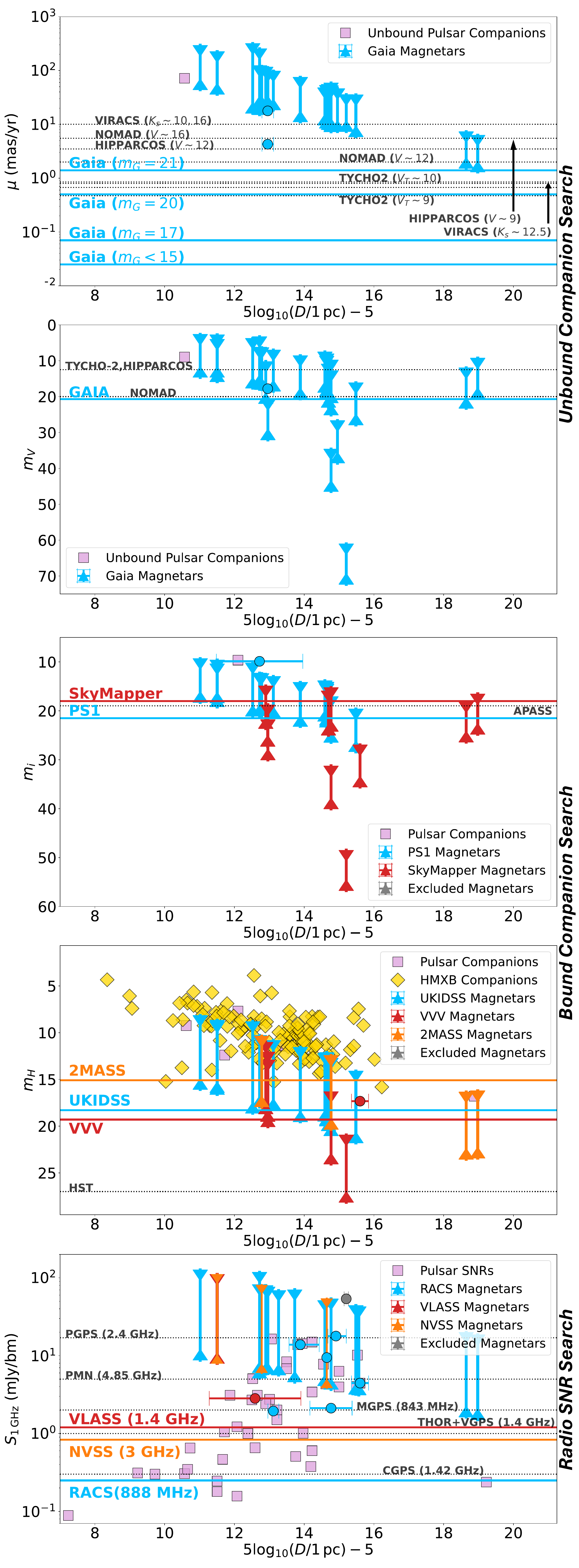}
 \caption{Completeness plots versus distance modulus for (\textit{rows 1 and 2}) the \textit{Gaia} search, (\textit{row 3}) the optical and (\textit{row 4}) IR searches, and (\textit{row 5}) the radio search. Circular, square, and diamond markers with errorbars indicate known companions or remnants of magnetars, pulsars, and HMXBs, respectively. Proper motion, apparent magnitude, and 1\,GHz flux ranges, are shown as vertical lines, with color relating to the survey used for the search. The sensitivity limits of each survey are indicated by horizontal lines. We expect most magnetars with SNR shells or stellar companions would have been detected if present. Sensitivity limits for past searches are shown as black dotted lines for comparison (see Appendix~\ref{app:pastsurveys} for references).}
 \label{fig:sensitivity}
\end{figure}

\noindent By combining these constraints with the distance, extinction, and proper motion of each magnetar (see Table~\ref{table:magnetars}), we estimate the expected range of apparent magnitudes ($V-$, $i-$, and $H-$band), SNR flux at 1\,GHz, and relative proper motion $\Delta \mu$ between the magnetar and an unbound companion. Figure~\ref{fig:sensitivity} and Table~\ref{table:results} summarize the results. Four magnetars have no distance measurements and are not included in this analysis.

For the \textit{Gaia} search we observe that the relative proper motion of each magnetar and a walk/runaway companion should be detectable down to $M_G\sim21$ given the proper motion uncertainty $\sigma_{\mu}\approx1.4$\,mas\,yr$^{-1}$ \citep{gaiacollaboration2022gaia}. Only the SMC magnetar CXOU\,J010043.1-721134 is partially incomplete for faint stars due to its large distance. However, four magnetars lie along high-extinction sightlines with $A_V\gtrsim17$ (CXOU\,J171405.7-381031, 1E\,1547.0-5408, PSR\,J1622-4950, SGR\,1627-41) making unbound companions undetectable at \textit{Gaia}'s $M_V<20.7$ sensitivity. 19 magnetars are partially complete, meaning that our search may miss fainter B-star companions. These results are more promising than indicated by the simulations of \citet{chrimes2023searching}, which determined that without the aid of next generation deep-field surveys, \textit{Gaia} may recover only a small fraction of unbound neutron star companions. This demonstrates that incorporating a trajectory search into our Monte Carlo analysis can make significant improvements to our search sensitivity.

The bound search is complete or partially complete for most magnetars covered by the IR survey; only SGR\,1627-41, which has a large extinction $A_V\approx55.87\pm11.17$, and the two magnetars in Magellanic Clouds, CXOU\,J010043.1-721134 and SGR\,0526-66 are incomplete. The optical search, on the other hand, is incomplete for six magnetars due to their high extinction. This is in agreement with \citet{chrimes2022magnetar}, which concluded that IR searches are more effective for OB stars due to the $\sim4\times$ lower extinction in $H-$band compared to $i-$band \citep[e.g.][]{green2019revised}. 

Finally, the SNR search is complete in all cases for typical radio luminosities and angular sizes in the free- or Sedov-Taylor expansion phases \citep[e.g.][]{truelove1999evolution,salpeter1955luminosity}. We note, however, that a bright synchrotron radio shell may not be produced if the magnetar inhabits a low-density local environment or lacks a significant internal energy source \citep[e.g.][]{frail1995does,chevalier1985wind,nomoto1987evolution,leung2020electron}. Future work may expand on this sensitivity analysis with detailed forward modelling of SNRs, but this is beyond the scope of this work. Table~\ref{table:results} indicates whether each magnetar's search is considered `partially or fully complete' (\checkmark) or `incomplete' (\xmark) in the above analysis. We conclude that our search for companion stars and SNRs is reasonably sensitive along sightlines with typical extinction and distances $< 10$\,kpc.



\begin{table*}
 \begin{center}
 \tiny
 \caption{Search Results and Completeness Ratios}
 \label{table:results} 
 \begin{tabular}{c||c|c||c|c||c|c||c}
 \hline
  \textbf{Magnetar} & \begin{tabular}{@{}c@{}} \textbf{SNR} \\ \textbf{Candidate} \\ \end{tabular} & \begin{tabular}{@{}c@{}} \textbf{Radio SNR}\\ \textbf{Complete?} \\ \end{tabular} & \begin{tabular}{@{}c@{}} \textbf{Bound} \\ \textbf{Companion} \\ \textbf{Candidate} \\ \end{tabular} & \begin{tabular}{@{}c@{}} \textbf{Optical/IR } \\ \textbf{Bound} \\  \textbf{Complete?}\end{tabular} & \begin{tabular}{@{}c@{}} \textbf{Unbound} \\ \textbf{Companion} \\ \textbf{Candidate} \\ \end{tabular} & \begin{tabular}{@{}c@{}} \textbf{\textit{Gaia}} \\ \textbf{Unbound} \\ \textbf{Complete?} \\ \end{tabular} & \begin{tabular}{@{}c@{}} \textbf{Prior} \\ \textbf{Association} \\ \textbf{Candidate(s)} \\ \end{tabular} \\
  \hline
    CXOU\,J0100 &  \textit{ None } & \checkmark (R) & None & \checkmark (S,T) & -- & \checkmark & -- \\
    4U\,0142 & \textit{None} & \checkmark (N) & \textit{None}  & \checkmark (P,T) & \textit{None} & \checkmark & --\\
    SGR\,0418 & \textit{None} & \checkmark (N) &  \textit{None} & \checkmark (P,U) & \textit{None} & \checkmark & --\\
    SGR\,0501 & \textit{None} & \checkmark (V) & \textit{None} & \checkmark (P,U) & \textit{None} & \checkmark & HBH\,9 (1) \\
    SGR\,0526 &  N49 \{p<0.1\%\}  & \checkmark (R) & \textit{None} &  \xmark (S,U)& \textit{None} & \checkmark &  N49 (2)\\
    1E\,1048 & \textit{None} & \checkmark (R) & \textit{None} &  \checkmark (S,T)& \textit{None} 
    & \checkmark & --\\
    1E\,1547 & G327.2-00.1 \{p=1.5\%\} & \checkmark (R) & \textit{None} &  \checkmark (S,V) & -- & \xmark & G327.2-00.1  (3) \\
    PSR\,J1622 & G333.9+00.0 \{p<0.1\%\}  & \checkmark (R) & \begin{tabular}{@{}c@{}} \textit{SMSS\,J162244.99-495055.7 \{p<0.1\%\}} \end{tabular} &  \checkmark (S,V) & -- & \xmark & G333.9+00.0 (4) \\
    SGR\,1627 & \textit{None} & \checkmark (R) & \textit{None} &  
    \xmark (S,V)& -- & \xmark & G337.0-00.1 (5) \\
    CXOU\,J1647 & \textit{None} & \checkmark (R) & \textit{None} &  \checkmark (S,V) & \textit{None} & \checkmark &  WR\,77F (6) \\
    1RXS\,J1708 & \textit{None} & \checkmark (R) & \textit{None} &  \checkmark (S,V) & \textit{None} & \checkmark & -- \\
    CXOU\,J1714 & CTB\,37B \{p<0.1\%\}& \checkmark (R) & \textit{None} &  \checkmark (S,V)& \textit{None} & \xmark & \begin{tabular}{@{}c@{}} \{CTB\,37B, \\ VIRAC\,384090103 \\(7,8)\} \\ \end{tabular} \\
    SGR\,J1745 & \begin{tabular}{@{}c@{}} \{\textit{0FGL J1746.0-2900} \textit{\{p<0.1\%\}} \\ GC Bipolar Radio Lobes \{p<0.1\%\} \\ \textit{NNM2015 \{p<0.1\%\}}\}\end{tabular} & \checkmark (R) & \textit{None} &  \checkmark (P,V)& \textit{None} & \checkmark & \begin{tabular}{@{}c@{}} GC Bipolar Radio \\ Lobes (13,14)\end{tabular} \\
    SGR\,1806& \textit{Fermi\,J1808.2-2029 \{p<0.1\%\}} & \checkmark (R) & \textit{None} &  \checkmark (S,V)& \textit{None} &  \checkmark & \begin{tabular}{@{}c@{}} \{G10.0-0.3 (9) \\ Fermi\,J1808.2-2029 (21)\} \end{tabular}\\
    XTE\,J1810& \textit{None} & \checkmark (R) & \textit{None} &  \checkmark (P,U) & \textit{None} & \checkmark & G11.0-0.0 (22)\\
    Swift\,1818 & \textit{None} & \checkmark (R) & \textit{None} &  \checkmark (P,U) & \textit{None} & \checkmark & --\\
    Swift\,J1822 & \textit{None} & \checkmark (R) & \textit{ \begin{tabular}{@{}c@{}} \{2MASS\,J18221794-1604259 \{p<0.1\%\} \\UKIDSS\,J182217.94-160425.9 \{p<0.1\%\} \\UKIDSS\,J182217.94-160426.0 \{p<0.1\%\} \\ 2MASS\,J18221839-1604241 \{p<0.1\%\}\\ UKIDSS\,J182218.39-160424.1 \{p<0.1\%\} \\ UKIDSS\,J182218.39-160424.0 \{p<0.1\%\} \\ UKIDSS\,J182217.97-160436.0 \{p=2.0\%\} \\ UKIDSS\,J182217.97-160435.9 \{p=1.0\%\} \} \end{tabular}} &  \checkmark (P,U) & \begin{tabular}{@{}c @{}}GDR3\,4097832458955594880 \{p<0.1\%\} \end{tabular} & \checkmark & --\\
    SGR\,1833 & \textit{None} & -- (R) & \textit{None} & -- (P,T) & \textit{None} & -- & -- \\
    Swift\,J1834 & W41 \{p<0.1\%\} & \checkmark (R) & \textit{None} &  \checkmark (P,U) & \textit{None} & \checkmark& W41 (10) \\
    1E\,1841-045 & \begin{tabular}{@{}c @{}}{4C-04.71 \{p<0.1\%\}} \end{tabular}   & \checkmark (R) & \textit{None} &  \checkmark (P,U) & \textit{None} & \checkmark & 4C-04.71 (11)\\ 
    3XMM\,J1852 & Kes\,79 \{p<0.1\%\} & \checkmark (R) & \textit{None} &  \checkmark (P,U) & \begin{tabular}{@{}c @{}}\textit{\{DR3\,4266311617807818624 \{p=1.08\%\},} \\ \textit{DR3\,4266313683745630080 \{p<1.0\%\},} \\ \textit{DR3\,4266315504811807488 \{p<1.0\%\},} \\ \textit{DR3\,4258978601120897664 \{p<1.0\%\}} \\ CXOU\,J185238.6+004020$^{*}$ \{p=0.02\%\}\}\end{tabular}  & \checkmark & Kes\,79 (17) \\
    SGR\,1900 &   \textit{S24-RACS\,J190707+091737\{p<0.1\%\}} & \checkmark(R) & \textit{None} &  \checkmark (P,U) &\textit{None}& \checkmark & G042.8+00.6 (12)\\
    SGR\,1935 & G057.2+00.8  \{p<0.1\%\} & \checkmark (R) & \textit{None} &  \checkmark (P,U) & \textit{None} & \checkmark & G057.2+00.8 (13)\\
    1E\,2259 & CTB\,109 \{p<0.1\} & \checkmark (V) 
    & \textit{None} &  \checkmark (P,U)& \textit{None} & \checkmark & G109.1-01.0 (14) \\
    \hline
    \hline
    SGR\,0755 & \begin{tabular}{@{}c @{}}\{ \textit{S24-RACS\,J075552-293709 \{p<0.1\%\}} \\ \textit{S24-RACS\,J230208+585020\{p<0.1\%\}} \\ \textit{S24-RACS\,J075520-293531 \{p<0.1\%\}} \\ \textit{S24-RACS\,J075523-292928 \{p<0.1\%\} }\\ \textit{S24-RACS\,J075603-292541 \{p<0.1\%\}} \\ \textit{S24-RACS\,J075613-293630 \{<0.1\%\}} \\ \textit{NVSS\,J075522-293937 \{p<0.1\%\}} \\ \textit{NVSS\,J075529-294201 \{p<0.1\%\}} \} \end{tabular} & \checkmark (R) & \begin{tabular}{@{}c @{}}\{ PSO J118.9270-29.5649  \{p<0.1\%\} \\ 2MASS\,J07554248-2933535 \{p<0.1\%\} \\ \textit{PSO J118.9228-29.5694 \{p=2.0\%\}} \\ \textit{PSO J118.9189-29.5686 \{p=4.0\%\}} \\ \textit{2MASS\,J07554146-2934098 \{p<0.1\%\}} \\ \textit{2MASS\,J07554053-2934071 \{p<0.1\%\}} \\ \textit{2MASS\,J07554295-2934093 \{p=2.0\%\}} \\ \textit{2MASS\,J07554149-2933368 \{p<0.1\%\}}\}\end{tabular}  &  \checkmark (P,T) & \textit{None}  & \checkmark & CPD-29 2176 (8)\\
    SGR 1801 &  \begin{tabular}{@{}c @{}}\{W28 \{p<0.1\%\} \\ \textit{18 sources$^\ddagger$ \{p<0.1\%\}}\} \end{tabular} & -- (R) & \textit{None} & --(P,U)& \textit{None} & -- & W28 (20)\\
    SGR\,1808 &  \begin{tabular}{@{}c @{}}\{G009.7-00.0 \{p<0.1\%\} \\ \textit{S24-RACS\,J180842-202849 \{p<0.1\%\}} \\ \textit{28 sources$^\ddagger$ \{p<0.1\%\}}\} \end{tabular}& -- (R) & \textit{None} & --(P,U) & \textit{None} & -- & --\\
    \textit{AX\,J1818.8-1559} & \textit{None} & -- (R) & \textit{None} & --(P,U) & \textit{None} & -- & --\\
    \textit{AX\,J1845-0258} & G029.6+00.1  \{p<0.1\%\} & \checkmark (R) & \textit{None} &  \checkmark (P,U) & \textit{None} & \checkmark &  G029.6+00.1 (15)\\
    \textit{SGR\,2013+34} & \textit{None} & \checkmark (N) & \textit{None} &  \checkmark (P,U) & \textit{None} & \checkmark & -- \\
    \textit{PSR\,J1846-0258} & Kes\,75 \{p=1.0\%\} & \checkmark (R) & \textit{None} &  \checkmark (P,U) & \textit{None} & \checkmark &  Kes\,75 (16) \\
  \hline
 \end{tabular}
\flushleft
\scriptsize
For each magnetar (written with its abbreviated name), the SNR candidates, bound companion candidates, and unbound companion candidates with $p<5\%$ are given, with p-values in brackets. Candidates suspected to be false positives are given in italics. Sources identified `by-eye' which are not confidently associated with any known object are labelled using the format, `S24-\textit{survey}-J\textit{hhmmss$\pm$ddmmss}', where \textit{survey} is the survey used and the italicized portion is the J2000 RA and declination of the centroid. For each search, the completeness along each magnetar's sightline is indicated by a \checkmark, meaning the search is either partially or fully sensitive to stars or remnants, or by an \xmark, meaning that even the brightest expected stars or remnants would not be detectable. The radio search completeness is labelled with `R', `V', or `N' to indicate whether the RACS, VLASS, or NVSS survey is used. The optical/IR bound search is considered partially or fully complete if either the optical or IR search are complete. These are labelled with `P' or `S', to indicate whether the PS1 or SkyMapper survey is used, and `T', `U', or `V' to indicate whether the 2MASS, PS1, or VVV survey is used. The \textit{Gaia} search is considered partially or fully complete if both the $V-$band magnitude and expected relative proper motion are detectable. Four magnetars or candidates have no distance measures and the completeness cannot be determined; these are marked with `--'.Candidates for SNR, bound, or unbound companions from previous analyses are provided in the last column with references matched to the following: (1) \citet{gaensler2008sgr} (2) \citet{2004ApJ...609L..13K} (3) \citet{2007ApJ...667.1111G} (4) \citet{2012ApJ...751...53A} (5) \citet{1999ApJ...526L..29C} (6) \citet{clark2014vlt} (7) \citet{2010ApJ...710..941H} (8) \citet{chrimes2022magnetar} (9) \citet{corbel2004connection} (10) \citet{2012ApJ...748...26K} (11) \citet{1997ApJ...486L.129V} (12) \citet{2000ApJ...533L..17V} (13) \citet{gaensler2014grb} (14) \citet{1981Natur.293..202F} (15) \citet{1999ApJ...526L..37G} (16) \citet{2000ApJ...542L..37G} (17) \citet{2014ApJ...781L..16Z} (18) \citet{ponti2015xmm} (19) \citet{yalinewich2017evolution} (20) \citet{2000ApJ...531..407C} (21) \citet{2018A&A...612A..11H} (22) \citet{ding2020magnetar}

$^\dagger$For the \textit{Gaia} search for unbound companions, we consider the proper motion sensitivity in four magnitude bins as described in \citet{gaiacollaboration2022gaia}: $m_G > 15$ ($\mu_{\rm min} = 0.025$\,mas\,yr$^{-1}$), $m_G = 17$ ($\mu_{\rm min} = 0.07$\,mas\,yr$^{-1}$), $m_G = 20$ ($\mu_{\rm min} = 0.5$\,mas\,yr$^{-1}$), and $m_G = 21$ ($\mu_{\rm min} = 1.4$\,mas\,yr$^{-1}$). 

$^\ddagger$For conciseness, we omit the names of the 17 and 28 false positives found for SGR\,1801-23 and SGR\,1808-20, resectively. For the former, the false positives are rejected in favor of W28 identified by \citet{2000ApJ...531..407C}. Detailed discussion for the latter is provided in Section~\ref{sec:racsresults}.

$^*$CXOU\,J185238.6+004020 is an X-ray pulsar associated with Kes\,79. We explore the scenario that the progenitors of 3XMM\,J1852 and this pulsar were bound in Section~\ref{sec:racsresults}.
 \end{center}
\end{table*}

\subsection{The Magnetar Binary and Massive Star Merger Fractions}\label{sec:stats}

\subsubsection{Markov-Chain Monte Carlo Analysis of Magnetar Binary Fractions}\label{sec:mcmcstats}


With more well defined estimates of our search completeness, we next determine the implications of our results on the CCSN model. We adapt the method of \citet{kochanek2023search} to constrain the binary parameters of the magnetar progenitor population using a Markov-Chain Monte Carlo (MCMC) simulation. Uniform priors are chosen to estimate the fraction of systems which were not binaries at death, $f_n$; the fraction that were unbound at death, $f_u$; and the fraction of interacting binaries remaining bound after the CCSN, $f_i$. We relate these to the total fraction of binaries remaining bound after the CCSN, $f_b$, and the fraction of non-interacting binaries remaining bound after the CCSN, $f_p$, by $f_n = 1 - f_u - f_b$ and $f_b = f_i + f_p$. 

The likelihood is the joint multinomial probability of the observed samples; we define the sample for this analysis incorporating the completeness ratios in Table~\ref{table:results}. Of the 31 magnetars and candidates, three systems, CXOU\,J164710.2-455216, 3XMM\,J185246.6+003317, and SGR\,J1822.3-1606, were proposed as unbound binary systems. The candidate companion of CXOU\,J171405.7-381031 from \citet{chrimes2022magnetar} is unconfirmed based on the optical and IR search results. For the purpose of our analysis, we will accept SGR\,0755-2933 and CPD-29\,2176 as a bound interacting binary system based on the results of \citet{richardson2023high}. Of the remaining sample, three magnetars are constrained to be not currently bound, two are constrained not to have unbound companions, and the remaining 21 are considered not to be binaries at the time of the CCSN. With this, the likelihood can be written as:




\begin{equation}
    \mathcal{P}_{\rm i} = f_n^{21} f_u^3 f_i^1 (1-f_u)^2 (f_n + f_u)^3
\end{equation}

\noindent where $(1-f_u)$ is the fraction that are not unbound systems and $(f_n + f_u)$ is the fraction that are not bound systems.
The sampled posteriors are used to estimate the median values and 90\% confidence intervals for $f_n$, $f_u$, and $f_i$.

In keeping with the notation of \citet{kochanek2023search}, we define the case above as `SNR incomplete', where we assume that all magnetars in the sample originate in CCSNe. While this is likely the primary formation channel, it is possible that other processes may form magnetars at lower rates. Considering this, we define the `SNR complete' case, in which magnetars without confirmed or proposed SNR associations are not formed via CCSNe, and therefore do not contribute to the binary fractions above. In this case, the likelihood becomes:

\begin{equation}
    \mathcal{P}_{\rm c} = f_n^{10}f_u^1f_{nc}^{15} (f_n + f_u)^2 (1-f_u)^2 (1-f_{nc})^1
\end{equation}

\noindent where $f_{nc}$ is the fraction of magnetars which do not originate in CCSNe. This modifies the definition for $f_n$ to $f_n = 1- f_u - f_b - f_{nc}$. Note this case is more restrictive, as SGR\,0755+2933, SGR\,J1822.3-1606, and 13 other magnetars for which the radio search is $>75\%$ complete have no confirmed SNR in this or any other work. A final case, `SNR/unbound complete' is defined if we also take unbound magnetar binaries as evidence of CCSNe; in this case, we include CXOU\,J164710.2-455216, SGR\,J1822.3-1606, and 3XMM\,J185246.6+003317 as unbound systems, and the likelihood becomes:

\begin{equation}
    \mathcal{P}_{\rm uc} = f_n^{10}f_u^3f_{nc}^{13} (f_n + f_u)^2 (1-f_u)^2 (1-f_{nc})^1
\end{equation}

Results for the three cases are summarized in Table~\ref{table:magstat}. In the `SNR incomplete' case, we find a median bound binary fraction $f_b \approx9^{+6}_{-4}\%$, and unbound binary fraction $f_u \approx 12^{+6}_{-5}\%$, where the errors are $1\sigma$ uncertainties. While the bound fraction is marginally consistent with the predictions from \citet{chrimes2022magnetar}, \citet{kochanek2019stellar}, and \citet{renzo2019massive} population synthesis results, the unbound fraction significantly disagrees with their prediction of $f_u\sim38-56\%$. Furthermore, we find that a median of only $1-f_n \approx 22^{+8}_{-7}\%$ of magnetars were binaries at death, which disagrees with the estimate of $\sim50-60\%$ from population synthesis with $90\%$ confidence. 

In the `SNR complete' case, we estimate a median unbound fraction $f_u\approx 3^{+3}_{-2}\%$, and a $90\%$ upper limit on the bound fraction of $f_b\lesssim12\%$, the former of which also disagrees with the \citet{chrimes2022magnetar}, \citet{kochanek2019stellar}, and \citet{renzo2019massive} population synthesis estimates. We find a median $1-f_n\approx63^{+9}_{-8}\%$ of magnetars undergo CCSNe while in binaries, which more closely agrees with population synthesis than the `SNR incomplete' case. Most significantly, we constrain the fraction of magnetars formed by channels other than CCSNe to $f_{nc} \approx 53^{+8}_{-10}\%$, with a 90\% confidence interval from $37-66\%$. This would contradict the view that CCSNe are the dominant formation pathway \citep[e.g.][]{schneider2019stellar,white2022origin,keane2008birthrates}. Furthermore, no magnetars have been confirmed to form from non-CCSN channels. 

A similar result is obtained in the `SNR/unbound complete' case; notably, the larger unbound fraction $f_u\approx8^{+5}_{-3}\%$ allows a lower non-CCSN fraction $f_{nc}\approx46\pm9\%$, while the fraction of magnetars from binaries $1-f_n\approx 61\pm9\%$ still agrees with population synthesis results. While the `SNR/unbound complete' and `SNR complete' cases do not differ in their underlying assumptions about alternate formation channels, they demonstrate that the detection of unbound magnetar binaries allows for tighter constraints to be placed on $f_{nc}$. Furthermore, these cases motivate discussion of alternate formation channels, such as Accretion Induced Collapse (AIC) of magnetized White Dwarfs, electron-capture supernovae of less massive stars, or binary Neutron Star mergers, to which we return in Section~\ref{sec:altchannels} \citep{ruiter2019formation,fryer1999can,giacomazzo2013formation,giacomazzo2015producing,miyaji1980supernova}.

\begin{table*}
 \caption{Constraints on magnetar binaries from this work}
 \label{table:magstat}
 \begin{tabular}{lccccccc}
 \hline
   \textbf{Sub-Sample}$^{\dagger}$ & \textbf{Symbol} & \multicolumn{2}{c}{\textbf{SNR incomplete}} & \multicolumn{2}{c}{\textbf{SNR complete}}  & \multicolumn{2}{c}{\textbf{SNR/unbound complete}}\\
    &  &  Median & 90\% Confidence &  Median & 90\% Confidence & Median & 90\% Confidence \\
  \hline
    \textbf{Not binary at death} & $f_n$ & $78^{+7}_{-8}\%$ & $64-88\%$  & $37^{+9}_{-8}\%$ & $24-52\%$ & $39\pm9\%$ & $25-54\%$ \\
    \textbf{Bound binary} & $f_b$ & $9^{+6}_{-4}\%$ & $3-20\%$ & -- & $<12\%$ & -- & $<13\%$ \\
    \textit{Interacting binary} & $f_i$ & $6^{+5}_{-3}\%$ & $1-15\%$ & -- &$<8\%$ & -- &$<8\%$ \\
    \textit{Non-interacting binary} & $f_p$ & -- & $<8\%$ & -- & $<7\%$ & -- & $<7\%$ \\
    \textbf{Unbound binary} & $f_u$ & $12^{+6}_{-5}\%$ & $5-24\%$ & $3^{+3}_{-2}\%$ &$1-10\%$ & $8^{+5}_{-3}\%$ &$3-16\%$ \\
    \textbf{Not CCSN Progenitor} & $f_{nc}$ & -- & -- & $53^{+8}_{-10}\%$ & $37-66\%$ & $46\pm9\%$ & $31-60\%$\\
    \hline
    Pre-CCSN Merger$^\ddagger$ & $f_{m}$ & $70^{+10}_{-12}\%$ & $48-86\%$  & $88^{+5}_{-8}\%$ & $73-96\%$ & $82^{+7}_{-9}\%$ & $65-92\%$ \\
  \hline
 \end{tabular}
\flushleft
\small
\textit{Note:} We define two broad cases based on the completeness of the RACS SNR search: `SNR incomplete' assumes that all magnetars originate in CCSNe, whether one is detected or not. `SNR complete' assumes that magnetars with RACS search completeness ratios $>75\%$ which had no SNR detected did not form through CCSNe. We also consider the `SNR/unbound complete' case in which unbound magnetar binaries are evidence of the CCSN progenitor channel as well. Median values are reported with $1\sigma$ errors.

$^\dagger$$f_n$, $f_b$, $f_u$, and $f_{nc}$ are defined such that $f_n + f_b + f_u = 1$ in the `SNR incomplete' case, and $f_n + f_b + f_u + f_{nc} = 1$ in the `SNR complete' and `SNR/unbound complete' cases. $f_i$ and $f_p$ sum to the total binary fraction $f_p + f_i = f_b$. The merger rate $f_m$ is the fraction of binary stars that merge, while the other fractions are taken out of the total number of CCSNe. 

$^\ddagger$$f_m$ is estimated assuming the progenitor stars are early B-type stars ($9M_\odot < M_* < 16M_\odot$) with binary fraction $F_0=84\pm9\%$ \citep{moe2017mind}. 
\end{table*}

\subsubsection{Implications for Massive OB Star Binary and Merger Fractions}

Many models of magnetar formation attribute their strong magnetic fields to a dynamo process driven by accretion from ablated stellar companions \citep[e.g.][and references therein]{popov2020high,revnivtsev2016magnetic,fuller2022spins}. The low $f_{b}$ and $f_{u}$ fractions in Table~\ref{table:magstat} may indicate that other magnetic field growth mechanisms are more common, such as the merger of massive stars. In this scenario, the magnetic field of the primary star is first amplified up to $\lesssim10^8\,$G through accretion before further amplification through merger of the cores, producing a highly magnetized massive star such as $\tau\,$Sco and the Wolf-Rayet star in the HD 45166 binary \citep{schneider2019stellar,Shenar_2023}. If magnetic flux is conserved during the CCSN event, a neutron star remnant with a magnetar-like surface magnetic field ($\sim10^{14}\,$G) could be achieved \citep[e.g.][and references therein]{popov2020high, schneider2019stellar, obergaulinger2014magnetic}. A high frequency of such pre-CCSN mergers could potentially explain the lack of magnetars with bound or unbound companions. 

\begin{figure*}
 \includegraphics[width=0.99\linewidth]{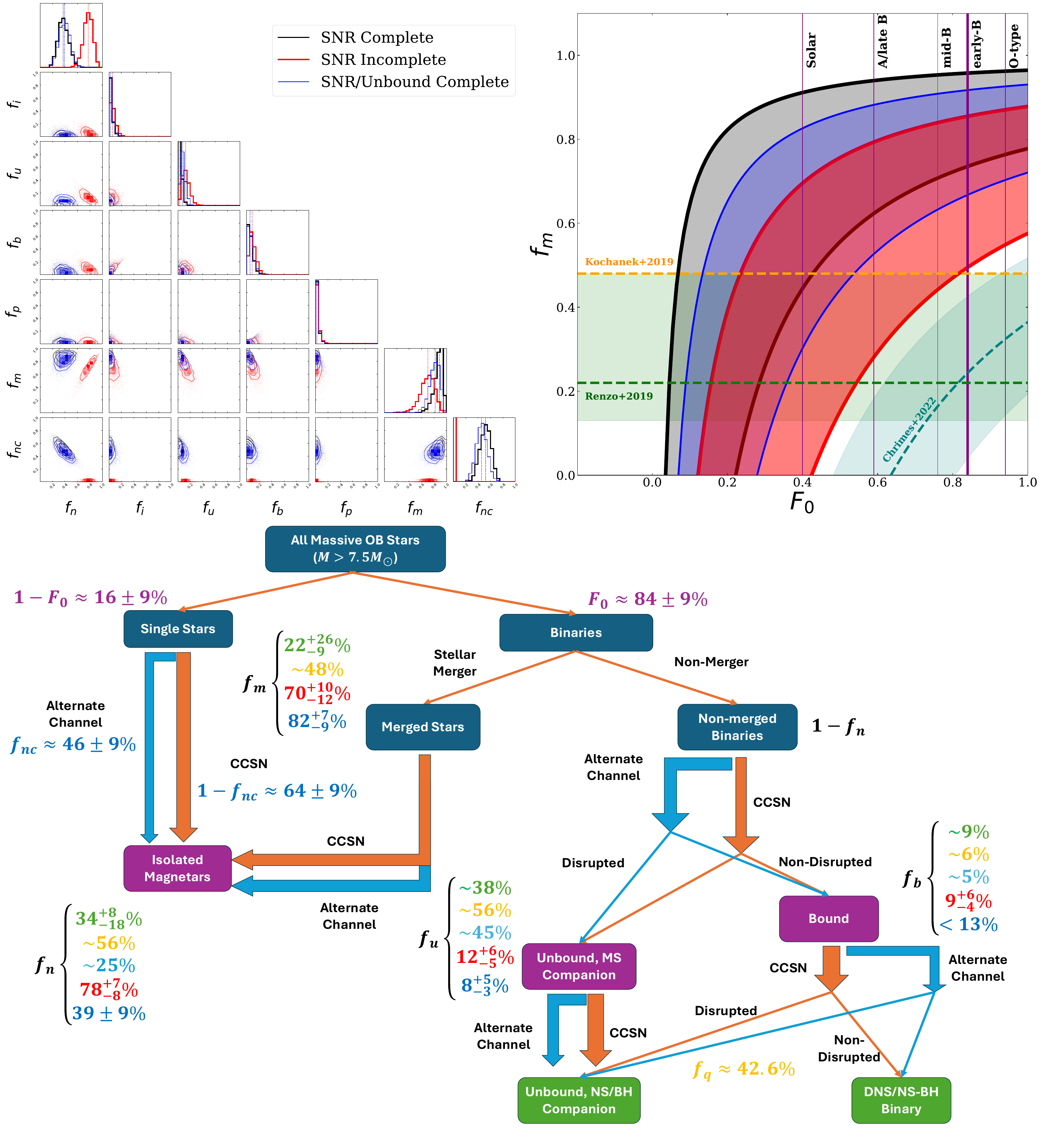}
 \caption{(\textit{Upper Left:}) Corner Plots from the Magnetar Binary Fraction Simulation in Section~\ref{sec:stats}. The `SNR complete', `SNR incomplete', and `SNR/unbound complete' cases as defined in Section~\ref{sec:discussion} and Table~\ref{table:magstat} are shown in black, red, and blue, respectively. Vertical lines indicate the median values from Table~\ref{table:magstat} in each case. The `SNR complete' and `SNR/unbound complete' cases would suggest that $f_{nc}\approx45-55\%$ of magnetars to descend from non-CCSN channels, whereas the `SNR incomplete' case suggests a higher stellar merger rate $f_m\approx 70\%$ among OB stars is sufficient to produce the observed sample. The merger rates displayed assume a massive star binary fraction $F_0=84\pm9\%$ \citep{moe2017mind}. (\textit{Upper Right:}) 90\% Confidence Regions for the merger fraction $f_m$ and OB star binary fraction $F_0$. The colors for each region correspond to the cases in the contour plot, and vertical purple lines show $F_0$ predicted for different stellar types through population synthesis \citep{moe2017mind}. The green and yellow dashed lines indicate the median merger fractions from \citet{renzo2019massive} and \citet{kochanek2019stellar}; the teal dashed line indicates the $f_m$ vs. $F_0$ curve inferred from $f_b$ and $f_u$ in \citet{chrimes2022magnetar}. The green and teal shaded regions indicate the $1\sigma$ errors for the \citet{renzo2019massive} and \citet{chrimes2022magnetar} cases. Each of these appear to underestimate the true fraction for early B-type stars. (\textit{Bottom:}) Flowchart showing the progression of massive stars into magnetars with constraints from population synthesis and observations in this work. CCSNe are indicated by orange arrows while alternate formation channels are indicated by blue arrows. Constraints on $f_m$, $f_n$, $f_u$, $f_b$, and $F_0$ (see Section~\ref{sec:MCresults}) are shown from simulations by \citet[][purple]{moe2017mind}, \citet[][green]{renzo2019massive}, \citet[][yellow]{kochanek2019stellar}, and \citet[][teal]{chrimes2022magnetar}. Observational constraints from this work are shown in red for the `SNR incomplete' case and blue for the `SNR/unbound complete' case. A lower binary fraction would allow for a lower merger fraction for magnetar progenitors, but would require lower mass progenitors.} 
 \label{fig:mcmc}
\end{figure*}

The fraction of massive stellar binaries that merge prior to CCSNe, $f_m$, can be inferred by assuming an initial massive star companion fraction $F_0 \approx 84\pm9\%$, derived in \citet{moe2017mind}, for early B-type stars. \footnote{Note that this fraction includes binary, tertiary, and higher order systems with primary early B-type stars ($9M_\odot<M_*<16M_\odot$), but we will collectively refer to them as `binaries' \citep{moe2017mind}. Any higher order system can lead to a compact remnant with bound or unbound companion stars, regardless of multiplicity.} While some simulations suggest that CCSNe can form from mergers of intermediate mass stars ($5-8M_\odot$), we focus this discussion on binaries where both stars are $>7.5M_\odot$ \citep[e.g.][]{zapartas2017delay}. Relating the initial companion fraction to the fraction of systems in binaries at death:

\begin{equation}
    \frac{f_u + f_b}{1-(f_b + f_u) f_q} = F_0(1 - f_m)
\end{equation}

\noindent where $f_q =0.426$ is the fraction of binaries with a secondary star less massive than the primary \footnote{This is derived by integrating a Salpeter-like ($\alpha=-2.35$) initial mass function \citep[IMF;][]{salpeter1955luminosity,moe2017mind}; see Section 1 and Equation 1 of \citet{kochanek2019stellar}, and references therein. Note that in \citet{kochanek2019stellar} Equation 1, $f_b$ is defined as the fraction of CCSNe that occur in stellar binaries, whereas we here define $f_b$ as the fraction of systems for which the binary remains bound following the CCSN.}. $f_q$ effectively estimates the fraction of CCSNe from explosions of the secondary star, which are excluded from this analysis since we consider CCSN of stars with \textit{stellar} companions\footnote{We assume that magnetar 3XMM\,J185246.6+003317 was formed from the first explosion, as is implied from our analysis in Section~\ref{sec:bound}.} \citep[see][for detailed discussions]{kochanek2009stellar,kochanek2019stellar,kobulnicky2007new}. If we adopt $f_b$ and $f_u$ from Table~\ref{table:magstat} and use a Gaussian distribution\footnote{
More rigorously, $F_0$ follows a binomial distribution, but we find this makes little difference to the results. The Gaussian method is adopted to more easily incorporate the uncertainty on $F_0$.} for $F_0$, we obtain median values $f_m\approx70^{+10}_{-12}\%$ for the `SNR complete' case, $f_m\approx88^{+5}_{-8}\%$ for the `SNR incomplete' case, and $f_m\approx82^{+7}_{-9}\%$ for the `SNR/unbound complete' case. In all three cases, $f_m$ is significantly higher than the merger fractions estimated for neutron star progenitors in population synthesis from \citet{kochanek2019stellar} ($\sim48\%$) and \citet{renzo2019massive} ($\sim22_{-9}^{+26}\%$). It is also possible that magnetar progenitors have an above average merger rate compared to other neutron star species. Corner plots summarizing this analysis are shown in the left-hand plot of Figure~\ref{fig:mcmc}.


The `SNR incomplete' case does not allow for alternate formation channels, suggesting instead that a large fraction $f_n\approx80\%$ were formed from single stars. For a fixed binary fraction $F_0$, this requires an inflated merger rate $f_m\approx75\%$. The `SNR/unbound complete' case is one step removed, taking unbound companions as evidence of a disruption event (CCSN) but also allowing for $f_{\rm nc}\approx45\%$ to form from other formation channels (e.g. AIC, neutron star merger). For this reason, $f_n\approx40\%$ is lowered, but the unbound fraction $f_u\approx10\%$ is relatively unchanged and the merger rate $f_m\approx80\%$ increases only marginally ($<1\sigma$ change). The `SNR complete' case, however, excludes both SGR\,1822.3-1606 and CXOU\,J164710.2-455216, which do not have SNR detections, from the unbound sample. SGR\,0755-2933, the only proposed bound system, is excluded for the same reason. Therefore, both the merger fraction $f_m\approx90\%$ and the fraction from alternate channels $f_{\rm nc}\approx 55\%$ increase to accommodate the lowered unbound fraction $f_u\approx5\%$. In all three cases, the merger fraction $f_m$ is larger than expected, requiring us to reconsider the conditions that lead to massive OB star mergers. However, a more strict interpretation (`complete' cases) is consistent with $f_{\rm nc}\approx31-66\%$ forming from alternate channels.

Alternatively, we can allow both $F_0$ and $f_m$ to vary, leaving the progenitor mass and stellar type unconstrained. Under this assumption, the right-hand plot of Figure~\ref{fig:mcmc} shows the $90\%$ confidence regions for $F_0$ and $f_m$ in each case. This demonstrates that a wider range of merger rates can be tolerated if magnetars can form from the mergers of less massive stars \citep{moe2017mind}. This would marginally support the analysis of \citet{zapartas2017delay}, which identified a population of CCSNe from the mergers of intermediate mass stars ($5-7.5M_\odot$). However, a reliable comparison requires a broader search that targets both massive and intermediate mass stars, which we leave to future work. Figure~\ref{fig:mcmc} summarizes the results of this analysis in a flowchart showing the evolution of stellar progenitors to magnetar remnants.

\subsubsection{Limitations in Comparison to Population Synthesis Results}\label{sec:limitations}

While we compare our resulting binary, merger, and CCSN fractions to population synthesis simulations, a discussion of the limitations of such analysis is warranted. We focus this discussion on the \citet{chrimes2022magnetar} simulations, though similar comparisons can be made with the \citet{kochanek2019stellar} and \citet{renzo2019massive} simulations. \citet{chrimes2022magnetar} utilized Binary Population and Spectral Synthesis (BPASS) models and imposed constraints on the total mass ($>2.0M_{\odot}$), CO core mass ($>1.38M_{\odot}$), and remnant mass ($1.38 < M_{\rm rem} < 3.0$) to identify stellar models that evolve into magnetars. One potential source of error is the Initial Mass Function (IMF), for which \citet{chrimes2022magnetar} use the two-component \citet{10.1093/mnras/262.3.545} broken power law model with index $\alpha=-1.30$ for masses $0.1-0.5\,M_\odot$ and $\alpha=-2.35$ for masses $0.5-300\,M_{\odot}$. Alternatively, some analyses have proposed a smoothly varying IMF with a mass-dependent $\alpha$ which is steeper for high mass stars \citep[e.g.][]{kroupa2002initial,kroupa2003galactic,miller1979initial,chabrier2001galactic,weidner2005variation}. This could lead to a larger ratio of low mass to high mass stars. The relationship of the IMF to the binary fraction also has a complex dependency on the mass-ratio ($M_2/M_1$) distribution. While a uniform distribution is assumed to compute $f_q\approx0.426$ \citep[e.g.][]{kochanek2018cas,kochanek2019stellar,2005A&A...430..137K}, \citet{renzo2019massive} find that a negative power law distribution results in a larger merger fracton $f_m$. This demonstrates that both the IMF and mass ratio distribution can have a significant impact on the massive star binary fraction, and motivates detailed consideration in future studies.


An additional caveat is that relatively loose constraints were used to gather the magnetar sample for this work in comparison to previous analyses. For example, \citet{chrimes2022magnetar} excludes magnetars in Magellanic Clouds, as well as those without distance estimates, imaging data, or photometric data. With the small 31-magnetar sample, we have chosen not to place arbitrary limits beyond the completeness constraints discussed in Section~\ref{sec:complete}. The reported p-values provide quantitative measures of confidence that incorporate large distances, errors, and accuracy of imaging data. Furthermore, the irregularity of transient magnetar outbursts could have marginal effects on our analysis; for example, both SGR\,1935+2154 and PSR\,J1846-0258 have recently entered X-ray outburst phases which may tighten constraints on their ages \citep{ibrahim2024xray,2020GCN.28193....1L,2020ATel13913....1K,sathyaprakash2024longterm}. SGR\,0755-2933 and SGR\,1808-20, contrarily have only one burst detection, each \citep{lamb2003sgr,2016ATel.8831....1B}. Future tests may benefit from partitioning the sample based on their distance, proper motion, extinction properties, or outburst phase to characterize its affect on the search.

\subsubsection{Comparing Magnetar Binary Statistics to the Neutron Star Population}

The apparent differences between the magnetar and neutron star formation processes limit the generalization of our results to the full neutron star population. Magneto-hydrodynamic (MHD) plasma simulations of CCSNe demonstrate that the magneto-rotational instability (MRI) or a dynamo action can amplify the progenitor stars' fossil fields during collapse \citep[e.g.][]{akiyama2003magnetorotational,obergaulinger2014magnetic,obergaulinger2018core,raynaud2020magnetar,2009A&A...498..241O}. However both MRI and dynamos require either high progenitor magnetic fields \citep[$\gtrsim 10^{8-11}\,$G core magnetic field;][]{obergaulinger2018core,schneider2019stellar} or high specific angular momentum \citep[$\gtrsim 9\times10^{14}$\,cm$^{2}$\,s$^{-1}$;][]{obergaulinger2018core}. Such conditions were considered at a high level by \citet{popov2006progenitors}, whose population synthesis experiment estimates that $\sim8-9\%$ of neutron stars would be magnetars, though the details of MRI and dynamo evolution were not incorporated. Thus magnetars and neutron stars probe different progenitor star populations that are not well-defined. Such a partition of the progenitor star population would also disfavor models presenting magnetar emission as a continuum among neutron stars.

Furthermore, some evidence suggests that highly magnetized massive stars can be formed through the merger of stars with standard surface magnetic field strengths \citep[$\sim 1-100\,$G; e.g.][]{schneider2019stellar}. This is presented as a possible origin of blue straggler stars with $\sim1$\,kG surface fields \citep[e.g. $\tau$\,Sco;][]{schneider2016rejuvenation,donati2006surprising}, which are thought to be magnetar progenitors. The high merger rate found in this work, $f_m\sim70-80\%$, supports this scenario since it implies neutron stars with typical field strengths may have a smaller progenitor merger rate. This may also explain the discrepancy between our work and the $\sim 22^{+26}_{-9}\%$ merger rate suggested by \citet{renzo2019massive}'s population synthesis. Without a more detailed understanding of the partitioning of magnetar and pulsar progenitors, our results are confined to the magnetar population and cannot be confidently applied to neutron stars as a whole.


\subsection{Alternate Magnetar Formation Channels}\label{sec:altchannels}

In this section, we consider the implications of the `SNR complete' and `SNR/unbound complete' cases which require large contributions $f_{nc}\sim45-55\%$ from non-CCSN formation channels. First, consider how this alters the birth rate assumptions for the neutron star population. \citet{keane2008birthrates} predicted that the Galactic CCSN rate ($1.9\pm1.1$\,century$^{-1}$) is too low to source the observed population of magnetars, pulsars, XDINS, and RRATs (total $10.8^{+7.0}_{-5.0}$\,century$^{-1}$). If we incorporate the results of the `SNR complete' and `SNR/unbound complete' cases, which predict $f_{nc}\approx55\%$ and $f_{nc}\approx45\%$ of magnetars are formed by alternate channels, respectively, then the total neutron star birth rate from CCSNe is decreased to $\sim5$\,century$^{-1}$ and $\sim6$\,century$^{-1}$. This is still inconsistent with the CCSN rate, and suggests that alternate channels are unlikely to wholly explain the observed binary fractions. 

It is possible that alternate formation channels are partially responsible for our observations. One such channel is Accretion Induced Collapse (AIC), in which ONe White Dwarfs accrete from Helium burning giant companions before collapsing \citep[e.g.,][]{ruiter2019formation,fryer1999can}. Following AIC, either a bound neutron star-white dwarf binary is formed, or the donor star is ejected as a runaway. While the ejected mass from AIC is expected to be lower than that of a CCSN ($\sim10^{-3}-0.05M_\odot$), low mass white dwarf companions would require less energy to become unbound than massive stars \citep{ruiter2019formation,el2023fastest,2018ApJ...865...15S}. 

Direct evidence for AIC has not yet been observed, making them unlikely to make up the required $f_{nc}\sim45-55\%$; however, \citet{Sharma_2023} observed that long delay times and high-mass host galaxies for two Fast Radio Bursts (FRBs) support origins from neutron stars or magnetars formed through AIC \citep[see also][for comparisons of offset and host galaxy flux distributions between FRBs and CCSNe]{woodland2023environments,gordon2023demographics}. More unique formation processes such as neutron star mergers require gravitational wave detection and followup, but are unlikely to supercede CCSNe in magnetar production \citep[e.g.][]{giacomazzo2013formation,giacomazzo2015producing}. Electron-capture supernovae, which occur in less massive ($8-11M_\odot$) stars, also are unlikely to produce magnetars as they impart weaker explosion kicks to their companions, resulting in a higher surviving binary fraction, $f_b$ \citep{poelarends2017electron,miyaji1980supernova,nomoto1987evolution,leung2020electron}. Therefore, the results of this search continue to support CCSNe as the most common magnetar formation channel, and we expect future improvements on radio survey sensitivity to recover any remaining SNRs with sufficient local density and explosion energy \citep[e.g.][]{frail1995does}.

\section{Conclusion}\label{sec:conclusion}

In this work, we have searched optical, IR, and radio image and point source catalogs for evidence of CCSNe and massive OB stars associated with magnetars. Out of the 31 known magnetars and candidates, we identify two candidates for unbound magnetar binaries at $>95\%$ confidence: a massive OB star in association with SGR\,1822.3-1606, and an X-ray pulsar in association with 3XMM\,J185246.6+003317. We also recover the proposed Be star companion of SGR\,0755-2933 and marginally detect the proposed unbound companion of CXOU\,J164710.2-455216. Through the radio search we recover nine out of ten confirmed SNR associations and four of five unconfirmed associations; additional SNR candidates for SGR\,1808-20 are identified but require a distance and proper motion measurement for the magnetar to confirm. Through an MCMC analysis, we find only $5-24\%$ of magnetars have unbound massive star companions with 90\% confidence. This is much lower than is predicted by neutron star population synthesis simulations from \citet{chrimes2022magnetar}, \citet{kochanek2019stellar}, and \citet{renzo2019massive}. 

Our results support a high merger rate for massive OB star progenitors between $48-86\%$. It is unclear if only magnetars or neutron stars as a whole preferentially form from mergers. While alternate formation channels such as AIC and neutron star mergers may account for a small fraction of magnetars, it is unlikely that they are solely responsible for the lack of observed unbound binaries, as this would require $\sim45-55\%$ of magnetars to follow such channels and would not entirely solve the neutron star birth rate problem. 

Significant improvements to this search can be made with proper motion measurements for the 23 magnetars without constraints. This may be achieved with next-generation, high sensitivity infrared telescopes like the JWST and NGRST. Similarly, Very Long Baseline Interferometry (VLBI) parallax measurements are available for only two magnetars; targeting the remaining sample will provide independent distance measurements to better inform the association tests. Next generation radio telescopes like the Deep Synoptic Array (DSA-2000) and the SKA will also detect a large sample of FRBs to investigate their origins, potentially constraining the magnetar birth rate from an extragalactic perspective.




\section{Data Availability}\label{sec:data}

The data underlying this article, including the magnetar parameters in Tables~\ref{table:magnetars} and Tables~\ref{table:magtimeparams} as well as derived values and p-values for each optical, IR, and radio source included in the search, are available in machine-readable format in the \textit{CaltechDATA} repository, at \url{https://doi.org/10.22002/cfj2d-zbc71}.

\section{Acknowledgements}

The authors would like to thank the members of the Deep Synoptic Array (DSA-110) team for their support and insight with regards to this search effort. The authors thank staff members of the Owens Valley Radio Observatory and the Caltech radio group, including Kristen Bernasconi, Stephanie Cha-Ramos, Sarah Harnach, Tom Klinefelter, Lori McGraw, Corey Posner, Andres Rizo, Michael Virgin, Scott White, and Thomas Zentmyer. Their tireless efforts were instrumental to the success of the DSA-110. The DSA-110 is supported by the National Science Foundation Mid-Scale Innovations Program in Astronomical Sciences (MSIP) under grant AST-1836018. The authors would also like to thank Casey Law, Alexa Gordon, Alice Curtin, Maxim Lyutikov, and an anonymous referee for their insightful discussions and feedback on the early draft of this manuscript. This material is based upon work supported by the National Science Foundation Graduate Research Fellowship under Grant No. DGE‐1745301. 

The Pan-STARRS1 Surveys (PS1) and the PS1 public science archive have been made possible through contributions by the Institute for Astronomy, the University of Hawaii, the Pan-STARRS Project Office, the Max-Planck Society and its participating institutes, the Max Planck Institute for Astronomy, Heidelberg and the Max Planck Institute for Extraterrestrial Physics, Garching, The Johns Hopkins University, Durham University, the University of Edinburgh, the Queen's University Belfast, the Harvard-Smithsonian Center for Astrophysics, the Las Cumbres Observatory Global Telescope Network Incorporated, the National Central University of Taiwan, the Space Telescope Science Institute, the National Aeronautics and Space Administration under Grant No. NNX08AR22G issued through the Planetary Science Division of the NASA Science Mission Directorate, the National Science Foundation Grant No. AST-1238877, the University of Maryland, Eotvos Lorand University (ELTE), the Los Alamos National Laboratory, and the Gordon and Betty Moore Foundation. The UKIDSS project is defined in \citet{2007MNRAS.379.1599L}. The national facility capability for SkyMapper has been funded through ARC LIEF grant LE130100104 from the Australian Research Council, awarded to the University of Sydney, the Australian National University, Swinburne University of Technology, the University of Queensland, the University of Western Australia, the University of Melbourne, Curtin University of Technology, Monash University and the Australian Astronomical Observatory. SkyMapper is owned and operated by The Australian National University's Research School of Astronomy and Astrophysics. The survey data were processed and provided by the SkyMapper Team at ANU. The SkyMapper node of the All-Sky Virtual Observatory (ASVO) is hosted at the National Computational Infrastructure (NCI). Development and support of the SkyMapper node of the ASVO has been funded in part by Astronomy Australia Limited (AAL) and the Australian Government through the Commonwealth's Education Investment Fund (EIF) and National Collaborative Research Infrastructure Strategy (NCRIS), particularly the National eResearch Collaboration Tools and Resources (NeCTAR) and the Australian National Data Service Projects (ANDS).

UKIDSS uses the UKIRT Wide Field Camera (WFCAM; \citet{2007A&A...467..777C}). The photometric system is described in \citet{2006MNRAS.367..454H}, and the calibration is described in \citet{2009MNRAS.394..675H}. The pipeline processing and science archive are described in \citet{irwin2004vista} and \citet{2008MNRAS.384..637H}. The VISTA Data Flow System pipeline processing and science archive are described in \citet{irwin2004vista}, \citet{2008MNRAS.384..637H} and \citet{cross2012vista}. We have used data from the 5th data release of the Vista Variables in the Via Lactea (VVV) Survey, which is described in detail in \citet{nikzat2022vvv}, and the VVV Infrared Astrometric Catalogue, described in detail in \citet{smith2018virac}. This publication makes use of data products from the Two Micron All Sky Survey, which is a joint project of the University of Massachusetts and the Infrared Processing and Analysis Center/California Institute of Technology, funded by the National Aeronautics and Space Administration and the National Science Foundation. 

This work also uses data obtained from Inyarrimanha Ilgari Bundara / the Murchison Radio-astronomy Observatory. We acknowledge the Wajarri Yamaji People as the Traditional Owners and native title holders of the Observatory site. CSIRO’s ASKAP radio telescope is part of the Australia Telescope National Facility (https://ror.org/05qajvd42). Operation of ASKAP is funded by the Australian Government with support from the National Collaborative Research Infrastructure Strategy. ASKAP uses the resources of the Pawsey Supercomputing Research Centre. Establishment of ASKAP, Inyarrimanha Ilgari Bundara, the CSIRO Murchison Radio-astronomy Observatory and the Pawsey Supercomputing Research Centre are initiatives of the Australian Government, with support from the Government of Western Australia and the Science and Industry Endowment Fund. This paper includes archived data obtained through the CSIRO ASKAP Science Data Archive, CASDA (https://data.csiro.au). This work has utilized data from the LoTSS survey and image archive, which is described in detail in \citet{shimwell2022lofar}. This work has made use of data from the European Space Agency (ESA) mission
{\it Gaia} (\url{https://www.cosmos.esa.int/gaia}), processed by the {\it Gaia}
Data Processing and Analysis Consortium (DPAC,
\url{https://www.cosmos.esa.int/web/gaia/dpac/consortium}). Funding for the DPAC
has been provided by national institutions, in particular the institutions
participating in the {\it Gaia} Multilateral Agreement. This research has made use of the CIRADA cutout service at \url{cutouts.cirada.ca}, operated by the Canadian Initiative for Radio Astronomy Data Analysis (CIRADA). CIRADA is funded by a grant from the Canada Foundation for Innovation 2017 Innovation Fund (Project 35999), as well as by the Provinces of Ontario, British Columbia, Alberta, Manitoba and Quebec, in collaboration with the National Research Council of Canada, the US National Radio Astronomy Observatory and Australia’s Commonwealth Scientific and Industrial Research Organisation.


\bibliographystyle{mnras}
\bibliography{Refs}

\appendix

\section{\textit{Gaia} ADQL Query and Photometric Conversions}\label{app:ADQL}

Below is the ADQL query used to compile the initial sample of \textit{Gaia} sources:

\bigskip
\noindent\makebox[\linewidth]{\rule{\linewidth}{0.1pt}}
\begin{verbatim}
select * from (
    select *,phot_g_mean_mag_absolute - (-0.02704 + 
                0.1424*bp_rp_absolute - 
                0.2156*POWER(bp_rp_absolute,2) + 
                0.01426*POWER(bp_rp_absolute,3)) 
                AS phot_v_mean_mag_absolute from (
            select source_id,
            ra,
            ra_error,
            dec,
            dec_error,
            parallax,
            parallax_error,
            parallax_over_error,
            pm,
            pmra,
            pmra_error,
            pmdec,
            pmdec_error,
            ruwe,
            phot_g_n_obs,
            phot_g_mean_flux,
            phot_g_mean_flux_error,
            phot_g_mean_flux_over_error,
            phot_g_mean_mag,
            phot_bp_n_obs,
            phot_bp_mean_flux,
            phot_bp_mean_flux_error,
            phot_bp_mean_flux_over_error,
            phot_bp_mean_mag,
            phot_rp_n_obs,
            phot_rp_mean_flux,
            phot_rp_mean_flux_error,
            phot_rp_mean_flux_over_error,
            phot_rp_mean_mag,
            phot_bp_rp_excess_factor,
            bp_rp,
            bp_g,
            g_rp,
            radial_velocity,
            radial_velocity_error,
            teff_gspphot,
            teff_gspphot_lower,
            teff_gspphot_upper,
            logg_gspphot,
            logg_gspphot_lower,
            logg_gspphot_upper,
            mh_gspphot,
            distance_gspphot,
            distance_gspphot_lower,
            distance_gspphot_upper,
            azero_gspphot,
            azero_gspphot_upper,
            ag_gspphot,
            ag_gspphot_lower,
            ag_gspphot_upper,
            ebpminrp_gspphot,
            ebpminrp_gspphot_lower,
            ebpminrp_gspphot_upper,
            l,
            b,
            phot_g_mean_mag-ag_gspphot-(
                    5*LOG10(1000*ABS(parallax))-5) 
                    AS phot_g_mean_mag_absolute,
            bp_rp - ebpminrp_gspphot 
                    AS bp_rp_absolute 
                    from gaiadr3.gaia_source 
        AS m) 
    AS mm)
AS mmm
WHERE parallax+parallax_error>0.008
    and parallax-parallax_error<1
    and phot_v_mean_mag_absolute<1.18
    and b>-80
    and b<20
\end{verbatim} 
\noindent\makebox[\linewidth]{\rule{\linewidth}{0.4pt}}

\bigskip

\noindent The magnitude limit was re-applied, along with the color limit, after the initial query to incorporate the \textit{Bayestar19} extinction estimates. From \textit{Bayestar19} we obtain the $E_{B-V}$ color excess and convert it to $E_{G-V}$ using the Johnson-Cousins Relations\footnote{See the \textit{Gaia} DR3 Documentation at \url{https://gea.esac.esa.int/archive/documentation/GDR3/} Section 5.5.1 Table 5.9 for details.}:

\begin{dmath}\label{eq:deriv1}
    E_{G-V} = -0.04749 - 0.0124E_{B-V} - 0.2901E_{B-V}^2 + 0.02008E_{B-V}^3 + 0.04772E_{B-V}^4
\end{dmath}

\noindent and similarly the $G_{BP} - G_{RP}$ excess:

\begin{dmath}\label{eq:deriv2}
    E_{G_{BP}-G_{RP}} = 0.06483+1.575E_{B-V} - 0.7815E_{B-V}^2 + 0.5707E_{B-V}^3 - 0.176E_{B-V}^4 + 0.01916E_{B-V}^5
\end{dmath}

\noindent \citep[e.g.][]{fernie1983relationships,2000PASP..112..925S}. The $V$- and $G$- band extinction are then computed using:

\begin{equation}\label{eq:deriv3}
    A_V \approx 3.2E_{B-V}
\end{equation}

\begin{equation}\label{eq:deriv4}
    A_G \approx A_V + E_{G-V}
\end{equation}

\noindent \citep[e.g.][]{savage1979observed,1979MNRAS.187P..73S}. The $G$-band absolute magnitude is then:

\begin{equation}\label{eq:deriv5}
    M_G = m_G - A_G - (5{\rm log}_{10}(d) - 5)
\end{equation}

\noindent while the absolute color is:

\begin{equation}\label{eq:deriv6}
    G_{BP}-G_{RP} = m_{G_{BP}}-m_{G_{RP}} - E_{G_{BP}-G_{RP}}
\end{equation}

\noindent Finally, we derive the absolute $G-V$ color from the Johnson-Cousins relations:

\begin{dmath}\label{eq:deriv7}
    G-V = -0.027 + 0.01424(G_{BP}-G_{RP}) - 0.2156(G_{BP}-G_{RP})^2 + 0.07426(G_{BP}-G_{RP})^3
\end{dmath}

\noindent which is used to get the absolute $V$-band magnitude:

\begin{equation}\label{eq:deriv8}
    M_V = M_G - (G-V)
\end{equation}

\noindent For source with $\delta \le -30^\circ$, the extinction and color excess from \textit{Gaia} GSP-Aeneas photometry are used as described in Section~\ref{sec:gaia}.

\section{Definition of Association P-values}\label{app:pvals}

In this appendix, we explicitly define p-values for association of magnetars with SNRs, bound companions, and unbound companions from the radio, optical, IR, and \textit{Gaia} searches. The p-values are based on the criteria defined in Sections~\ref{sec:PS1}-\ref{sec:gaia}, and are estimated through Monte Carlo simulations.

\subsection{Optical Search for Bound Companions}\label{app:pvals_PS1}

Recall for the bound companion search, the null hypothesis (\textit{H0}) is that the magnetar and source are not bound in a binary system, and we use the position, distance, and difference between PSF and Kron magnitudes (`$grizy$' for PS1, `$uvgriz$' for SkyMapper) as association criteria. For each source, we draw 1000 samples of the RA and declination of the source and magnetar ($\alpha_{\rm PS1},\delta_{\rm PS1},\alpha_{\rm mag},\delta_{\rm mag}$), and the distance of the magnetar ($d_{\rm mag}$), assuming each are normally distributed. Using $\Delta \alpha = |\alpha_{\rm PS1}-\alpha_{\rm mag}|$, $\Delta \delta = |\delta_{\rm PS1}-\delta_{\rm mag}|$, we compute the search statistic $\hat{g}_1$ as the fraction of samples that fall below the 95$^{\rm th}$ percentiles of folded zero-mean normal distributions with $\sigma_{\alpha} = (\sigma_{\alpha,{\rm PS1}}^2 + \sigma_{\alpha,{\rm mag}}^2)^{1/2}$ and $\sigma_{\delta} = (\sigma_{\delta,{\rm PS1}}^2 + \sigma_{\delta,{\rm mag}}^2)^{1/2}$, respectively. The \textit{H0} distribution of $\hat{g}_1$ is estimated by repeating the simulation 100 times with mean $\Delta \alpha,\Delta \delta$ selected uniformly between $1-5\sigma_{\alpha/\delta}$, and the p-value $p_1$ is taken as the percentile of $\hat{g}_1$ given the resulting $\hat{g}_1$ distribution. 

$p_2$ is computed by a similar process; since the errors on apparent magnitudes are small, we assume they are negligible, and use $g-$ and $r-$band magnitudes to derive the apparent $V-$band magnitude:

\begin{equation}
    m_{V} = m_{r} + 0.005 + 0.462(m_g - m_r) + 0.013(m_g - m_r)^2
\end{equation}

\noindent \citep{tonry2012pan}. From \textit{Bayestar19} we obtain the $B-V$ color excess\footnote{Note that \textit{Bayestar19} reports reddening in arbitrary units and is converted to $E_{B-V}$ by multiplying by $0.995$ as described at \url{http://argonaut.skymaps.info/usage}\citep{schlafly2011measuring}. We proceed assuming this factor is approximately 1, which is accurate to the first decimal place.} along the source's sightline, $E_{B-V}$, for distances between $60\,$pc$-60\,$kpc, converting to the $V-$band extinction with $A_V \approx 3.2E_{B-V}$. If \textit{Bayestar19} extinction estimates are not available, we use the $N_H-A_V$ relation from \citet{predehl1995x} from Table~\ref{table:magnetars}. Then, after samples of the absolute magnitude are drawn from a uniform distribution between $-10 \lesssim M_V \lesssim 3$, linear interpolation is used to obtain a sample of the PS1 source distance, $d_{\rm PS1}$, from the difference $m_V - M_V$. Finally, we use the magnetar distance samples to get samples of $\Delta d = |d_{\rm PS1} - d_{\rm mag}|$, which is again compared to the 95$^{th}$ percentile of a folded zero-mean distribution and \textit{H0} samples between $1-10\sigma_{\Delta d}$ to get the statistic $\hat{g}_2$ and p-value $p_2$.

Finally, the Kron magnitude describes the flux of extended sources within a characteristic Kron radius \citep{kron1980photometry}. As described by \citet{farrow2014pan}, the difference $m_{x,{\rm Kron}} - m_{x,{\rm PSF}}$ can be used to distinguish between point sources and extended sources by imposing a cutoff $m_{x,{\rm cutoff}}(m_{x,{\rm Kron}})$. Point sources are expected to exceed this limit, and we adopt the \citet{farrow2014pan} fit equation for the cutoff:

\begin{dmath}
    m_{x,{\rm cutoff}}(m_{x,{\rm Kron}}) = -0.319 + 0.129(m_{x,{\rm Kron}} - 21) + 0.007(m_{x,{\rm Kron}} - 21)^2
\end{dmath}

\noindent Since the apparent magnitude error is negligible, we set the p-value $p_{3,x} = 0$ if $m_{x,{\rm Kron}} - m_{x,{\rm PSF}} > m_{x,{\rm cutoff}}(m_{x,{\rm Kron}})$ and $p_{3,x} = 1$ if $m_{x,{\rm Kron}} - m_{x,{\rm PSF}} \le m_{x,{\rm cutoff}}(m_{x,{\rm Kron}})$, combining results in all available passbands to get $p_3$:

\begin{equation}
    p_3 = 1 - \prod_{x}^{`uvgrizy'}(1-p_{3,x})
\end{equation}

\noindent For SkyMapper sources, multiple measurements for the Kron and PSF magnitudes are available in each band, each of which is used to compute $p_3$. We additionally exclude SkyMapper sources with stellar classification probability (\textit{class_star}) $<90\%$ to avoid extended objects \citep{wolf2018skymapper,onken2024skymapper}.

\subsection{Infrared Search for Bound Companions}\label{app:pvals_2MASS}

Like the optical search, the IR search uses the position, distance, and PSF $J-$ and $H-$band magnitudes as criteria for association, with \textit{H0} that the magnetar and source are not bound in a binary system. $p_1$ is computed in a similar way to the optical data for UKIDSS and VVV using their respective PSF FWHMs, $1\arcsec$ and $0.75\arcsec$, as errors in RA and declination. For 2MASS sources, the RA and declination are sampled as a bi-normal distribution using the reported major axis ($a_{\rm IR}$), minor axis ($b_{\rm IR}$), and position angle ($\phi_{\rm IR}$) of the error ellipse. The covariance matrix of the bi-normal distribution is given by:

\begin{dmath}
    \sigma_{\rm pos} =\begin{bmatrix}
        (a_{\rm IR}{\rm sin}(\phi_{\rm IR}))^2 + (b_{\rm IR}{\rm cos}(\phi_{\rm IR}))^2 & (a_{\rm IR}^2 - b_{\rm IR}^2){\rm cos}(\phi_{\rm IR}){\rm sin}(\phi_{\rm IR}) \\
        (a_{\rm IR}^2 - b_{\rm IR}^2){\rm cos}(\phi_{\rm IR}){\rm sin}(\phi_{\rm IR}) & (a_{\rm IR}{\rm cos}(\phi_{\rm IR}))^2 + (b_{\rm IR}{\rm sin}(\phi_{\rm IR}))^2 
\end{bmatrix}
\end{dmath}

\noindent We proceed as in the PS1 search to compute the search statistic $\hat{g}_1$ and its p-value $p_1$, utilizing the 95$^{\rm th}$ percentile contour as a threshold and simulating the \textit{H0} distribution of angular separation, $\Delta \theta$, using mean separations between $1-5\sigma_{\rm pos}\{1,1\}$ for RA and $1-5\sigma_{\rm pos}\{0,0\}$ for declination.

For the second condition we follow the same method as for PS1 sources using instead the $J-$\,band and $H-$\,band magnitudes, $m_J$ and $m_H$. Following \citet{chrimes2022magnetar}, we expect OB stars to have $H-$band magnitudes in the range $-2.76\lesssim M_H \lesssim 3.34$ and $J-H$ colors between $-0.14\lesssim J - H \lesssim 0.31$. After drawing samples of $M_H$ and $J-H$ from uniform distributions in these ranges, we compute $M_J = (J-H) + M_H$. The \textit{Bayestar19} extinction library is again queried, using $A_H \approx 0.4690E_{B-V}$ and $A_J \approx 0.7927E_{B-V}$, and the inferred distance is estimated from both the $J-$ and $H-$ bands, $d_{\rm IR,J}$ and $d_{\rm IR,H}$. The search statistic is taken as the fraction of sources for which either $d_{\rm IR,J}$ or $d_{\rm IR,H}$ falls outside the 95$^{\rm th}$ percentile of the \textit{H1} distribution.

\subsection{Radio Search for SNR Shells}\label{app:pvals_RACS}

The radio search uses the position, proper motion, and angular size to define association p-values, which cover two independent cases: first, that the magnetar resides within the SNR shell, and second, that the magnetar was kicked from the center of the SNR. \textit{H0} in both cases is that the magnetar did not originate at the center of the SNR. $p_1$ is derived similarly to the IR case; when the angular size of the SNR is known, we assume the semi-major ($a_{\rm SNR}/2$) and semi-minor axes ($b_{\rm SNR}/2$) are the $3\sigma$ half-widths of the error ellipse. If not, the error ellipse is used to derive the binormal distribution. We again derive the search statistic $\hat{g}_1$ using the 95$^{\rm th}$ percentile contour, and the p-value $p_1$ by comparing to the \textit{H0} distribution of $\hat{g}_1$, which is derived using mean angular separations between $1-5\sigma_{\alpha/\delta}$ of the origin.

The second criteria is similar to that above, but instead compares the angle of the line connecting the magnetar location and the SNR, $\rho_{\rm SNR} = {\rm tan}^{-1}(\Delta \delta / \Delta \alpha)$, and the direction of the magnetar's proper motion, $\rho_{\rm mag} = {\rm tan}^{-1}(\mu_\delta/(\mu_\alpha {\rm cos}(\delta)))$. The \textit{H0} distribution of $\hat{g}_2$ is computed using mean angular differences $\Delta \rho = |\rho_{\rm SNR} - \rho_{\rm mag}|$ between the 95$^{\rm th}$ percentile of the observed distribution and $180^\circ$. $p_3$ uses the magnetar proper motion and the median magnetic field decay age $\mathcal{T}_{\rm decay}$ computed from the X-ray luminosity \citep[see Appendix~\ref{app:decay} and][]{ferrario2008origin} to derive samples of the expected angular separation between magnetar and SNR, $\Delta \theta_{\rm mag} = |\overrightarrow{\mu}|/\mathcal{T}_{\rm decay}$. If the magnetar has no X-ray luminosity or magnetic field estimate, the spin-down age $\mathcal{T}_{\rm rot}$ from the McGill magnetar catalog is used instead. Samples of the true angular separation, $\Delta \theta_{\rm SNR}$ are derived using samples of the RA and declination of the magnetar and SNR. The search statistic $\hat{g}_2$ is the fraction of samples for which the difference in angular separation, $\Delta(\Delta \theta) = |\Delta \theta_{\rm SNR} - \Delta \theta_{\rm mag}|$ is outside the 95$^{\rm th}$ percentile from the \textit{H1} case. $p_3$ is then derived from the \textit{H0} distribution of $\hat{g}_3$, which assumes mean angular separations between the 95$^{\rm th}$ percentile and $1\arcmin$. 

Finally, we estimate $p_4$ based on the angular size $\Delta \theta$ of the SNR; we impose a lower limit on the angular size required to be associated with the magnetar given the magnetar's distance and age $\mathcal{T}_{\rm decay/rot}$. By sampling the magnetar distance $d$ we compute samples of the physical radius of the SNR as $r_{\rm SNR} = d_{\rm mag}\Delta \theta$ and estimate the corresponding dynamical age $\mathcal{T}_{\rm dyn}$ from either free-expansion or Sedov-Taylor phase expansion as described in Appendix~\ref{app:plasma} \citep{sedov2018similarity}. For this we also sample the progenitor envelope mass from a uniform distribution between $2.75-120M_\odot$\footnote{This range implicitly assumes that mass transfer to a binary companion or lost to stellar winds are negligible, which is not necessarily the case \citep{nomoto1995evolution,thielemann1996core}.} and use an explosion energy $10^{51}$\,ergs \footnote{\citet{nomoto1995evolution} suggest that gravitational binding energy $\sim10^{53}$\,ergs can increase the energy of supernovae. However, we use the observational constraint of $\sim10^{51}$\,erg from Type II CCSN lightcurves from \citep{morozova2017unifying}. Since Type II supernovae make up $\sim90\%$ of CCSNe, we find this assumption justified.}\citep{morozova2017unifying}. We then draw samples of the magnetar age $\mathcal{T}_{\rm decay/rot}$ and compute the statistic $\hat{g}_3$ as the fraction of samples for which $\mathcal{T}_{\rm decay/rot} > \mathcal{T}_{\rm dyn}$. We impose only a loose lower limit since magnetar spindown and decay ages are uncertain. The p-value $p_3$ is taken as the percentile of $\hat{g}_3$ within the \textit{H0} distribution which assumes mean ages between $\mathcal{T}_{\rm dyn}$ and the 95$^{th}$ percentile of the \textit{H1} distribution. $p_1$, $p_2$, and $p_3$ are combined using Equations~\ref{eq:pvaltotradio1} and \ref{eq:pvaltotradio2}. 

\subsubsection{`By-Eye' Identification and Fitting of Candidates}\label{app:RACSbyeye}

Optical, infrared, and radio sources identified `by eye' which do not have known counterparts in public catalogs considered here were fit with ellipses parameterized by the RA and declination of the center ($\alpha_c$, $\delta_c$), major and minor axes ($a$, $b$), and position angle measured North of East ($\chi$). Shifting the center of the ellipse to 0, the fit equation for the radius $r$ as a function of polar angle $\theta$ is given by:

\begin{dmath}
    r(\theta) = ab((b^2{\rm cos}^2(\chi) + a^2{\rm sin}^2(\chi)){\rm cos}^2(\theta) + \\ (b^2{\rm sin}^2(\chi) + a^2{\rm cos}^2(\chi)){\rm sin}^2(\theta) + \\ 2(b^2 - a^2){\rm cos}(\chi){\rm sin}(\chi){\rm cos}(\theta){\rm sin}(\theta))^{-1/2}
\end{dmath}

\noindent For each source, the contour at one-half the maximum value within one arcminute (for radio sources; one arcsecond for optical and IR sources) of the identified center. The mean $\alpha$ and $\delta$ were recomputed from this contour prior to fitting with an ellipse using the equation above. The major and minor axes and position angle were then used identically to those of SNRs for the association test (see Section~\ref{app:pvals_RACS}. Note that apparent magnitudes were not estimated for these sources, and p-values could not be estimated for the optical and IR searches. Any point sources of interest were queried in SIMBAD and public catalogs to determine the likelihood of association.

\subsection{\textit{Gaia} Search for Unbound Companions}\label{app:pvals_GAIA}

Criteria for association of unbound \textit{Gaia} sources with magnetars is based on their current positions, proper motion magnitudes and directions, and the consistency of the magnetar travel time with age estimates. \textit{H0} is that the magnetar and source positions are not consistent with having an origin at the same location. For $p_1$, we assume the \textit{Gaia} DR3 proper motion, parallaxes, and positions are Gaussian (or 2-sided Gaussian, where applicable) distributed, with standard deviation equal to the uncertainty. To account for bias in the conversion from parallax to distance ($d\approx\omega^{-1}$) we adopt the photo-geometric distance from the \citet{bailer2023estimating} catalog.

For each magnetar, samples of the distance $d$, RA ($\alpha$) and declination ($\delta$) are drawn from their respective distributions to get samples of the position vector, $\overrightarrow r_{\rm mag,i}$ (the subscript $i$ indicates the quantity is Monte Carlo sampled). Magnetar proper motion measurements are often unavailable due to their reliance on Very Long Baseline Interferometry (VLBI) \citep[e.g][]{tendulkar2012proper}. In such cases, we draw samples of the magnetar velocity magnitude from a Maxwellian distribution with scale parameter $\sigma = 265$\,km\,s$^{-1}$ as described in \citet{2005MNRAS.360..974H}. The velocity direction is drawn from a 3D uniform distribution to obtain $\overrightarrow v_{\rm mag,i}$. If proper motions are known, the total velocity is still drawn from the truncated \citet{2005MNRAS.360..974H} distribution, and the radial velocities are then computed from the velocity and proper motion samples with the sign chosen from a uniform distribution. Samples of the age $\tau_i$, defined here as the time since the magnetar supernova would have occurred, are drawn from a uniform distribution from $1-10$\,kyr. A sub-sample of \textit{Gaia} sources within $5^\circ$ of the magnetar-under-test are selected from the full sample to test for association. In a similar method to that above, samples of position $\overrightarrow r_i$ and velocity $\overrightarrow v_i$ are drawn for each \textit{Gaia} source; here we sample the empirical walk/runaway velocity distribution from the fiducial simulation of \citet{renzo2019massive} for which $\sim95\%$ of simulated ejected massive stars have velocities $\lesssim30$\,km\,s$^{-1}$ \citep[see also][]{blaauw1961origin,tetzlaff2010catalogue,sayers1996factor,feast1965kinematics}. We choose not to utilize the \textit{Gaia} radial velocity measurements since they have not been confirmed through spectroscopic follow-up, and many ($\sim10\%$) of sources within our sample do not have radial velocity measurements. This also prevents us from over-constraining the velocity samples while we impose the \citet{renzo2019massive} distribution.

Next, the magnetar and source trajectories are traced for each sample from their current positions $\overrightarrow r_{\rm mag,i}$ and $\overrightarrow r_i$ to their locations at time $\tau_i$ in the past:

\begin{equation}
    \overrightarrow r_{{\rm o,mag,i}} = \overrightarrow r_{\rm mag,i} - \tau_i \overrightarrow v_{{\rm mag,i}}
\end{equation}

\begin{equation}
    \overrightarrow r_{o,i} = \overrightarrow r_i - \tau_i \overrightarrow v_i
\end{equation}

\noindent We have utilized the \textit{astropy} library's \textit{apply\_space\_coord} method to perform this propagation, and use the \textit{astropy.SkyCoord} object for conversion between Cartesian and Spherical coordinate systems. The initial RA and declination of source and magnetar, $\alpha_{o,i}$,$\delta_{o,i}$ and $\alpha_{o,{\rm mag},i}$,$\delta_{o,{\rm mag},i}$, are derived from $\overrightarrow r_{o,i}$ and $\overrightarrow r_{{\rm o,mag,i}}$, and are used to compute the angular separation at the time of the supernova, $\Theta_i$. We compute this using \textit{SkyCoord.angular\_separation} method, which implements the Vincenty formula for the Great-circle distance between two points on a spherical surface \citep{doi:10.1179/sre.1975.23.176.88}. 
Additionally, using the G-band magnitude for each source, the $G-V$ color derived through Johnson-Cousins relations, and the distance samples which are converted to distance moduli, we also derive samples of the V-band absolute magnitude $M_{V,i}$:

\begin{equation}
    M_{V,i} = (m_G - A_G - (5{\rm log}_{10}(d_i) - 5)) - (G-V) 
\end{equation}

\noindent $\Theta_i$ and $M_{V,i}$ are used as the parameters of interest and compared to a threshold contour in $\{\Theta,M_V\}$ space. The contour is computed by injecting 100 positively-associated simulated sources with an OB-star-like absolute magnitude, $-10 \le M_V \le +3$. The $90^{\rm th}$ percentile contour of the 2D histogram is used as a threshold, and we define the search statistic $\hat f$ as the fraction of Monte Carlo samples falling outside the contour. A separate threshold contour is determined for each magnetar, using the position, distance, and proper motion of the magnetar as available. Critical values $\hat f_c$ were derived through a Receiver Operating Characteristic (ROC) analysis for each magnetar, which is summarized in Appendix~\ref{app:ROC}. By comparing the final $\hat{f}$ to samples in the \textit{H1} hypothesis (obtained during the ROC test), we estimate a p-value, $p_1$, as the percentile of $\hat{f}$ within the \textit{H1} sample.

To further narrow down the candidate \textit{Gaia} sources with $\hat{f} < \hat f_c$, we iteratively compute the true intersection points $\overrightarrow{r}_{\bigotimes,i}$ of the sampled trajectories with those of the magnetars, eliminating any samples with non-physical solutions. We then use the Monte Carlo sampled proper motions and positions to estimate the 2-dimensional travel times for each magnetar-candidate pair:

\begin{equation}
    \tau_{{\rm 2D},i} \approx \frac{\Delta \Phi_{\bigotimes,i}}{\sqrt{\mu_{\alpha,i}^2 {\rm cos}^2(\delta_i) + \mu_{\delta,i}^2}}
\end{equation}

\begin{equation}
    \tau_{{\rm 2D, mag},i} \approx \frac{\Delta \Phi_{\bigotimes,{\rm mag},i}}{\sqrt{\mu_{\alpha,{\rm mag},i}^2 {\rm cos}^2(\delta_{{\rm mag},i}) + \mu_{\delta,{{\rm mag},i}}^2}}
\end{equation}

\noindent where $\Delta \Phi_{\bigotimes}$ is the angular separation between the intersection point and the current coordinate. Since we expect $\tau_{{\rm 2D, mag},i}=\tau_{{\rm 2D},i}$, we estimate a $\chi^2$ statistic comparing the $\tau_{{\rm 2D, mag},i}$ vs. $\tau_{{\rm 2D},i}$ curve to a line with slope 1, taking the p-value $p_2$ as another measure of association. Similarly, we compute the physical travel time for the magnetar assuming it is at the \textit{Gaia} source distance:

\begin{equation}
    \mathcal{T}_{{\rm 3D,mag},i} \approx \frac{|\overrightarrow{r}_{{\rm mag},i} - \overrightarrow{r}_{\bigotimes,i}|}{|\overrightarrow{v}_{{\rm mag},i}|}
\end{equation}

\noindent We expect each magnetar's travel time to be less than its age; if an SNR candidate with an age measurement is confidently associated with the magnetar, we use the SNR age, $\mathcal{T}_{\rm SNR}$. Five magnetars in our sample have   $\mathcal{T}_{\rm SNR}$ estimates, four of which are measured assuming Sedov-Taylor plasma dynamics and the other which is computed kinematically using the magnetar proper motion \citep{sedov2018similarity,dohm1996young,anderson2012multi,blumer2019x,borkowski2017expansion,lyman2022fast,gaensler1999new}. If no SNR association is known or its age is unconstrained, we estimate the age $\mathcal{T}_{\rm decay}$ from from continuum X-ray luminosity measurements using the crustal field decay model \citep[see][and Appendix~\ref{app:decay}]{ferrario2008origin}. This applies to 18 magnetars in our sample.  If no SNR association is known and the X-ray luminosity or surface magnetic field strength are unknown, the spin-down age $\mathcal{T}_{\rm rot}$ reported in the McGill magnetar catalog is used in its place \citep{olausen2014mcgill}. 5 magnetars in the sample require spin-down ages to be used, decreasing confidence in their p-values since $\mathcal{T}_{\rm rot}$ is predicted to over-estimate magnetar ages \citep[e.g.][]{kaspi2017magnetars}. For 5 magnetars, none of the three age estimates are known, and the final p-value is computed from $p_1$ and $p_2$ alone.

For $p_4$, we use samples of the magnetar and source distances from the initial Monte Carlo search to estimate the distribution of the difference, $\Delta d_i=d_i - d_{\rm mag,i}$. The width of this distribution is dominated by the magnetar's distance error. We estimate the \textit{H1} distribution, for which the magnetar and source are at the same distance, by shifting the $\Delta d$ samples to have median $0$. We then compute the search statistic as the fraction of $\Delta d$ samples within the $3\sigma$ width ($0.15-99.85^{\rm th}$ percentiles) of the \textit{H1} distribution. We compare this to 100 realizations of the $H0$ distribution which are shifted to have medians between $1-6\sigma_d$, where $\sigma_d$ is the width of the \textit{H1} distribution, and compute the search statistic for each trial. The p-value $p_4$ is computed as the percentile of the observed $\Delta d$ statistic among the $H0$ distribution of statistics.

\subsubsection{Receiver Operating Characteristic Analysis}\label{app:ROC}

\begin{table}
 \begin{center}
 \small
 \caption{Critical Thresholds for Magnetar Association Tests}
 \label{table:ROC}
 \begin{tabular}{c|c}
 \hline
  \textbf{Magnetar} & $\mathbf{\hat{f}_c}$ \\
  \hline
    CXOU\,J010 & 72.4\%  \\
    4U\,0142 & 5.7\%  \\
    SGR\,0418 & 87.2\%  \\
    SGR 0501 & 91.2\%  \\
    SGR 0526 & 80.4\%  \\
    1E\,1048 & 94.4\%  \\
    1E 1547 & 8.4\% \\
    PSR J1622& 98.5\% \\
    SGR 1627 & 11.4\% \\
    CXOU\,J1647 & 93.4\%  \\
    1RXS\,J1708 & 99.2\%  \\
    CXOU\,J1714 & 96.4\% \\
    SGR\,J1745 & 20.4\% \\
    SGR\,1806 & 2.8\%  \\
    XTE\,J1810 & 74.5\%   \\
    Swift\,1818 & 36.3\%  \\
    SGR\,J1822& 83.3\%  \\
    SGR\,1833 & 90.0\% \\
    Swift\,J1834 & 54.4\% \\
    1E\,1841 & 14.3\% \\
    3XMM\,J1852 &89.3\% \\
    SGR\,1900 & 40.3\%  \\
    SGR\,1935 & 12.4\%   \\
    1E\,2259 & 54.5\%  \\
    \hline
    \textit{SGR\,0755} & 63.7\% \\
    \textit{SGR\,1801} & 11.2\% \\
    \textit{SGR\,1808} & 64.4\% \\
    \textit{AX\,J1818} & 96.7\% \\
    \textit{AX\,J1845} & 91.0\% \\
    \textit{SGR\,2013} & 84.3\%  \\
    \textit{PSR\,J1846}& 39.3\% \\
  \hline
 \end{tabular}
\end{center}
\end{table}

The optimal critical threshold for the \textit{Gaia} Monte Carlo search, $\hat{f}_c$, was derived through an ROC analysis. To compute the true positive rate (TPR) for trial thresholds from 0.1 to 1, 10000 injection sources were simulated with randomly chosen ages between 1-10 kyr and velocity drawn from the fiducial simulation described in \citet{renzo2019massive}. A test magnetar was simulated with velocity drawn from the \citet{2005MNRAS.360..974H}, and both magnetar and source were propagated to present-day positions, starting from the same start position ($P_0 = P_{0,{\rm mag}}$). Each simulated source was given an OB-star-like absolute magnitude $M_V = -1.5$. The Monte Carlo search was conducted with 1000 trials for each simulated source and the simulated magnetar. Search statistics $\hat{f}$ were computed for each source. The TPR for each trial $\hat{f}_c$ was computed as the fraction of sources with $\hat{f} < \hat{f_c}$. The false negative rate (FNR) is computed as FNR = 1-TPR.

To compute the false positive rate (FPR) for a given magnetar location, we tile the region nearby with $10\arcmin$ pointings, starting at $10\arcmin$ from the magnetar. For each tile, we pass all sources through the Monte Carlo association test, using the tile's center as the coordinates of the `fake' magnetar.  Since no magnetar is in the query region, these test samples are used as true negative injections. The FPR for each trial $\hat{f}_c$ was computed as the fraction of total sources (from all tiles) with $\hat{f} < \hat{f_c}$.

We expect differences in each magnetar's local environment, both in source density and, potentially, stellar types. Therefore, the ROC analysis is carried out for each magnetar individually. Note that in most cases, the FPR is very low; the FPR is higher for SGR\,1801-23 because of its lack of distance measure and large position error. We choose the threshold for each magnetar where the TPR is maximized and the FPR is minimized; in cases where the FPR is consistently 0\%, the threshold is $\hat{f}_c = 100\%$. In these cases, a \textit{Gaia} source is a candidate for association if any Monte Carlo sample falls within the $90^{\rm th}$ percentile contour in $\{\Theta,M_V\}$ space. Similarly, in cases where the FNR is consistently 0\% (which occurs when the proper motion, position, and distance are well-constrained), the threshold is $\hat{f}_c = 0\%$. In these cases, a \textit{Gaia} source is a candidate for association only if all Monte Carlo samples fall within the $90^{\rm th}$ percentile contour in $\{\Theta,M_V\}$ space. Overall, we find this has little effect on the results, as most \textit{Gaia} sources have either all or none of their samples within the contour region.  We list $\hat{f}_c$ for each magnetar in Table~\ref{table:ROC}.



\section{False Positive Analyses}

\subsection{\textit{Gaia} Search False Positives}\label{app:GAIAFPs}

In this section we discuss in detail the reasons that the candidates for unbound stellar companions of 3XMM J185246.6+003317 were rejected as false positives.

\subsubsection{3XMM J185246.6+003317}

For 3XMM\,J185246.6+003317, four candidates, \textit{Gaia}\,DR3\,4266311617807818624 (Source\,1), \textit{Gaia}\,DR3\,4266313683745630080 (Source\,2), \textit{Gaia}\,DR3\,42663155048110807488 (Source\,3), and \textit{Gaia}\,DR3\,4258978601120897664 (Source\,4), are identified with $p<5\%$, all requiring the magnetar to have proper motion $\mu_\alpha{\rm cos}(\delta)\sim -5$ and $\mu_\delta \sim -3.5$\,mas\,yr$^{-1}$. For Sources\,1 and 2, the stellar velocities are consistent with walkaway stars with $|\overrightarrow{v}|\approx28^{+8}_{-5}$\,km\,s$^{-1}$ for Source\,1 and $|\overrightarrow{v}|\approx30^{+8}_{-5}$\,km\,s$^{-1}$ for Source\,2; the travel times are consistent with the magnetar's at the $1\sigma$ and $2\sigma$ levels, respectively. Higher velocities more consistent with runaway stars are required for Sources\,3 and 4: $|\overrightarrow{v}|\approx68^{+10}_{-15}$\,km\,s$^{-1}$ and $|\overrightarrow{v}|\approx143^{+6}_{-10}$\,km\,s$^{-1}$, respectively. Without uncertainties on the distance of 3XMM\,J185246.6+00317, it is difficult to evaluate if each source is consistent with the magnetar's distance. However, the magnetar's estimated distance $\sim7.1$\,kpc far exceeds the four Sources' parallax distances, which are between $1-3$\,kpc. 3XMM\,J185246.6+00317's distance was estimated from the nearby SNR Kes\,79 based on their similar hydrogen column densities $N_H\approx1.5\times10^{22}$\,cm$^{-2}$ \citep{2014ApJ...781L..16Z}. This is questionable given that Kes\,79 is confirmed in association with a separate X-ray pulsar, CXOU\,185238.6+004020 \citep{seward2003compact}, although we explore a scenario in which both the pulsar and magnetar may be associated with the SNR in Section~\ref{sec:gaia}. It is also known that some X-ray sources have significant local extinction that can bias the foreground $N_H$ estimate \citep[e.g.][]{chaty2012broadband,predehl1995x}. However, the lack of notable IR emission from 3XMM\,J185246.6+003317 distinguish it from sources with significant local $N_H$ such as HMXBs. Therefore, it is likely that $N_H$ is indeed from the foreground and its similarity to that of Kes\,79 suggest they are at the same distance $\sim7.1$\,kpc. If this is the case, then all five sources are in the foreground and cannot be associated with 3XMM\,J185246.6+003317. A precise proper motion measurement and independent distance measurement would allow more confident conclusions to be drawn.

\subsection{Radio SNR Search False Positives}\label{app:RACSFPs}

In this section we justify why candidates for SNR association with SGR\,0755-2933 were rejected as false positives.

\subsubsection{SGR 0755-2933}


Eight radio sources were identified with $p<5\%$ as potential SNRs associated with SGR\,0755-2933 \citet{richardson2023high} and \citet{doroshenko2021sgr} \citep[see also][]{2016ATel.8943....1S,archibald2013anti,2016ATel.8831....1B}. VLASS image cutouts with $1\arcsec$ pixels were obtained for each candidate, and four are ruled out as point sources. Two other sources are found to be coincident with known stellar objects within $1'$ (an eclipsing binary and a Carbon star). The remaining two sources are each $\sim14\arcsec$ across, or $\sim0.22$\,pc at SGR\,0755-2933's distance of $3.5\pm2$\,kpc. Assuming the free-expansion phase, this corresponds to an SNR age range of $80-300$\,years, which is marginally consistent with magnetar ages in general. If we accept the \citet{doroshenko2021sgr} association of SGR\,0755-2933 with the X-ray source 2SXPS\,J075542.5-293353, we can use the X-ray luminosity $L_X\approx 10^{34}$\,erg\,s$^{-1}$ to estimate $\mathcal{T}_{\rm decay}$ for a typical magnetic field strength $B\sim10^{14}$\,G (see Appendix~\ref{app:decay}). From this method we estimate $\mathcal{T}_{\rm decay}\approx60$\,kyr which is much larger than the free-expansion ages for the two SNR candidates. The detection of a subsequent burst resulting in more precise localization could confidently rule out one source which lies outside the $3\sigma$ error region of SGR\,0755-2933. With the information at hand, we conclude that both sources are likely to be false positives based on the large position uncertainty and decay age estimate.



\subsection{Bound Companion Search False Positives}\label{app:PS1FPs}

In this section we justify why the bound companion candidates for PSR\,J1622-4950 and SGR\,J1822.3-1606 were rejected as false positives.

\subsubsection{PSR J1622-4950}

Through the optical search, the SkyMapper source SMSS\,J162244.99-495055.7 was recovered with p-value $p=1\%$ as a potential bound companion of PSR\,J1622-4950. This magnetar lies along a highly extincted sightline, with $A_V\approx 30.17^{+8.94}_{-7.82}$ from the \citet{predehl1995x} $N_H-A_V$ relation. From the $g-$ and $r-$band apparent magnitudes, we therefore estimate an absolute magnitude $M_V\sim-23.4$, which would be extremely bright for an OB star. From the mass-luminosity relation, this would correspond to a $\sim10^{13}M_\odot$ star, which is unphysical. We also note that the candidate has a large angular separation $\sim3.31\arcsec$, corresponding to $0.14$\,pc at the magnetar's $9\pm1.35$\,kpc distance. This is impractical for the orbit of a neutron star ($\sim1.4M_\odot$) and massive star ($\lesssim120M_\odot$), suggesting an orbital period of $\sim10^{18}$\,years. Therefore, we conclude that this source is most likely a false positive. \citet{chrimes2022new} identify a potential infrared counterpart for PSR\,J1622-4950 using HST data, but this is not recovered in either the optical or IR search in this work. This is consistent with their claim in \citet{chrimes2022magnetar} that this is more likely to be surface emission from the neutron star than a companion star.

\subsubsection{SGR J1822.3-1606}


Three distinct IR sources nearby SGR\,J1822.3-1606 are found to have $p<5\%$: 2MASS\,18221839-1604241 ($p<0.1\%$; UKIDSS\,J182218.39-160424.0, UKIDSS\,J182218.39-160424.1), 2MASS\,18221794-1604259 ($p<0.1\%$; UKIDSS\,J182217.94-160425.9, UKIDSS\,J182217.94-160426.0), and 2MASS\,18221797-1604360 ($p\approx5\%$; UKIDSS\,J182217.97-160436.0, UKIDSS\,J182217.97-160435.9). The first is unlikely based on its angular separation $\sim6.7\arcsec\pm0.7\arcsec$; at the magnetar's distance $1.6\pm0.3$\,kpc, this corresponds to $\sim0.05$\,pc which is too large for a bound orbit. While this could be an unbound companion, it was not recovered in the \textit{Gaia} search, and the absolute $J-H$ color is inconsistent with OB stars at the $5\sigma$ level when placed at the magnetar distance, with $J-H\approx0.85\pm0.10$ \citep{chrimes2022magnetar}. The latter source was identified with $p<5\%$ only for the UKIDSS counterparts ($p=1\%,2\%$), while the 2MASS source had a marginal $p\approx 5\%$. Again, we find it unlikely this is associated with SGR\,J1822 based on its angular separation $\sim9.2\arcsec\pm0.7\arcsec$ and color $J-H\approx1.69\pm0.14$. We conclude that both 2MASS\,18221839-1604241 and 2MASS\,18221797-1604360 are false positives for bound companions.

The second source, 2MASS\,18221794-1604259, is much closer to the magnetar at $\sim0.01\arcsec\pm0.7\arcsec$, or $\sim20$\,au at the magnetar's distance $\sim1.6\pm0.3$\,kpc, though this is still too distant for a bound system. This source was similarly identified as evidence that SGR\,J1822.3-1606 may be a Be star X-ray binary, like SGR\,0755-2933, when initially discovered by e.g. \citet{2011GCN.12260....1H}, \citet{2011GCN.12170....1G}, \citet{2011ATel.3496....1G}. The source's \textit{Gaia} counterpart, \textit{Gaia}\,DR3\,4097826476059022848, has a negative parallax $\omega=-1.2\pm0.6$\,mas which excluded it from the \textit{Gaia} search (see Section~\ref{sec:gaia} and Appendix~\ref{app:ADQL}). Its parallax distance is $d\approx7.0^{+3.1}_{-2.8}$\,kpc, which agrees with the magnetar's distance within $1\sigma$. At the magnetar's distance, the intrinsic magnitude and color are $M_H\approx2.5\pm0.1$ and $J-H\approx1.5\pm0.1$, which are stellar in nature. However, the color is redder than expected for OB stars, and is more characteristic of an evolved supergiant star \citep{chrimes2022magnetar}. Notably, the star's proper motion from \textit{Gaia} is $\mu_\alpha{\rm cos}(\delta)=-5.6\pm0.7$\,mas\,yr$^{-1}$, $\mu_\delta=-8.2\pm0.6$\,mas\,yr. Applying the \textit{Gaia} Monte Carlo test results in a p-value $p=79\%$ due to the significant travel time difference $\Delta \tau_{\rm 2D} \approx 0.3^{+0.7}_{-0.2}$\,kyr required to trace trajectories to a common origin. While a magnetar proper motion measurement would motivate re-analysis, we conclude that the close proximity and companion's significant proper motion make it unlikely that the candidate is associated with SGR\,1822.3-1606.

\section{Magnetar Age Estimation}

Magnetars' magnetic field decay and irregular outbursting and `glitching' behavior make their characteristic spindown ages $\mathcal{T}_{\rm rot} = P(2\dot{P})^{-1}$ overestimates of their true age \citep{nakano2015suzaku,kaspi2017magnetars,olausen2014mcgill,ferrario2008origin}. In this section we provide a brief overview of two alternative estimates based on the X-ray luminosity and the dynamical plasma age of an associated SNR.

\subsection{Derivation of Magnetar Age X-ray Luminosity Decay}\label{app:decay}

\begin{figure*}
 \includegraphics[width=\linewidth]{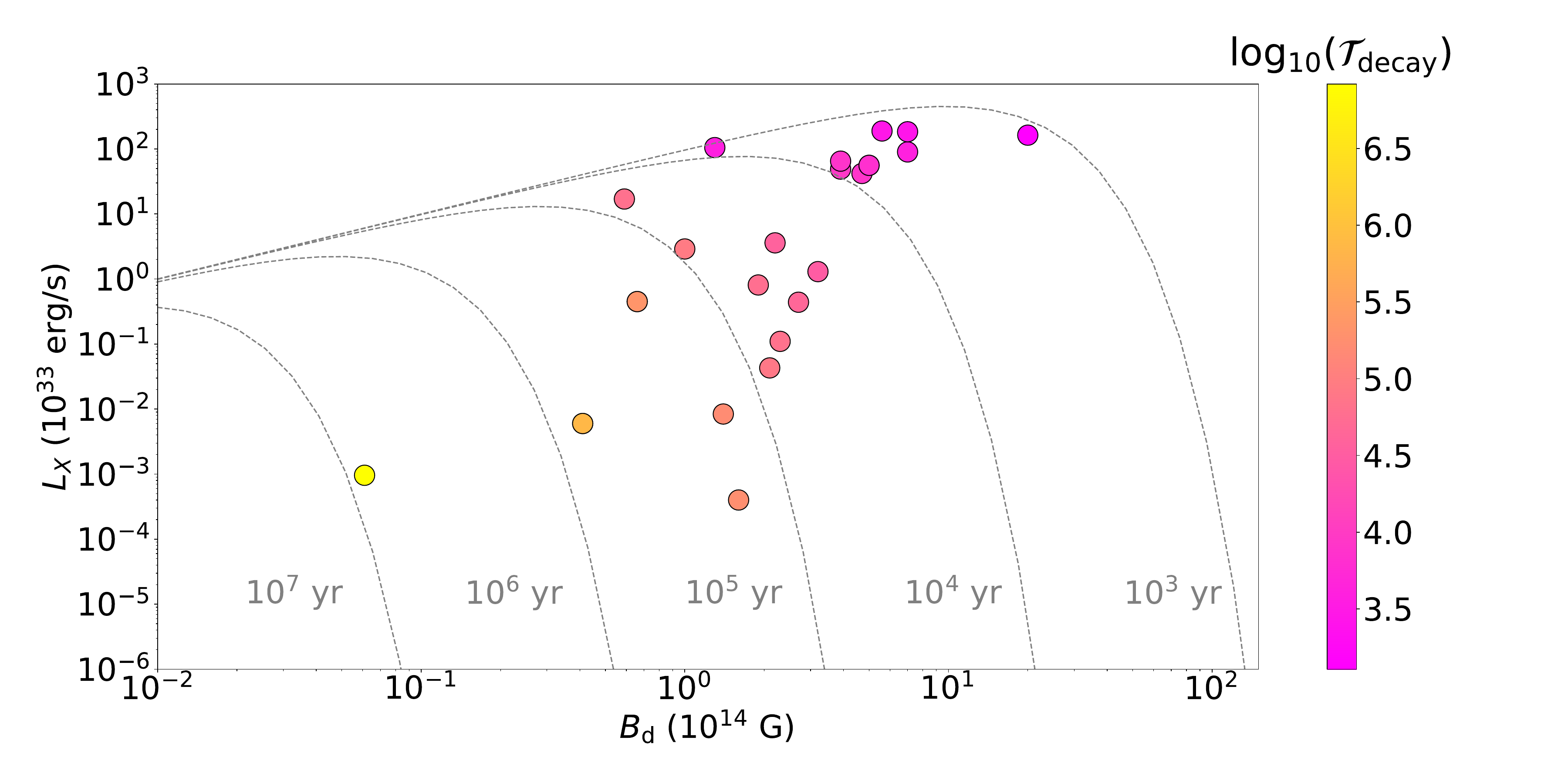}
 \caption{X-ray Luminosity ($L_X$) vs. Dipole Magnetic Field Strength ($B_d$) for Known Magnetars using the Decay Model of \citet{ferrario2006modelling,ferrario2008origin}. The magnetic field estimate is derived from the pulse period $P$ and spindown rate $\dot{P}$ using $B_d\approx(3.2\times10^{19} G)\sqrt{P\dot{P}/1\,{\rm s}}$ as reported in the McGill catalog \citep{olausen2014mcgill}. The X-ray luminosity is derived from the $2-10$\,keV flux and reported distance, also from the McGill catalog. The colorbar corresponds to the decay age $\mathcal{T}_{\rm decay}$ derived from Equation~\ref{eq:age}, and lines of constant age are shown in grey. This estimate is preferred to the spindown age, $\mathcal{T}_{\rm rot}$, which likely overestimates magnetar ages through neglect of field decay and glitching \citep[e.g.][]{nakano2015suzaku,kaspi2017magnetars}. } 
 \label{fig:decay}
\end{figure*}

We define $\mathcal{T}_{\rm decay}$ as the age of the magnetar estimated from the decaying crustal X-ray luminosity, $L_x$. This model is described in detail in \citet{ferrario2008origin} and \citet{ferrario2006modelling} based on \citet{pons2007magnetic}. In this model, the dipole magnetic field component is responsible for spindown energy loss, but does not decay on timescales below $\sim100$\,Myr. The toroidal component decay on timescales $\sim1-10$\,kyr drives energy loss through quiescient crustal X-ray emission $L_X$. We assume the luminosity is modelled by an exponential decay following the initial cooling phase:

\begin{equation}\label{eq:Lx}
    L_{X}(t) = L_{0}\Biggl(\frac{B_d}{10^{13}\,{\rm G}}\Biggr)e^{-t/\tau_d}
\end{equation}

\noindent where $B_d$ is the dipole field strength, $L_0=10^{33}$\,erg\,s$^{-1}$, and $\tau_d$ is the decay timescale. The decay timescale is inversely proportional to the magnetic field, and is modelled as a power law dependence with $\delta=1.3$:

\begin{equation} \label{eq:tau}
    \tau_d = \tau_{d0}\Biggl(\frac{B_d}{10^{13}\,{\rm G}}\Biggr)^{-\delta}
\end{equation}

\noindent where $\tau_{d0}=500$\,kyr. Plugging this into Equation~\ref{eq:tau} and solving for $t=\mathcal{T}_{\rm decay}$, which we take to be the present-day:

\begin{equation}\label{eq:age}
    \mathcal{T}_{\rm decay} = \tau_{d0}\Biggl(\frac{B_d}{10^{13}\,{\rm G}}\Biggr)^{-\delta}\,{\rm ln}\Biggl(\frac{L_X(t=\mathcal{T}_{\rm decay})}{L0}\frac{10^{13}\,{\rm G}}{B_d}\Biggr)
\end{equation}

\noindent where $L_X(\mathcal{T}_{\rm decay})$ is the X-ray luminosity at present-day. The constants $L_0$, $\tau_{d0}$, and $\delta$ were determined through simulations by \citet{ferrario2008origin}. Figure~\ref{fig:decay} shows each magnetar's $\mathcal{T}_{\rm decay}$ as a function of $L_X$ and $B_d$.

\subsection{Derivation of Free-Expansion and Sedov-Taylor Expansion SNR Ages}\label{app:plasma}

In this section, we provide a brief description of the free-expansion and Sedov-Taylor expansion phases for SNRs as are relevant to this paper. We do not offer a detailed derivation, but refer the reader to \citet[][Chapters IV. and V.]{sedov2018similarity}, \citet{truelove1999evolution}, and \citet{suzuki2021quantitative} from which this section is summarized. See also \citet{taylor1950formation} and \citet{sedov1946propagation} for additional details. In this work, we are concerned with the age-radius relations for SNRs, and focus the discussion on this.

The earliest stage of a supernova explosion, which we refer to as `free expansion', is dominated by the momentum of the ejecta mass, $M_{\rm ej}$, rather than mass swept up from the surrounding medium \citep{truelove1999evolution}. This, coupled with the negligible radiative energy losses during this phase, allow the ejecta from the supernova to expand freely into the surrounding medium. During this phase, the radius of the expanding ejecta front is approximately linear with time:

\begin{equation}
    \frac{R_{\rm FE}}{1\,{\rm km}} \approx (7090\,{\rm km/s})\sqrt{\frac{E}{10^{51}\,{\rm ergs}}\frac{M_\odot}{M}}\frac{\mathcal{T}}{1\,{\rm s}}
\end{equation}

\noindent where $E$ is the energy of the supernova and $\mathcal{T}$ is the age of the remnant. The expansion velocity is lower than the sound speed in the surrounding medium, causing a `blast wave' shock to precede the ejecta shell and inducing a reverse shock to oppose the ejecta. This marks the start of the transition to the Sedov-Taylor expansion phase, and occurs roughly on the characteristic timescale:

\begin{equation}
    \mathcal{T}_{\rm ch} = \frac{M^{5/6}}{E^{1/2}(\mu n_{e,0})^{1/3}}
\end{equation}

\noindent where $\mu\approx 1.4m_p$ is the reduced mass of the ejecta and $n_{e,0}$ is the initial electron density, which increases with time through interaction with the ambient medium. The corresponding characteristic radius is

\begin{equation}
    R_{\rm ch} = \Biggl(\frac{M}{\mu n_{e,0}}\Biggr)^{1/3}
\end{equation}

\noindent \citep{truelove1999evolution,sedov2018similarity}.

Beyond this timescale and radius, the system enters the Sedov-Taylor expansion phase, in which the expansion is damped by the ambient medium. On late timescales, the ejecta expands as a power law with $R\propto\mathcal{T}^{2/5}$. To first order, we can form a piecewise model of $R(\mathcal{T})$ by joining the free expansion and Sedov-Taylor solutions at $R_{\rm ch}(\mathcal{T}_{\rm ch})$. Imposing this condition, we can derive the coefficient for the Sedov-Taylor solution:

\begin{dmath}
    \frac{R_{\rm ST}}{1\,{\rm km}} \approx\Biggl(\frac{R_{\rm ch}}{1\,{\rm km}}\Biggr)\Biggl(\frac{\mathcal{T}}{1\,{\rm s}}\Biggr)^{-2/5} =  (7090\,{\rm km/s})\sqrt{\frac{E}{10^{51}\,{\rm ergs}}\frac{M_\odot}{M}}\Biggl(\frac{\mathcal{T}_{\rm ch}}{1\,{\rm s}}\Biggr)^{-3/5}\Biggl(\frac{\mathcal{T}}{1\,{\rm s}}\Biggr)^{-2/5}
\end{dmath}

\noindent which we note is independent of mass \citep{truelove1999evolution,sedov2018similarity}. We use this age estimate as an upper limit on the required age of a candidate SNR when deriving $p_3$ in the radio SNR search (see Section~\ref{app:pvals_RACS}) and when considering the associations of SNRs in Section~\ref{sec:racsresults}. We do not use this approximation for the SNR age estimates in Table~\ref{table:magtimeparams}, and opt only to use $T_{\rm SNR}$ for those with age estimates in the literature.



\section{Past SNR and Neutron Star Companion Search Thresholds}\label{app:pastsurveys}

Here we list references for the previous surveys used for comparison in Figure~\ref{fig:sensitivity}. For past radio SNR searches \citep[see][]{green2019revised}, we include the Canadian Galactic Plane Survey \citep[CGPS;][\textit{bottom}]{gerbrandt2014search,reich1990radio,kothes2006catalogue}, the Parkes Galactic Plane Survey \citep[PGPS;][\textit{bottom left}]{duncan1997supernova}, the Parkes-MIT-National Radio Astronomy Observatory \citep[PMN;][\textit{bottom left}]{stupar2007statistical,condon1993parkes}, the 20\,cm VLA supernova remnant search \citep[VLA92;][\textit{bottom middle}]{1992AJ....104..704S}, the Multi-Array Galactic Plane Imaging Survey \citep[MAGPIS;][\textit{bottom middle}]{helfand2006magpis}, The HI, OH, Recombination line (THOR) and VLA Galactic Plane Survey (VGPS) combined survey \citep[THOR+VGPS;][\textit{bottom middle}]{2016A&A...588A..97B,anderson2017galactic}, and the Molonglo Observatory Synthesis Telescope (MOST) Galactic Plane survey \citep[MGPS;][\textit{bottom right}]{whiteoak1996most}. We include limits from the Naval Observatory Merged Astrometric Dataset \citep[NOMAD;][]{zacharias2004second,zacharias2005vizier} and Tycho-2 \citep[][]{hog2000construction} point source and proper motion catalogs (\textit{middle}), the former of which \citet{kochanek2018cas} use to search for stellar companions of the Crab and Cas A pulsars. We also include limits from the Asteroid Terrestrial-impact Last Alert System (ATLAS) Photometric All-Sky Survey \citep[APASS;][]{} and the HST used by \citet{kochanek2021supernovae} (in conjunction with 2MASS, UKIDSS, PS1, and \textit{Gaia}) and \citet{chrimes2022magnetar}, respectively, to search for pulsar and magnetar companions. 

\bsp	
\label{lastpage}
\end{document}